\documentclass[]{aa}

\newcommand{\myemaila}{c.walsh1@leeds.ac.uk}
\usepackage[fleqn]{amsmath}
\usepackage{amssymb}
\usepackage{url}
\usepackage{gensymb}
\usepackage{graphics}
\usepackage{subfigure}
\usepackage{wasysym}
\usepackage{textcomp}
\usepackage{xcolor}
\usepackage{multirow}
\usepackage[version=3]{mhchem}
\usepackage{txfonts}
\usepackage[draft]{hyperref}
\newcommand{\kms}{km~s$^{-1}$}
\newcommand{\perbeam}{beam$^{-1}$}

\begin{document}

\title{CO emission tracing a warp or radial flow within $\lesssim$100~au in the HD~100546 protoplanetary disk}
\author{Catherine Walsh\inst{1,2} \and
Cail Daley\inst{3} \and 
Stefano Facchini\inst{4} \and
Attila Juh\'{a}sz\inst{5}}

\institute{School of Physics and Astronomy, University of Leeds, Leeds, LS2 9JT, UK \email{\myemaila} \label{1} \and 
Leiden Observatory, Leiden University, P.~O.~Box 9531, 2300~RA Leiden, The Netherlands  \label{2} \and 
Astronomy Department, Wesleyan University, 96 Foss Hill Drive, Middletown, CT 06459, USA \label{3} \and 
Max-Planck-Institut f\"{u}r Extraterrestrische Physik, Giessenbackstrasse 1, 85748 Garching, Germany  \label{4} \and 
Institute of Astronomy, University of Cambridge, Madingley Road, Cambridge CB3 0HA, UK  \label{5}}

\date{Received 07 June 2017 / Accepted 28 September 2017} 

\titlerunning{Kinematics of the HD 100546 disk}
\authorrunning{Catherine Walsh et al.}

\abstract
{We present spatially resolved Atacama Large Millimeter/submillimeter Array
(ALMA) images of $^{12}$CO~$\mathrm{J}=3-2$ emission from the protoplanetary disk
around the Herbig Ae star, HD~100546. 
We expand upon earlier analyses of this data
and model the spatially-resolved kinematic structure of the CO emission. 
Assuming a velocity profile which prescribes a flat or flared 
emitting surface in Keplerian rotation,
we uncover significant residuals with a peak of $\approx7\delta v$, where 
$\delta v =  0.21$~\kms~is the width of a single spectral resolution element.
The shape and extent of the residuals reveal the possible 
presence of a severely warped and twisted inner disk extending to at most 100~au. 
Adapting the model to include a misaligned inner {gas} disk 
with (i) an inclination almost edge-on to the line of sight,  
and (ii) a position angle almost orthogonal to that of the outer disk  
reduces the residuals to $< 3\delta v$. 
However, these findings are contrasted by recent VLT/SPHERE, MagAO/GPI, and VLTI/PIONIER 
observations of HD~100546 that show no evidence of a severely misaligned inner {\em dust} disk 
down to spatial scales of $\sim 1$~au. 
An alternative explanation for the observed kinematics are fast radial flows mediated 
by (proto)planets. 
Inclusion of a radial velocity component at close to free-fall speeds and 
inwards of $\approx 50$~au results in residuals of $\approx 4 \delta v$.  
Hence, the model including a radial velocity component only does not reproduce the 
data as well as that including a twisted and misaligned inner gas disk. 
Molecular emission data at a higher spatial resolution (of order 10~au) 
are required to further constrain the kinematics within $\lesssim 100$~au.
HD~100546 joins several other protoplanetary disks 
for which high spectral resolution molecular emission shows that 
the gas velocity structure cannot be described by a purely Keplerian 
velocity profile with a universal inclination and position angle.
Regardless of the process, the most likely cause is the presence of 
an unseen planetary companion.}


\maketitle

\section{Introduction}
\label{introduction}

Observations of protoplanetary disks around nearby young stars offer unique 
insight into the initial conditions of planetary system formation.  
Resolved continuum observations spanning optical to cm wavelengths  
reveal the spatial distribution of dust across a range of grain sizes, 
which in turn, can highlight signposts of ongoing planet formation 
and/or as yet unseen massive companions/planets (e.g., cavities, gaps, rings, 
and spirals; see the recent reviews by 
\citealt{espaillat14}, \citealt{andrews15}, and \citealt{grady15}).  
Likewise, spectrally and spatially resolved observations of molecular 
line emission disclose the spatial distribution and excitation of various 
gas species, from which information on disk gas properties can be extracted
\citep[e.g.,][]{dutrey14a,sicilia-aguilar16}.

Second only to \ce{H2} in gas-phase molecular abundance, CO is a powerful
diagnostic of various properties including the disk gas mass, radial surface density,
and temperature. 
The primary isotopologue, $^{12}$CO, is optically thick and thus emits 
from the warm disk atmosphere; this allows the gas temperature 
in this region to be derived \citep[e.g.,][]{williams11,dutrey14a}.  
The rarer isotopologues ($^{13}$CO, C$^{18}$O, C$^{17}$O and $^{13}$C$^{18}$O) 
have progressively lower opacities and so enable penetration 
towards and into the disk midplane 
(see, e.g., recent theoretical studies by \citealt{bruderer13}, 
\citealt{miotello16}, and \citealt{yu16}). 
In observations with sufficiently high spatial resolution, now routine with ALMA, 
this allows a direct determination of the location of the CO snowline 
with high precision \citep[see, e.g.,][]{nomura16,schwarz16,zhang17}.  
However, it has been demonstrated recently that chemistry, in particular 
isotope-selective photodissociation \citep{visser09}, can complicate 
the extraction of disk gas masses from CO isotopologue emission 
\citep{miotello14,miotello16}.
Chemical conversion of CO to a less volatile form, e.g., \ce{CO2}, 
complex organic molecules, or hydrocarbons,  
is an alternative explanation for apparently low disk masses derived from 
CO observations \citep{helling14,furuya14,reboussin15,walsh15,eistrup16,yu17}.  

Because emission from $^{12}$CO (and often $^{13}$CO) at (sub)mm wavelengths 
is bright, it has historically been used as a tracer of disk kinematics allowing 
a dynamical determination of the mass of the central star \citep[e.g.,][]{simon00}.  
However, gas motion can deviate from that expected due to 
Keplerian rotation alone because of a variety of different physical effects 
that can be inferred from spatially-resolved observations. 
These include spiral density waves, a substantial (and thus measurable) 
gas pressure gradient, radial flows mediated by accreting planets 
across cavities, or a disk warp 
\citep[see, e.g.,][]{rosenfeld12,tang12,casassus13,dutrey14b,rosenfeld14,christiaens14,casassus15}.  
Spirals, radial flows, and warps can all signify the presence 
of (potentially massive) planetary companions; 
hence, perturbations from Keplerian motion traced in 
bright and spectrally- and spatially-resolved CO emission 
may expose unseen planets.  

Here, we present high signal-to-noise and spectrally-resolved 
ALMA Cycle 0 images of $^{12}$CO $\mathrm{J}=3-2$ emission from the protoplanetary 
disk around the nearby Herbig~Ae star, HD~100546.  
The HD~100546 disk has been proposed to host (at least) two 
massive companions \citep[see, e.g.,][]{acke06,quanz13,walsh14}. 
However, recent MagAO/GPI observations presented in \citet{rameau17} 
have raised doubt on the previous identification of the point source at 50~au 
as a (proto)planet by two previous and independent groups \citep{currie15,quanz15}.
In \citet{walsh14}, henceforth referred to as Paper I, 
we presented the $^{12}$CO ($\mathrm{J}=3-2$, $\nu=345.796$~GHz) first moment map 
and dust continuum emission (at 302 and 346~GHz).  
These data spatially resolved the CO emission and allowed direct 
determination of the radial extent of the molecular disk ($\approx 390$~au; 
see also \citealt{pineda14}). 
The continuum data analysed in Paper I showed that the (sub)mm-sized dust grains 
had been sculpted into two rings. 
\citet{pinilla15} showed that this 
dust morphology is consistent with dust trapping by two massive companions: 
one with mass $\approx 20 M_\mathrm{J}$ at 10~au, and one with mass 
$\approx 15 M_\mathrm{J}$ at 70~au.  
Hence, the ALMA data support the presence of an outer (proto)planet. 
Emission from $^{12}$CO ($\mathrm{J}=3-2$, $6-5$, and $7-6$) from HD~100546 
had been detected previously in single-dish observations with APEX \citep{panic10}.    
The APEX data revealed an asymmetry in the red and blue peaks 
in the double-peaked line profiles most apparent in the $\mathrm{J}=3-2$ 
and $6-5$ transitions. 
\citet{panic10} hypothesised that the asymmetry may arise 
due to shadowing of the outer disk by a warp in the inner disk.  
Using the same data set as here, \citet{pineda14} showed that the 
position-velocity diagram across the major axis of the disk is 
better described by a disk inclination of $\approx 30\degree$, 
rather than an inclination of $44\degree$ that best reproduces the 
aspect ratio of the disk as seen in continuum emission (Paper I).  

In this work we revisit the HD~100546 ALMA Cycle~0 data and conduct a deeper analysis 
of the spatially and spectrally resolved $^{12}$CO $\mathrm{J}=3-2$ emission.  
The focus of this work is the search for evidence of a warp in the inner 
regions of the disk, as suggested by the single dish data presented in 
\citet{panic10}. 
In Sect.~\ref{almaimages}, we outline the imaging presented 
in the paper, and in 
Section~\ref{modelkinematics} we describe the modelling techniques 
used and present the results.  
Sections~\ref{discussion} and \ref{conclusion} discuss the implications 
and state the conclusions, respectively.  

\section{ALMA~imaging~of~HD~100546 }
\label{almaimages}

HD~100546 was observed with ALMA on 2012 November 24 with 
24 antennas in a compact configuration, with baselines 
ranging from 21 to 375~m.  
The self-calibrated and phase-corrected 
measurement set, produced as described in Paper I, 
is used in these analyses.  
In this work, we adopt the revised distance to 
HD~100546 determined by Gaia \citep[$109\pm4$ pc,][]{gaia16a,gaia16b}, 
and a stellar mass of $2.4 M_{\odot}$, \citep{vandenancker98}

In Paper I, the integrated intensity
and first moment maps from the $^{12}$CO $\mathrm{J}=3-2$ rotational 
transition at 345.796~GHz ($E_\mathrm{up} = 33.19~\mathrm{K}$ 
and $A_\mathrm{ul} = 2.497\times 10^{-6}~\mathrm{s}^{-1}$) were presented.  
The data cube from which those maps were produced was itself produced 
using the CASA task \texttt{clean} with Briggs weighting (robust=0.5) 
at a spectral resolution of 0.15~\kms.  
The resulting channel maps had an rms noise of 19 mJy~\perbeam~channel$^{-1}$ 
and a synthesised beam of $0\farcs95\times0\farcs42~(38\degree)$.  
The $^{12}$CO was strongly detected with a signal-to-noise ratio 
(S/N) of 163 in the channel maps.   

\begin{figure*}[!t]
\includegraphics[width=\textwidth]{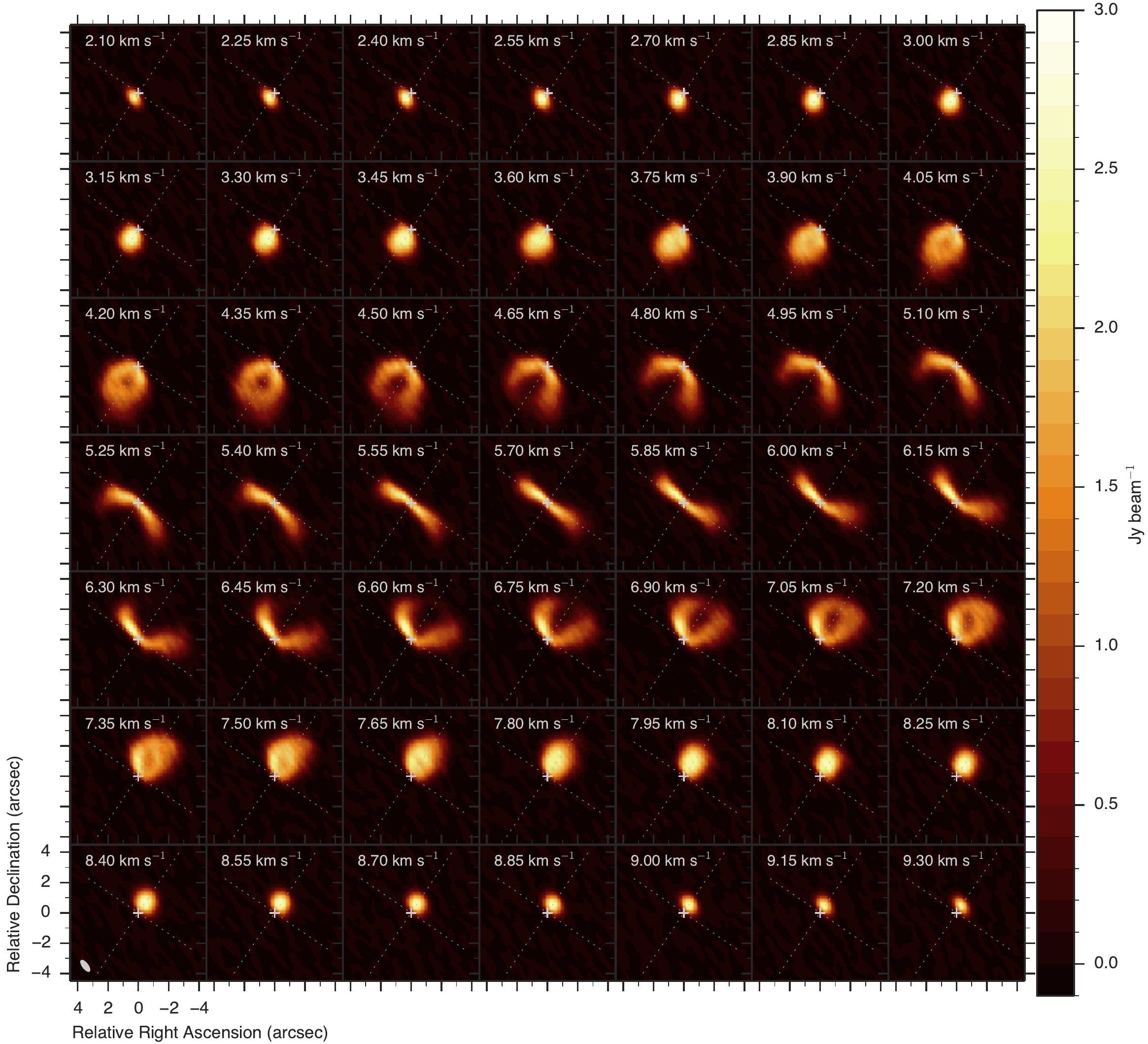}
\caption{Channel maps of the CO $\mathrm{J}=3-2$ line emission imaged at a 
velocity resolution of 0.15~\kms. 
Note that this is slightly over-sampled with respect to the native spectral 
resolution of the data (0.21~\kms). 
The dashed lines 
represent the disk major and minor axes determined from analysis of the 
continuum \citep{walsh14}.} 
\label{figure1}
\end{figure*}

Because of the high S/N the imaging is redone here 
using uniform weighting which results in a smaller beam (and improved  
spatial resolution) at the expense of sensitivity.  
The resulting channel maps have an rms noise of 26 
mJy~beam$^{-1}$~channel$^{-1}$, a S/N of 106, and 
a synthesised beam of $0\farcs92\times0\farcs38~(37\degree)$.   
The maps were created using a pixel size of 0\farcs12 to 
ensure that the beam is well sampled.
Figure~\ref{figure1} presents the channel maps.  
Emission is detected ($\ge 3\sigma$) across 111 channels: the central channel 
is centred at the source velocity of 5.7~\kms~as constrained 
previously by these data (see Paper I).  
The highest velocity emission detected is $\pm 8$~\kms relative 
to the source velocity.
Given that the disk inclination (as constrained by the outer disk) 
is 44\degree~and that the stellar mass is $2.4 M_\odot$, 
emission is detected down to a radius of 16~au from the central star. 
Using the estimate of $\approx 30\degree$ for the inclination of the 
inner disk from \citet{pineda14}, reduces this radius to 8~au.  

\begin{figure*}[!t]
\includegraphics[width=\textwidth]{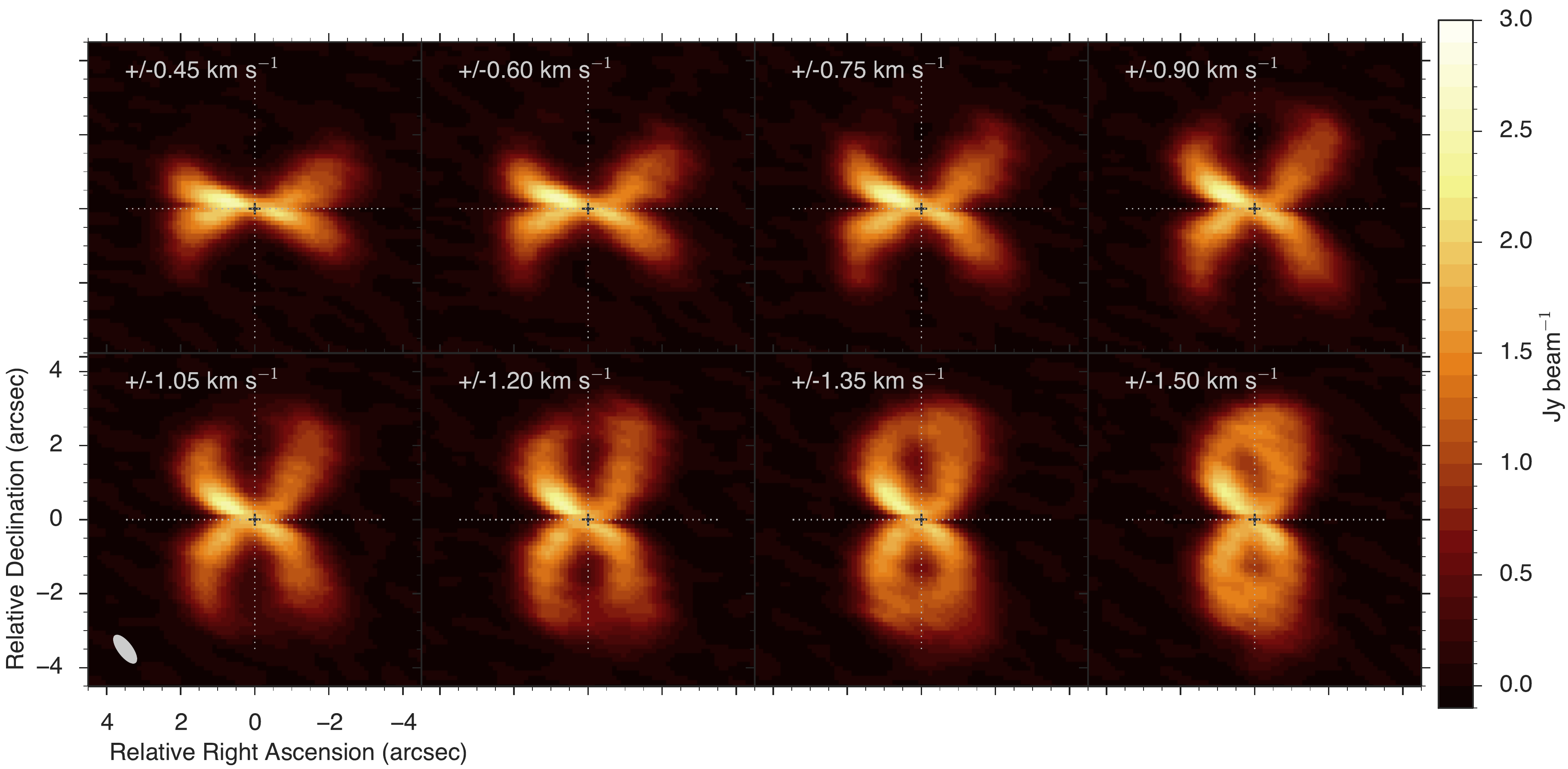}
\caption{Channel maps of the CO $\mathrm{J}=3-2$ line emission rotated in the counter-clockwise direction 
by 34\degree~to align the disk major axis in the vertical direction, and mirrored across the disk minor axis 
to highlight the asymmetry in emission across the disk minor axis (now orientated in the horizontal direction).} 
\label{figure2}
\end{figure*}

The channel maps in Fig.~\ref{figure1} reveal the classic ``butterfly'' morphology of 
spectrally- and spatially-resolved line emission from an inclined 
and rotating protoplanetary disk \citep[see, e.g.,][]{semenov08}.  
Compared with the resolved $^{12}$CO emission from 
the disk around the Herbig Ae star HD~163296 which has a similar inclination, 
there is no evidence of emission from the back side of the disk that is a 
signature of CO freezeout in the disk midplane coupled with 
emission from a flared surface \citep{degregorio13,rosenfeld13}.
The blue-shifted emission also appears mostly symmetric about the disk major axes  
(especially that from the south-east of the disk) 
indicating that it arises from a relatively ``flat'' surface.   
This is in contrast with emission from the disk around the Group I 
Herbig Ae disk HD~97048 \citep{walsh16,vanderplas17}.  
However, the emission is not wholly symmetric about the disk {\em minor} axis, 
with the red-shifted emission from the north-west quadrant appearing both 
fainter, and with a positional offset, relative to blue-shifted 
emission at the same velocity.  
In Fig.~\ref{figure2} the channel maps from $\pm 0.45$ to $\pm 1.5$~\kms~are 
shown, now rotated counter-clockwise by 34\degree~($180\degree - $P.A.) 
to align the disk major axis in the vertical direction, and 
mirrored across the disk minor axis.  
Exhibiting the data in this velocity range and in this manner highlights 
the described asymmetry in brightness across the disk minor axis, the flatness of the emission, and 
the positional offset of the red-shifted north-west lobe relative to 
its blue-shifted counterpart.  
The brightest lobe to the north east, an apparent CO ``hot spot'', is 
consistent in position angle with the proposed (proto)planetry companion 
seen in direct imaging \citep{quanz13,currie15,quanz15,rameau17}.  
Further, we have also recently detected emission from SO from HD~100546 
which has multiple velocity components.  
A clear blue-shifted component (-5 \kms~with respect to the source velocity) 
is coincident in position angle with both the CO ``hot spot'' and 
the proposed protoplanet which we 
attribute to a potential disk wind (see \citealt{booth17} for full details).     

\begin{figure*}[!ht]
\centering
\subfigure{\includegraphics[width=0.45\textwidth]{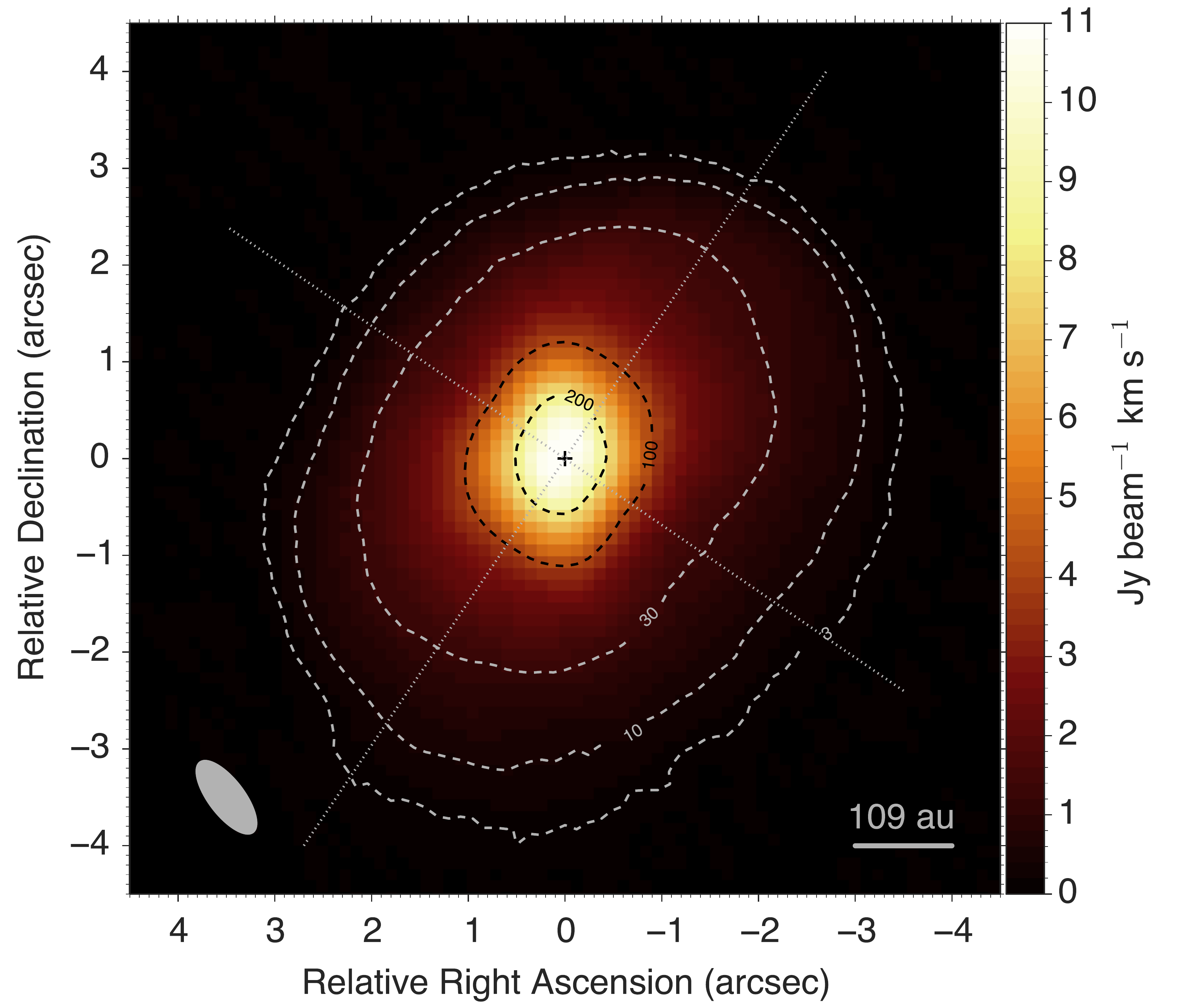}}
\subfigure{\includegraphics[width=0.45\textwidth]{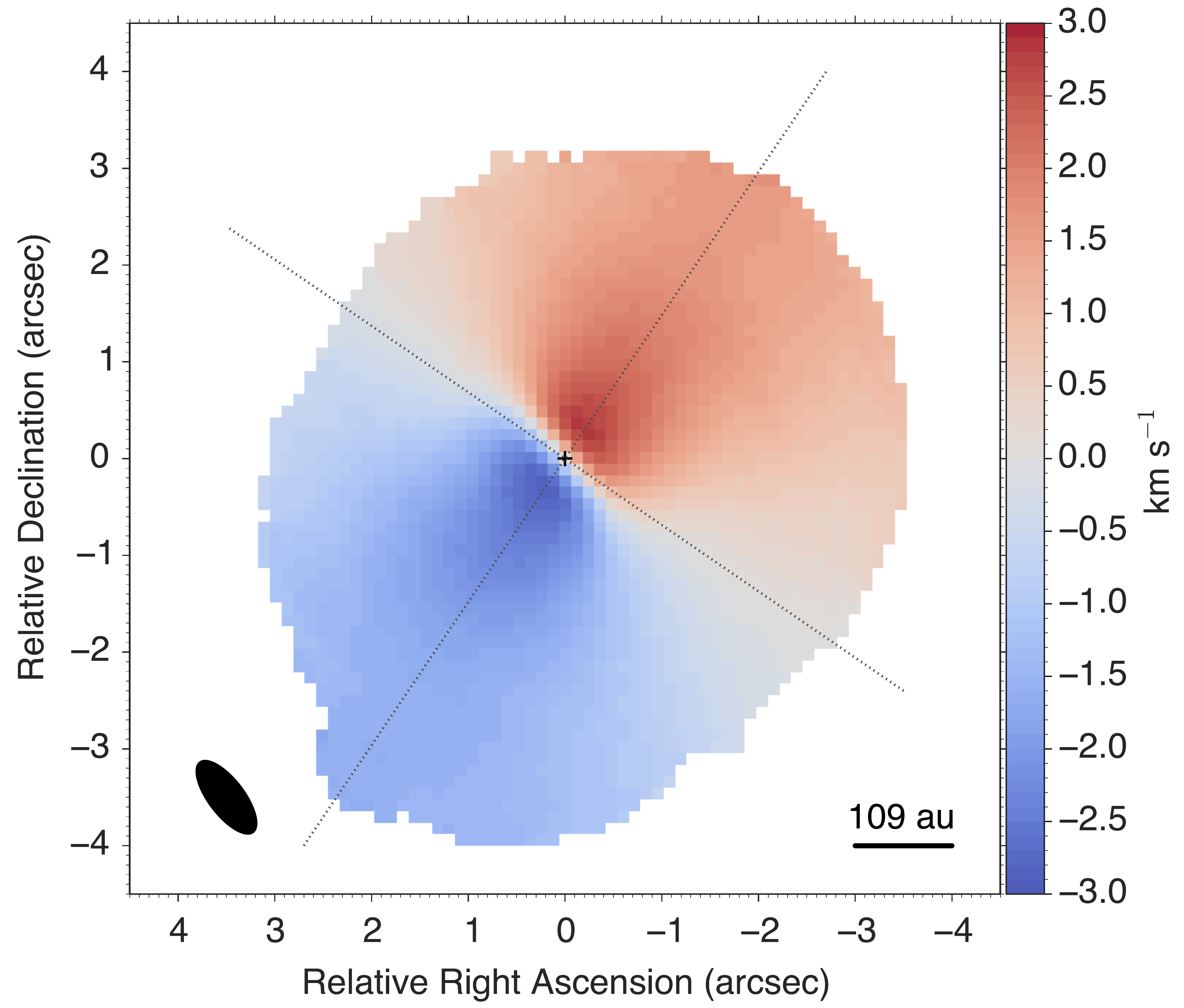}}
\subfigure{\includegraphics[width=0.45\textwidth]{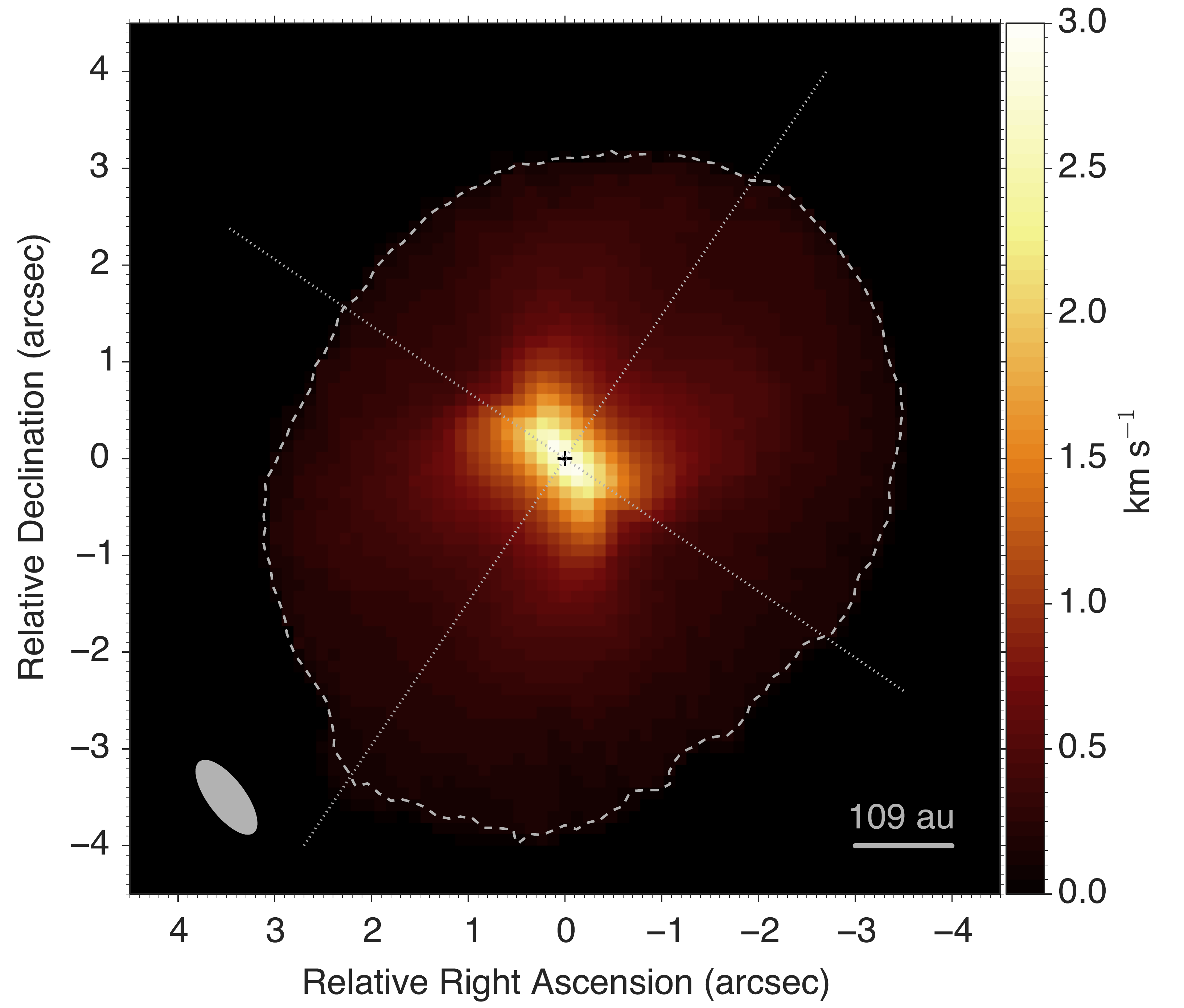}}
\subfigure{\includegraphics[width=0.45\textwidth]{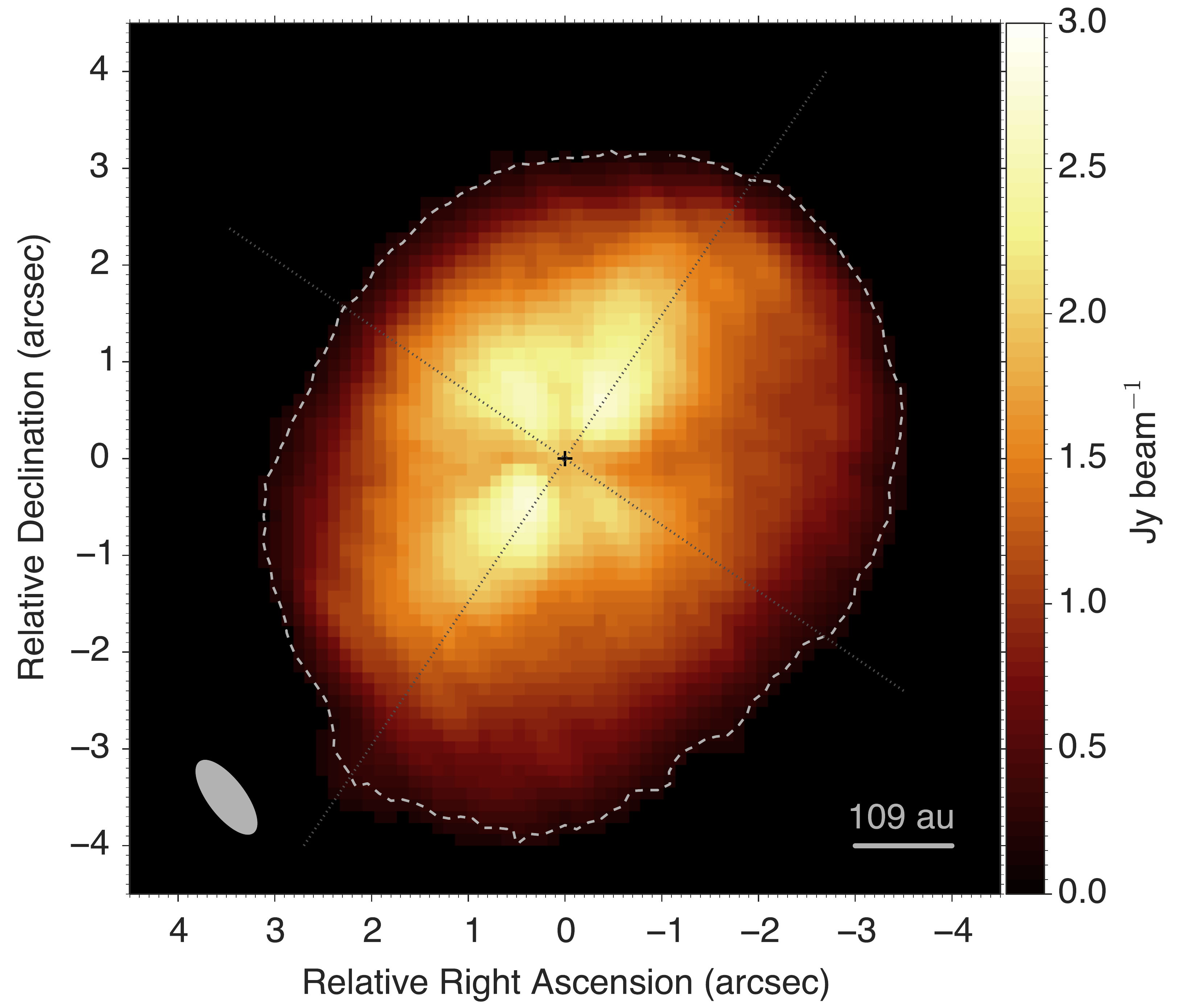}}
\caption{Moment maps for CO $\mathrm{J}=3-2$ line emission from HD 100546.  Clockwise from  top 
left: zeroth moment map (integrated intensity, Jy~\perbeam~\kms), first moment map (intensity-weighted 
velocity, \kms), eighth moment map (peak intensity, Jy~\perbeam), and second moment map 
(intensity-weighted velocity dispersion, \kms).  
The dashed contour in the second and eighth moment maps corresponds to the $3\sigma$ contour 
of the integrated intensity.} 
\label{figure3}
\end{figure*}

Figure~\ref{figure3} presents the moment maps 
(zeroth, first, second, and eighth).  
The zeroth moment map (integrated intensity) was produced using 
a $3\sigma$ rms noise clip.  
The first (intensity-weighted velocity), second (intensity-weighted velocity dispersion), 
and eighth (peak flux density) moment maps were produced using a more 
conservative clip of $6\sigma$.  
The integrated intensity appears relatively symmetric about the 
disk minor axis; however, the $^{12}$CO integrated emission extends 
further to the south-west than it does to the north-east.  
The asymmetry across the disk major axis is also evident in the eighth-moment map with the 
north-east side of the disk appearing brighter than the south-west side 
with a dark lane in the east-west direction.  This dark lane coincides
in position angle with a dark ``wedge'' seen in scattered 
light images with VLT/SPHERE \citep{garufi16}.
Both maps hint at emission from a flared disk which would lead to an 
asymmetry in integrated emission across the disk major axis (i.e., the axis of inclination).   
The ALMA data provide additional evidence that the 
far side of the disk lies towards the north east \citep{garufi16}.  
A flared disk possesses a geometrical 
thickness which increases with radius and leads to an extension in emission along the 
direction towards the near side of the disk and arising from emission from the disk outer ``edge''.
The first and second moment maps also hint at asymmetric emission, in particular, 
the emission at the source velocity through the inner disk is twisted relative to the 
disk minor axis determined from the continuum emission.  
The velocity dispersion in the inner disk is also not wholly symmetric across 
the disk minor axis.  
A ``by-eye'' inspection of the first- and eighth-moment maps, 
in particular, suggest the possible presence of a warp in the inner disk.  

\section{Modelling the kinematics}
\label{modelkinematics}

Modelling of the kinematics as traced by the $^{12}$CO emission is conducted 
using analytical models which describe the line-of-sight projected velocity.  
The model moment maps are convolved with the synthesised beam of the observations.  
The residuals ($\mathrm{data} - \mathrm{model}$) are in units of a single spectral 
resolution element ($\delta v = 0.21$~\kms).
This discretisation is necessary because features smaller than the native spectral 
resolution of the data cannot be fit. 
Note that this analysis will not address the asymmetry in CO brightness across 
the disk seen in the channel maps and zeroth and eighth moment maps.  
We leave such an analysis to future work when data for multiple 
CO transitions and isotopologues allow a robust extraction of 
the gas surface density and gas temperature.  

We explore several metrics of ``best fit'':
(i) the total number of pixels for which the analytical and smoothed 
projected velocity reproduces the data within one spectral resolution element,
(ii) the sum of the square of the residuals scaled by the total number of unmasked pixels, and 
(iii) the magnitude of the peak residual. 
The total number of unmasked pixels in the observed first moment map is 2562.  
The synthesised beam of the observed images corresponds to 
38 pixels in area; hence, distinguishment between models is possible for features larger 
than $\approx 0.5 \times$ the beam area or 0.75\% of the total number of pixels. 
Given the relatively small number of parameters for each model considered, 
the modelling approach is grid based, i.e. all possible grid combinations are explored. 

\subsection{A flat emitting surface}
\label{flat}

The simplest prescription for describing the first moment map of spectrally-resolved 
line emission from a disk 
is axisymmetric emission arising from a geometrically flat surface inclined to the line of sight.  
Assuming that the position angle of the disk is aligned with the 
y axis, the projected velocity on the sky relative to the observer is described by
\begin{equation}
v(x',y') = \sqrt{\frac{G M_\star}{\rho}} \sin i \sin \theta, 
\label{velocity}
\end{equation}
where $G$ is the gravitational constant, $M_\star$ is the mass of the central 
star, $\rho=\sqrt{x^2+y^2}$ is the radius, $i$ is the inclination, and 
$\theta = \arctan(y/x)$ \citep[e.g.,][]{rosenfeld13}.  
In this projection and for this particular orientation, 
$x = x'/\cos i$, $y = y'$, and $z = 0$. 
Model first moment maps for a flat disk with the same P.A.~as 
HD~100546 and inclinations of 30\degree, 45\degree, and 
60\degree, are shown in Fig.~\ref{figurea1} in the Appendix.  

The wide range of disk inclinations ($[20\degree,60\degree]$) and disk 
position angles ($[120\degree,170\degree]$) explored are motivated by previous analyses 
of the continuum data which suggested a P.A.~of $146\degree \pm 4\degree$  
and an inclination of $44\degree \pm 3\degree$ \citep[see Paper I and][]{pineda14}.  
Using the same CO dataset as here, \citet{pineda14} suggest that the inner disk may 
be better described with an inclination of $\approx 30\degree$; hence, we extend our explored range 
accordingly to ensure good coverage over the parameter space.  
First, a coarse grid with a resolution of 5\degree~is run over the full parameter space, 
followed by a zoomed in region with a resolution of 1\degree.

\begin{figure*}[]
\subfigure{\includegraphics[width=0.33\textwidth]{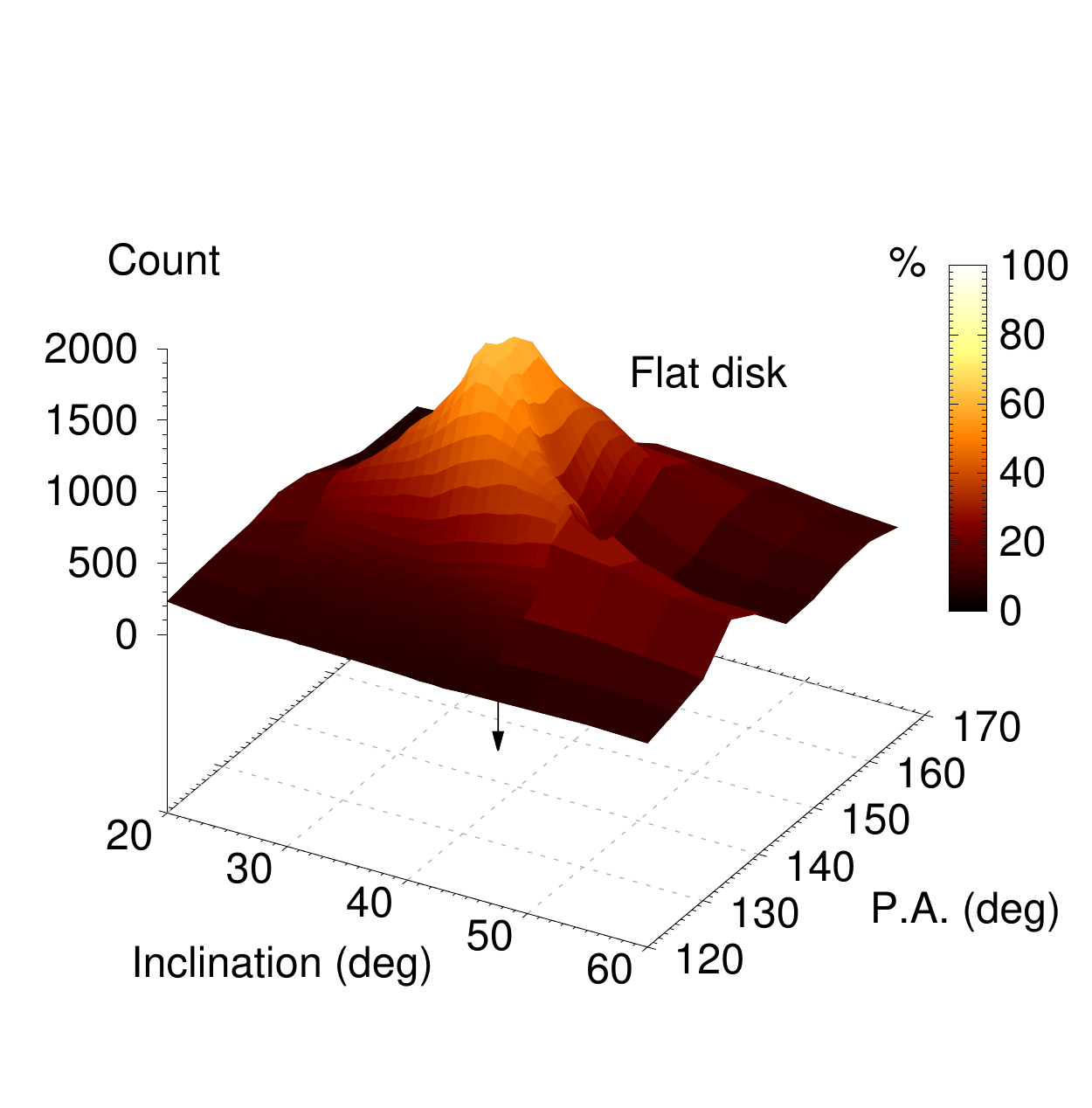}}
\subfigure{\includegraphics[width=0.33\textwidth]{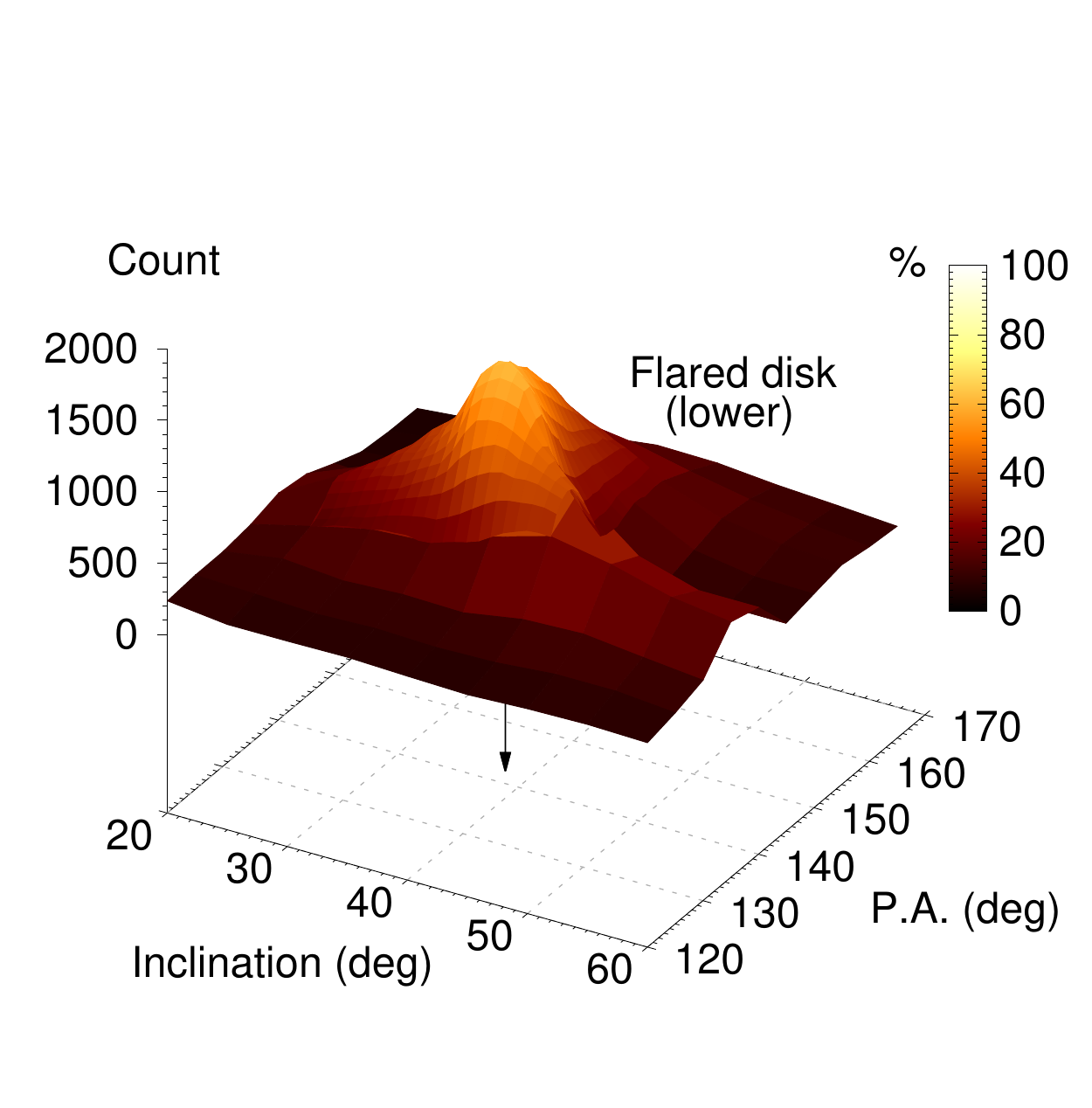}}
\subfigure{\includegraphics[width=0.33\textwidth]{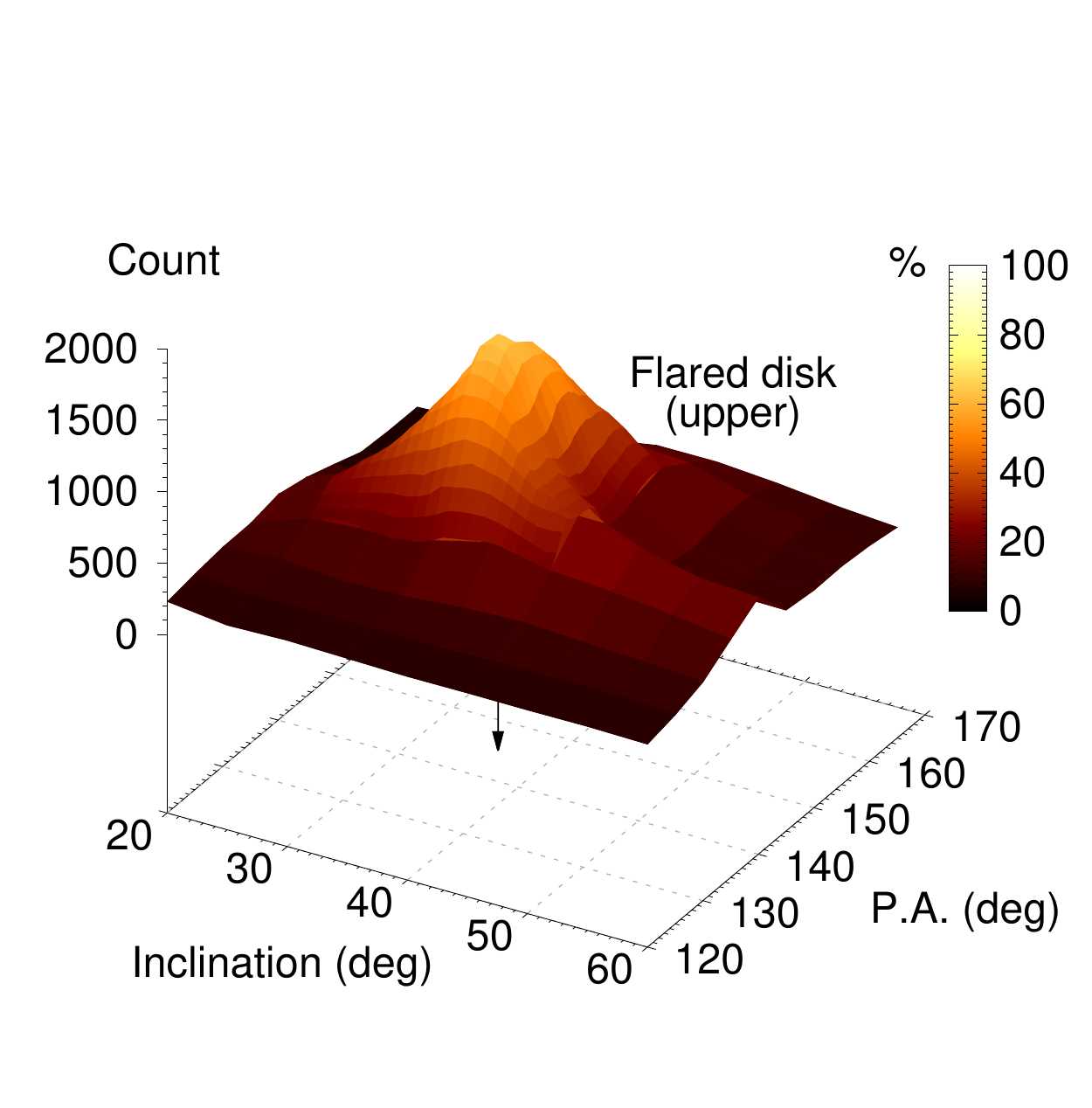}}
\caption{Distribution of model best-fit values using metric (i) as a function of 
inclination and position angle for the best-fit flat disk, flared disk (lower cone), 
and flared disk (upper cone), respectively. 
The best-fit opening angles, $\alpha$, of the flared disks (with respect to the 
disk midplane) are 13\degree~and 9\degree~for the lower and upper cones, respectively. 
The percentage scale corresponds to the full range of pixel values (from 0 to 2652).} 
\label{figure4}
\end{figure*}

The top-left panel of Fig.~\ref{figure4} presents a 3D plot showing the 
total number of pixels which fit the data velocity field within one spectral resolution 
element, $\delta v$, as a function of disk inclination and position angle. 
The distribution is strongly peaked: the best-fit flat disk model using this 
metric has an inclination of 36\degree~and a P.A.~of 145\degree~with 62.1\% of model pixels 
lying within one spectral resolution element of the data.  
These data are also listed in Table~\ref{table1}.  
The P.A.~is in excellent agreement with that derived from the continuum observations.  
The inclination, on the other hand, is lower and closer 
to the suggested inclination from \citet{pineda14}.

\begin{figure*}[]
\subfigure{\includegraphics[width=0.33\textwidth]{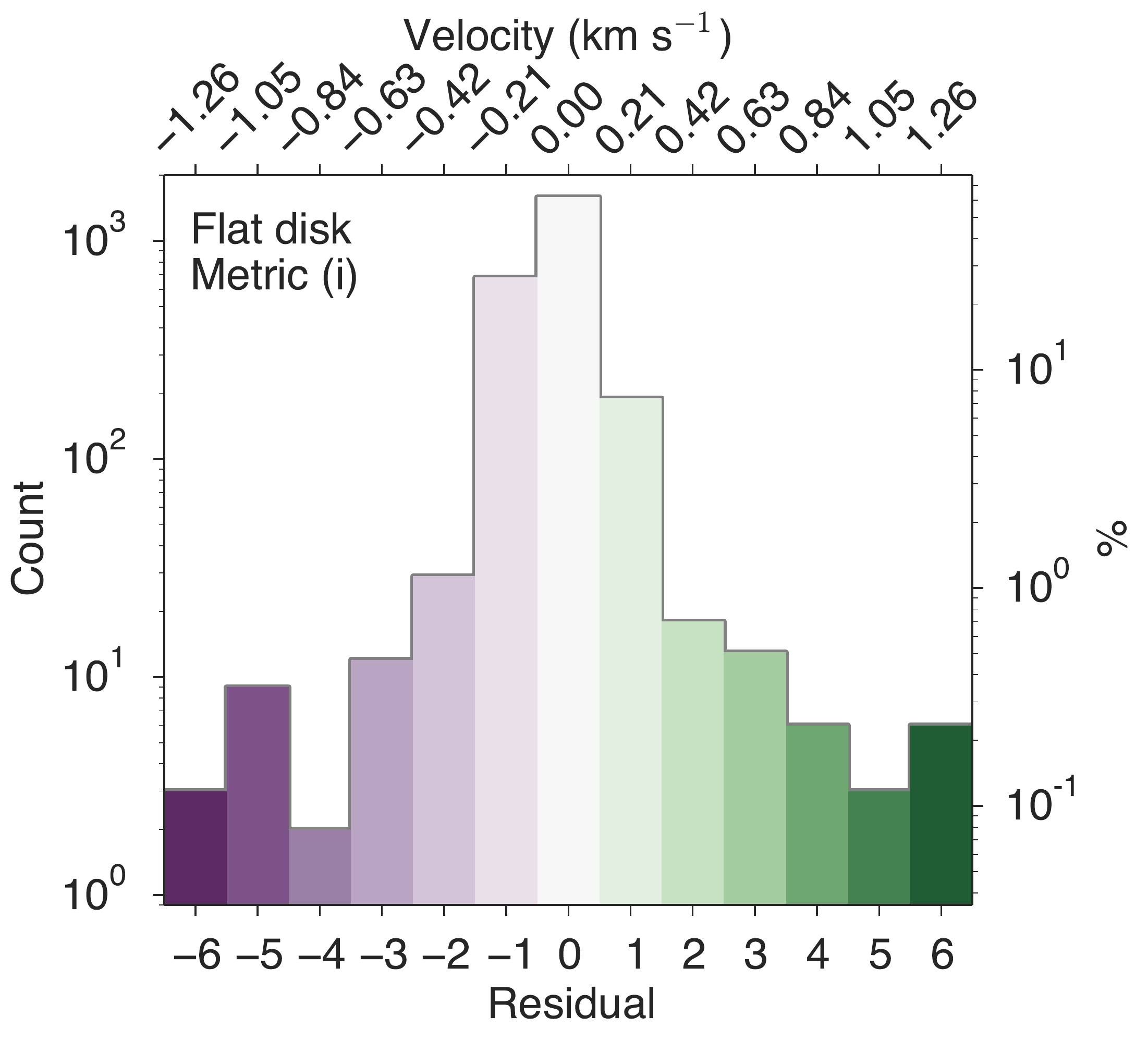}}
\subfigure{\includegraphics[width=0.33\textwidth]{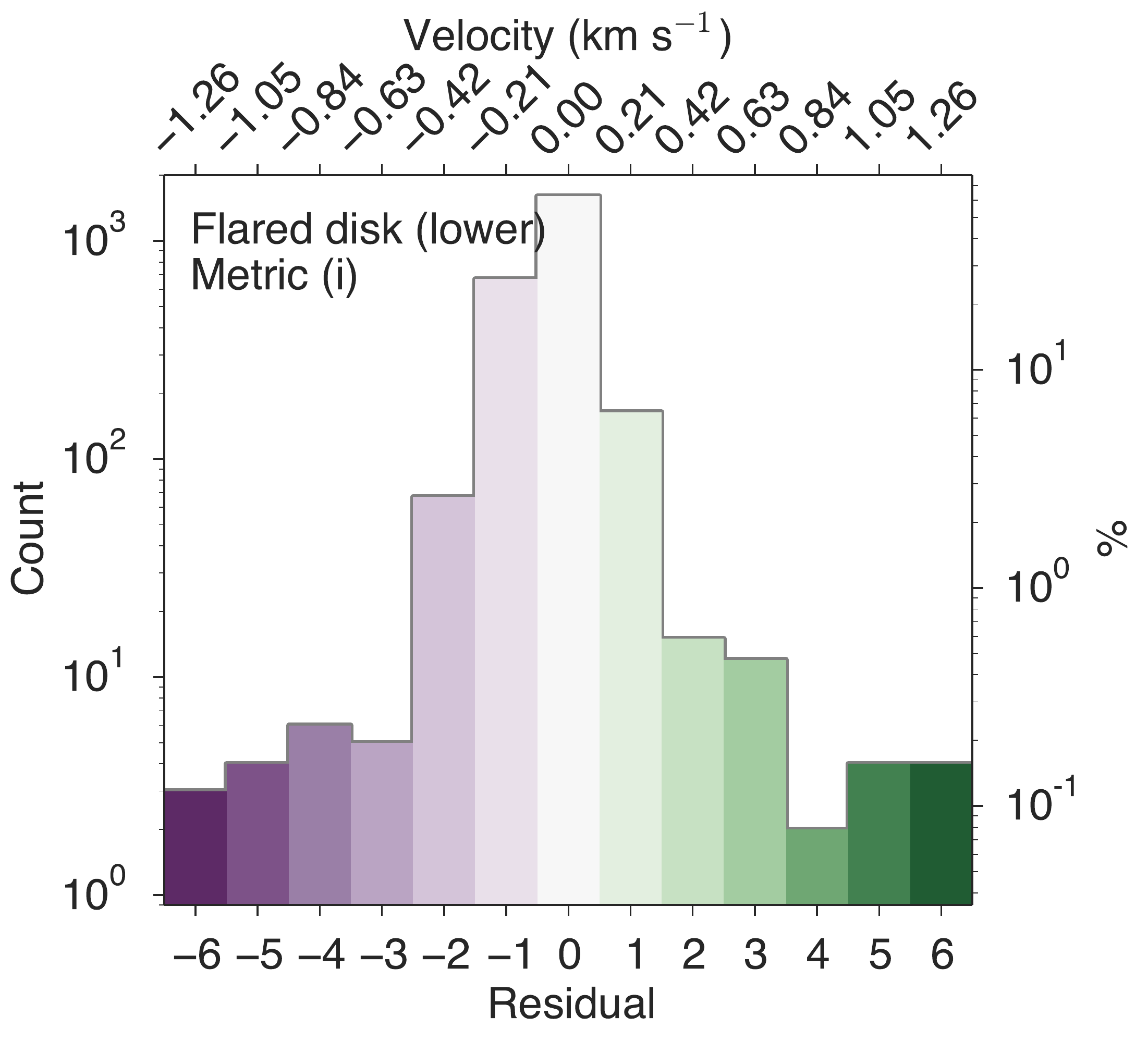}}
\subfigure{\includegraphics[width=0.33\textwidth]{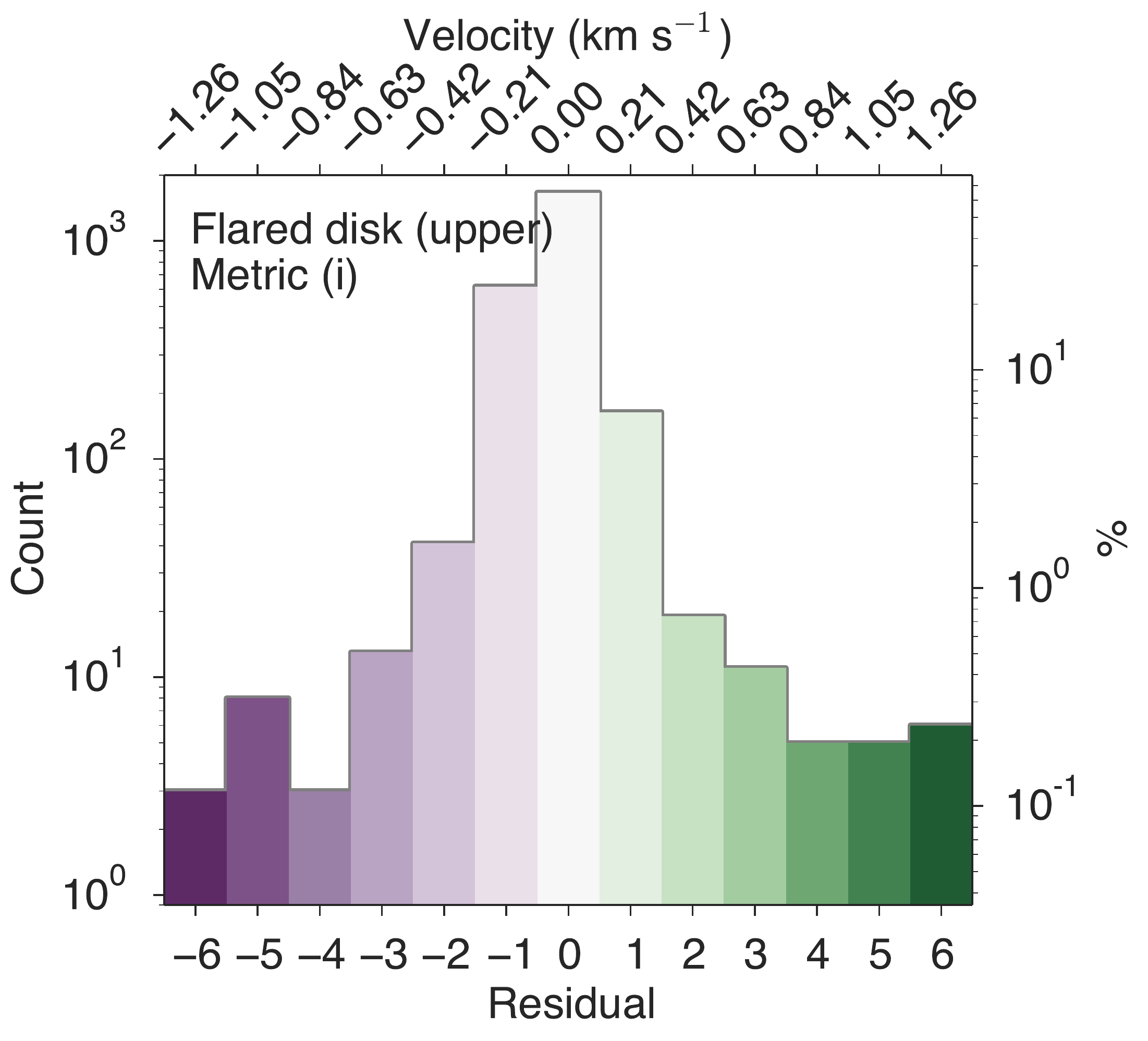}}
\subfigure{\includegraphics[width=0.33\textwidth]{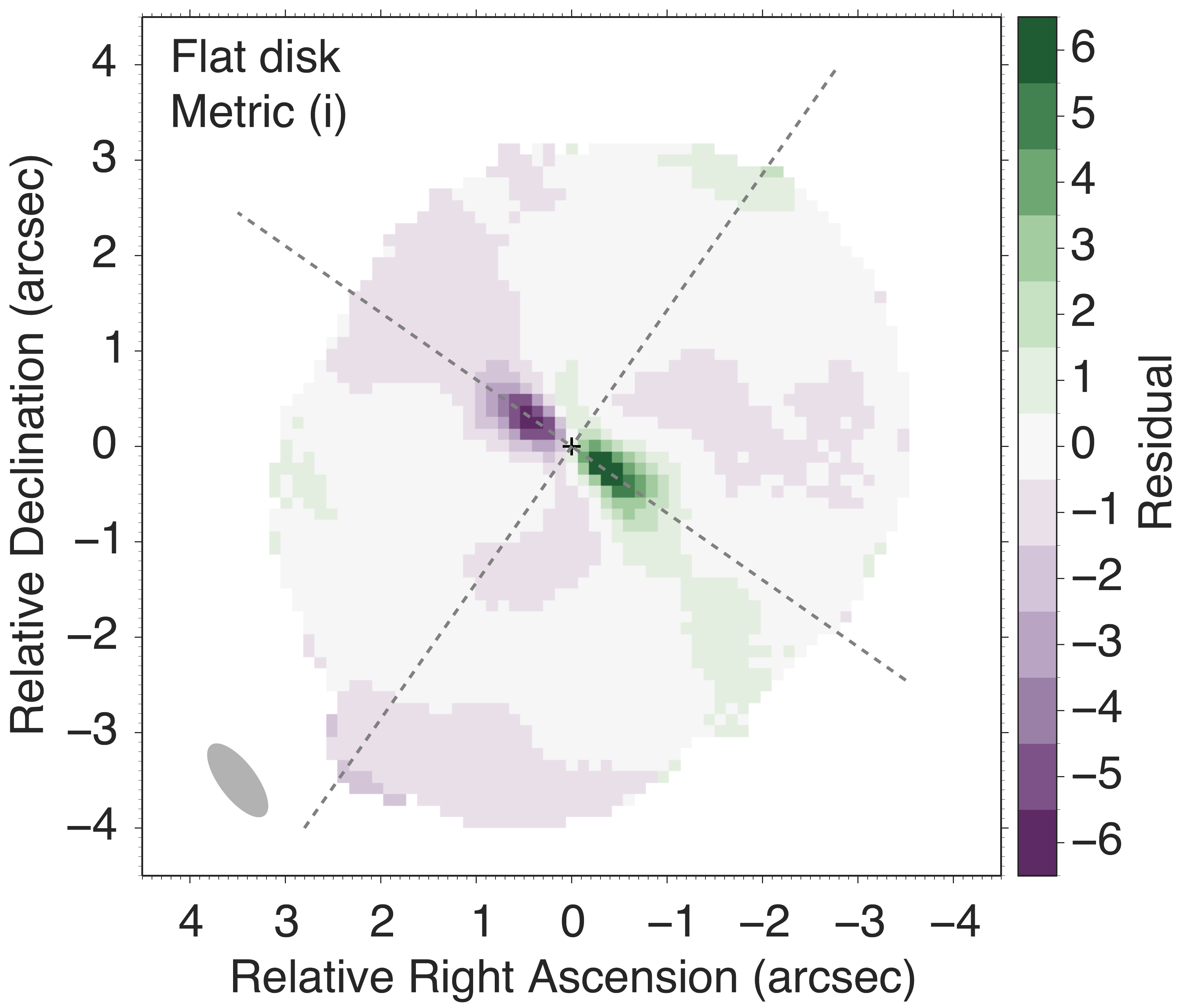}}
\subfigure{\includegraphics[width=0.33\textwidth]{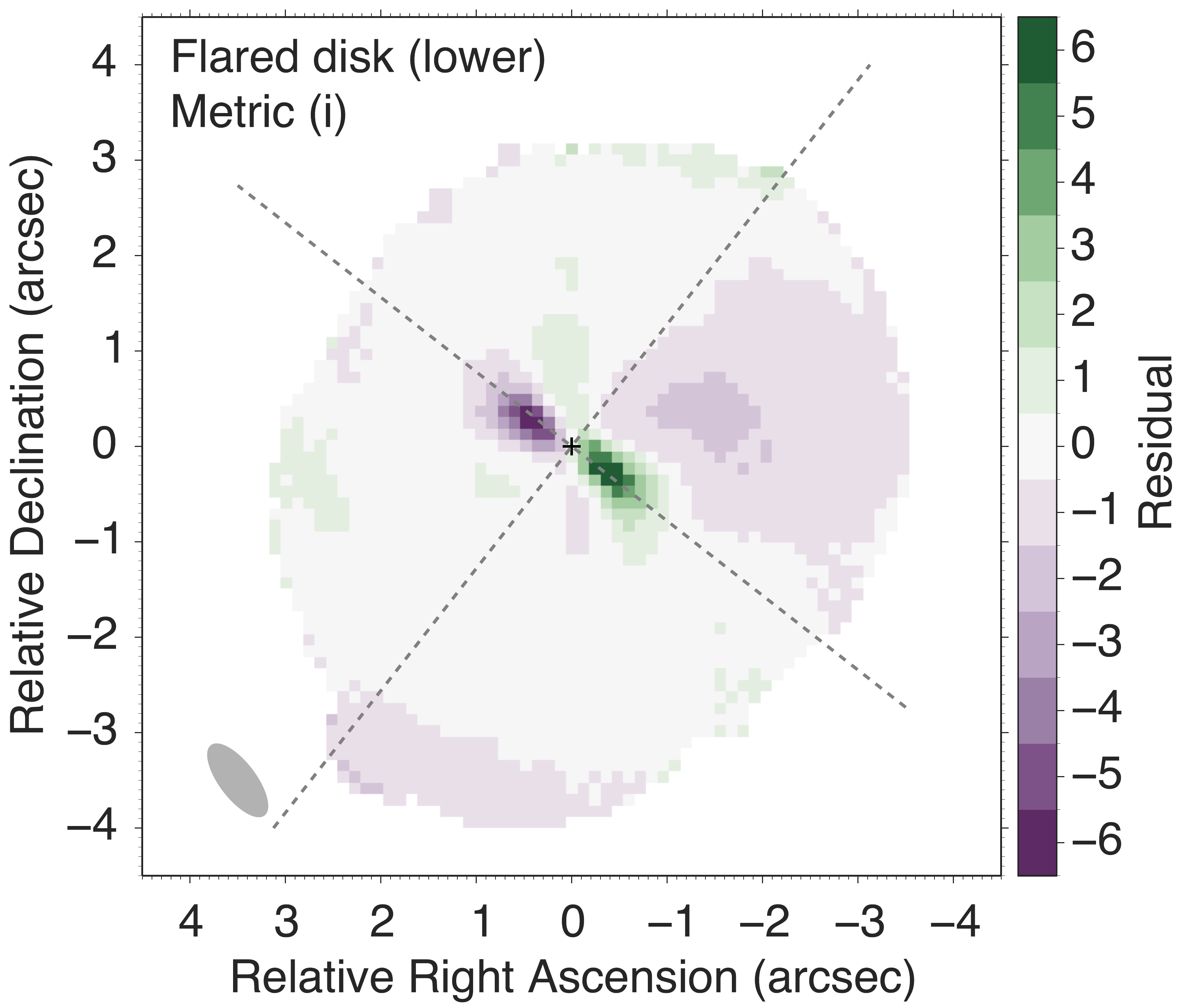}}
\subfigure{\includegraphics[width=0.33\textwidth]{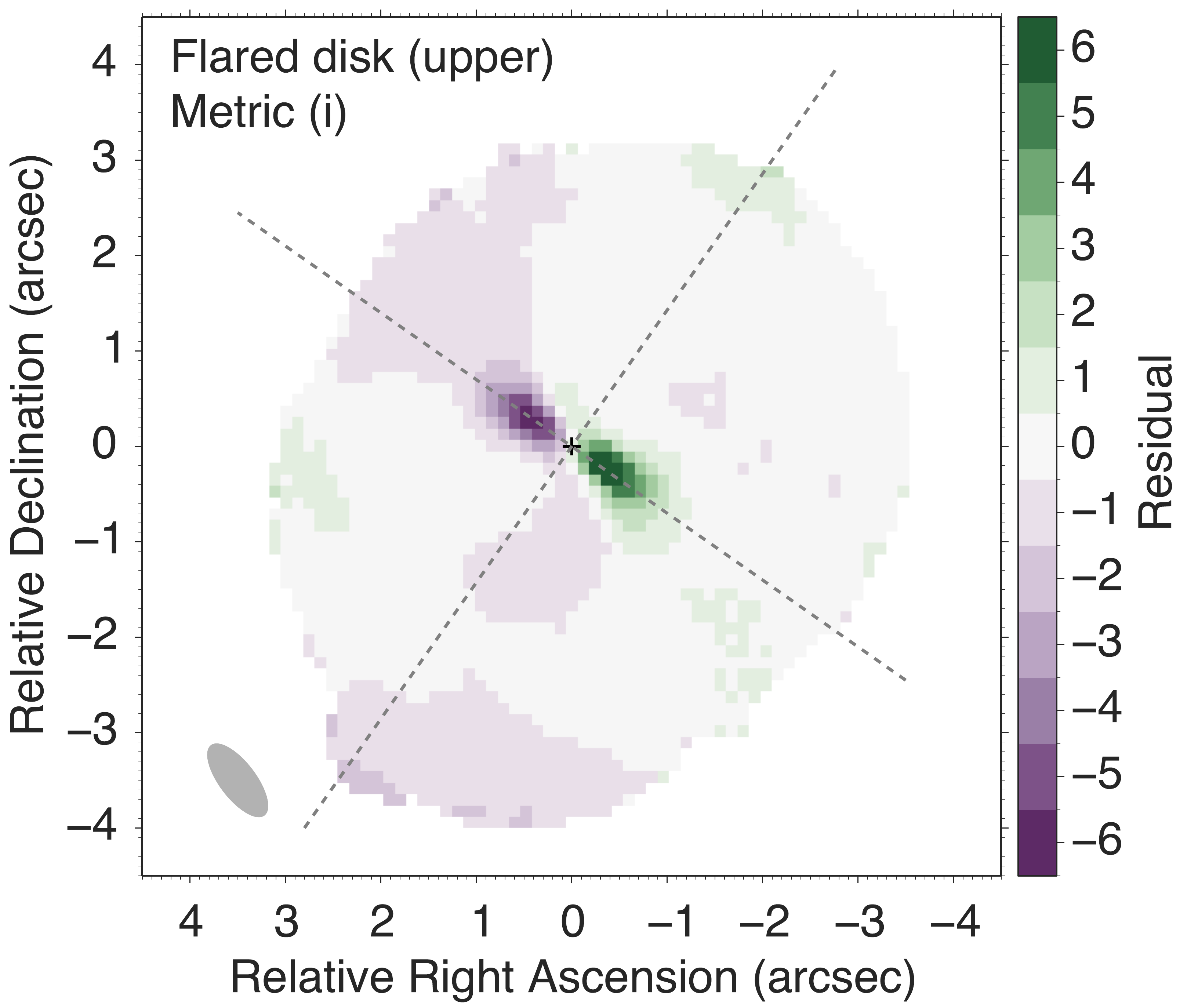}}
\caption{Residual histograms (top) and maps (bottom) 
using metric (i) as the measure of best fit for a geometrically flat disk (left), 
and the lower cone (middle) and upper cone (right) of a flared disk.  
The histograms are displayed on a log scale to emphasise the largest residuals.} 
\label{figure5}
\end{figure*}

The left hand plots of Fig.~\ref{figure5} show the distribution of residuals summed 
over the entire disk (top panel) and the residual first moment map 
(bottom panel).  
The histogram of residuals shows small dispersion about 0 with 96.0\% of pixels 
matching the data within $\pm0.315$ \kms.  
The residual map shows that a flat disk well reproduces the large-scale 
velocity field: the largest deviations from this model occur in the innermost 
disk where the model velocity field over-predicts 
(by up to $\approx 7\delta v$) the magnitude of the 
projected line-of-sight velocity along the minor axis of the disk.  
This leads to negative residuals in the north-east 
and positive residuals in the south-west.  
The morphology of the residuals suggests that the inner disk has an 
additional inclination along the minor axis of the outer disk, i.e., 
close to orthogonal to that of the outer disk.  

\subsection{A flared emitting surface}
\label{flared}

Although a geometrically flat disk well reproduces much of 
the velocity field, particularly for the outer disk, 
we test next whether emission from a flared surface can improve upon the flat disk fit.  
This is important to check because HD~100546 is classified as a Group I (i.e., flared) 
Herbig Ae star \citep{meeus01}, so one might expect the $^{12}$CO emission to arise from a layer 
higher up in the disk atmosphere.  
Indeed, thermo-chemical modelling of the disk around HD~100546 by \citet{bruderer12} 
suggests that the $^{12}$CO line emission arises from a layer $z/\rho \approx 0.2$. 

\citet{rosenfeld13} modelled the emission from the disk around 
the Herbig~Ae star HD~163296 by assuming that the emission arises from 
an inclined and flared surface with some opening angle, $\alpha$, relative 
to the $(x,y)$ plane (the disk midplane), i.e., a ``double-cone'' morphology.    
In this way, the front and back sides of the disk with the same projected 
line-of-sight velocity are spatially offset \citep[see also][]{degregorio13}.  
Here, a similar toy model is used; however, to determine the 
line-of-sight velocity, the radius is defined using spherical coordinates 
($r=\sqrt{x^2+y^2+z^2}$) rather than cylindrical coordinates 
($\rho=\sqrt{x^2+y^2}$; \citealt{rosenfeld13}).  
For small opening angles the two methods give similar results: 
the radii differ by no more than 10\% for $\alpha \le 25\degree$.  
A flared disk with this emission morphology has two possible orientations with either 
the lower or the upper face of the cone visible to the observer 
(see e.g., figure 3 in \citealt{rosenfeld13}).  
Model first moment maps for a flared disk with the same P.A.~and inclination as 
HD~100546, but with different opening angles (20\degree, 45\degree, and 
60\degree) are shown in Fig.~\ref{figurea2} in the Appendix.  

The range of surface opening angles ($[0\degree,20\degree]$) is motivated by 
previous thermo-chemical 
modelling of CO emission from HD~100546 
which suggest an opening angle $\alpha \approx 11\degree$ \citep[][]{bruderer12}.  
The symmetry in the channel maps (Fig.~\ref{figure3}) also suggests that the emitting layer 
lies relatively close to the midplane.  
As before, a coarse grid with a resolution of 5\degree~is initially 
run over the full parameter space, followed by a zoomed in grid 
with a resolution of 1\degree.

Figures~\ref{figure4} and \ref{figure5} present the statistics and residuals 
for the best-fit lower cone and upper cone of a flared disk.  
Using metric (i), the best-fit upper cone model fits the data marginally better 
(reproducing 65.0\% of the velocity field) than both the flat disk and the best-fit lower 
cone model (62.1\% and 62.6\%, respectively, see Table~\ref{table1}).  
The best-fit inclination, P.A., and opening angle are 36\degree, 
145\degree, and 9\degree, respectively (see Table~\ref{table1}).  
The opening angle of the $^{12}$CO-emitting surface agrees well with that 
suggested by thermo-chemical models of HD~100546 \citep{bruderer12}.

\begin{figure*}[]
\centering
\subfigure{\includegraphics[width=0.33\textwidth]{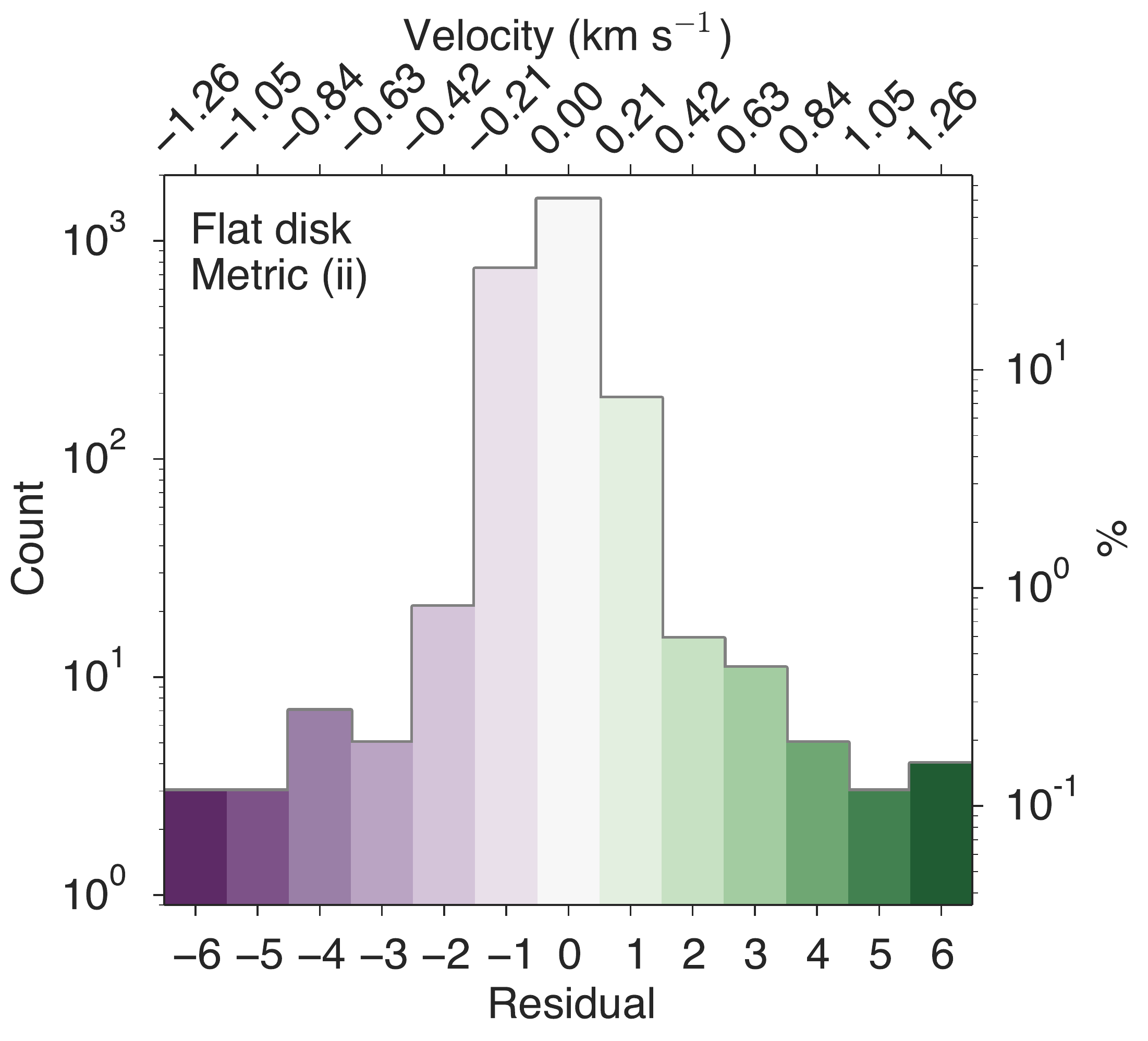}}
\subfigure{\includegraphics[width=0.33\textwidth]{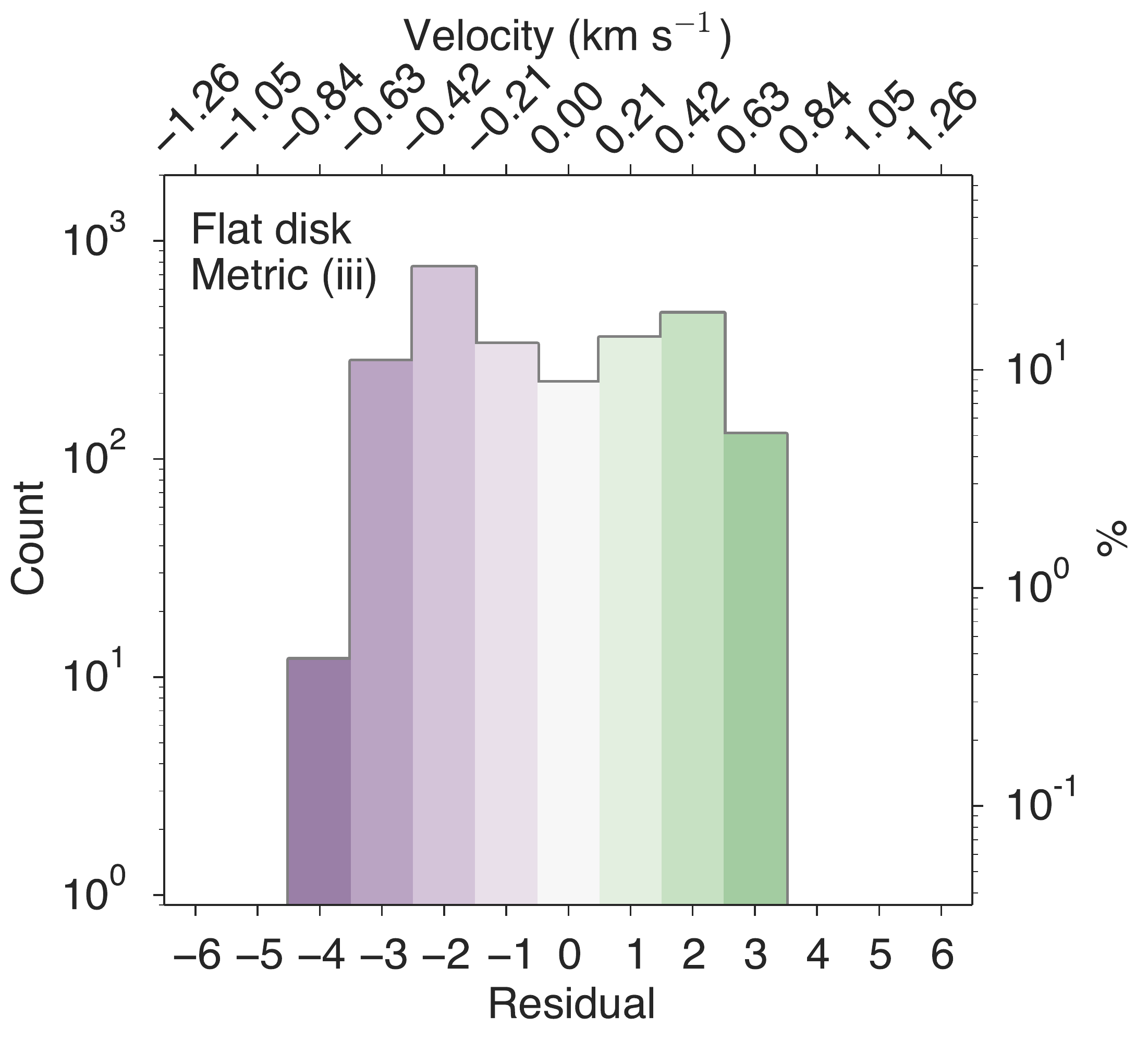}}\\
\subfigure{\includegraphics[width=0.33\textwidth]{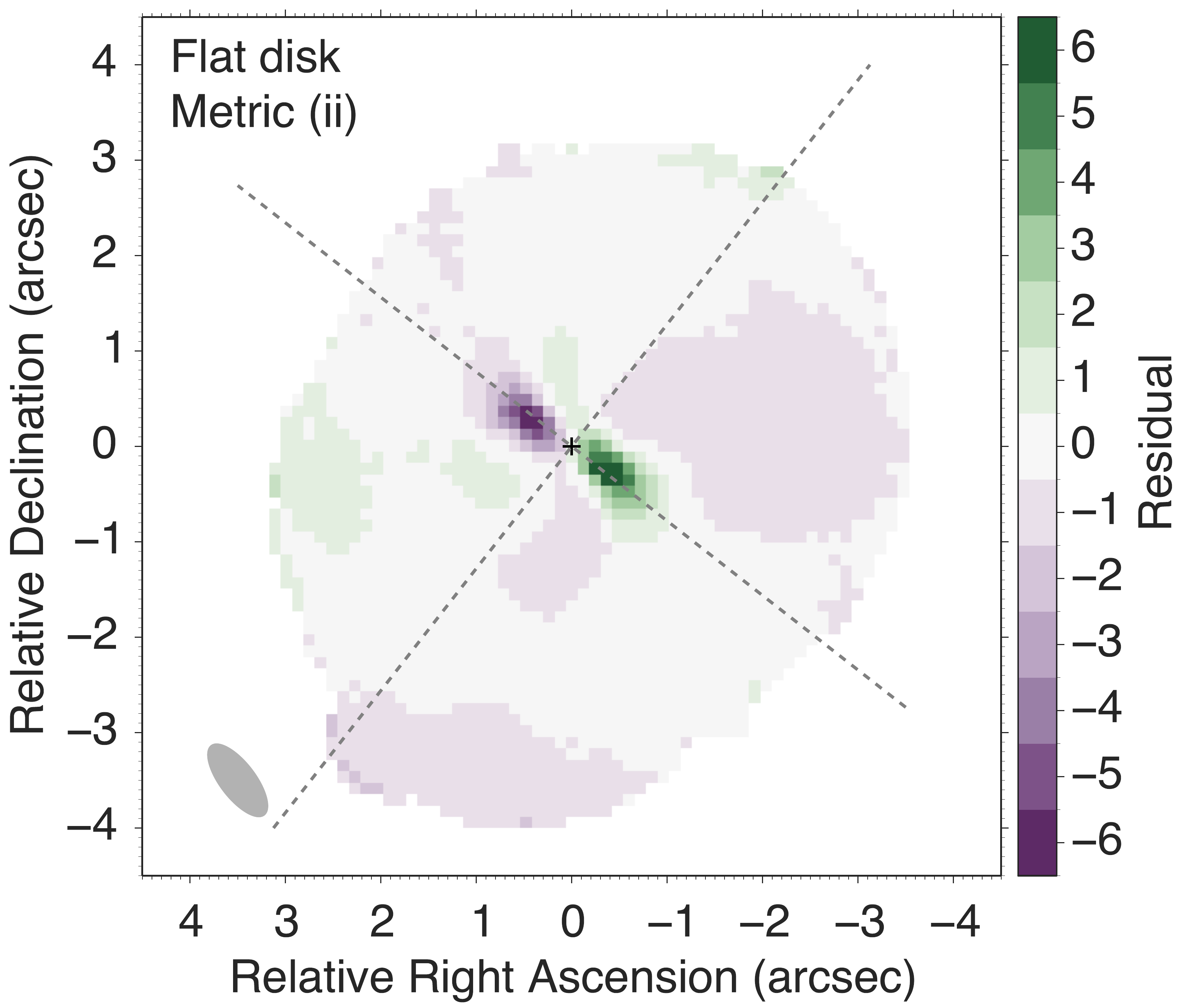}}
\subfigure{\includegraphics[width=0.33\textwidth]{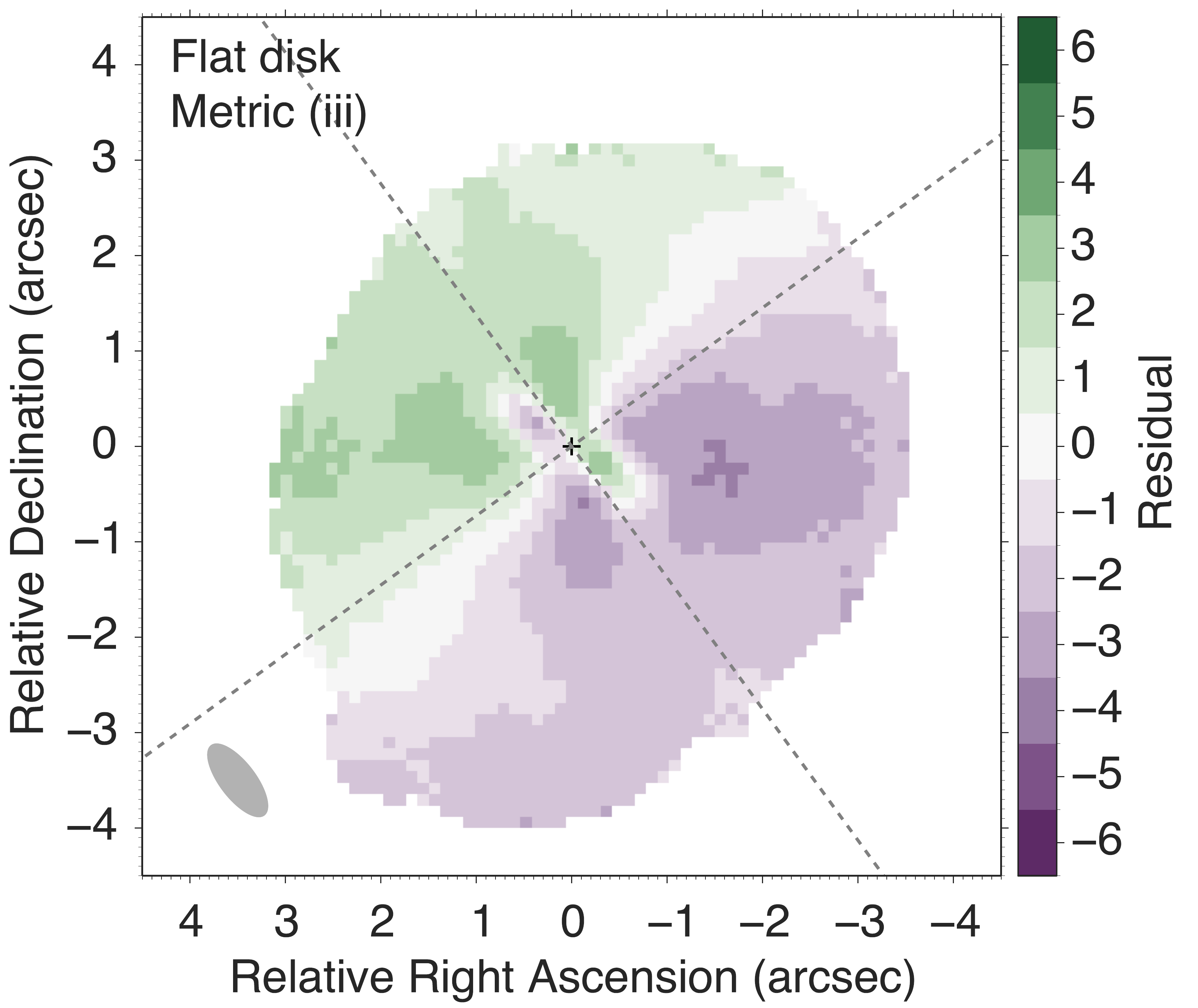}}
\caption{Residual histograms (top) and maps (bottom) for a geometrically flat, rotating disk, 
using metrics (ii) and (iii) as the measure of best fit.  
The histograms are displayed on a log scale to emphasise the largest residuals.} 
\label{figure6}
\end{figure*}

The best-fit lower cone model has an inclination of 38\degree, 
a P.A.~of 142\degree, and an opening angle of 13\degree.  
The inclination of this model lies closest to that derived from the 
continuum observations ($44\degree \pm 3\degree$).  
Despite resulting in a marginally worse fit to the data than 
the upper cone model (see Table~\ref{table1}), a ``by-eye'' 
examination of the residual map (bottom left panel of 
Figure~\ref{figure5}) shows that this 
morphology best reproduces the velocity field in all quadrants 
of the outer disk excepting the north-west quadrant for 
which the magnitude of the velocity field is over-estimated.  
Comparing this residual map to both the 
channel map (Fig.~\ref{figure1}) and the 
eighth moment map (bottom right panel of Fig.~\ref{figure3}) 
shows that emission from this quadrant appears less bright 
and exhibits a positional offset relative to 
that mirrored across the minor axis of the disk.  
However, that the upper cone model fits the data best using this metric
is in agreement with the moment maps presented in Fig.~\ref{figure3} and 
recent VLT/SHERE images that confirm that the the far side of the flared disk 
surface lies towards the north east \citep{garufi16}.  
The global best-fit across all three model as determined by metric (ii), i.e., 
sum of the squares of the residuals scaled by the total number of pixels, 
is a flat disk with an inclination of 37\degree~and a P.A.~of 142\degree.  
The best-fit model selected by the smallest peak residual, 
i.e., metric (iii), is also shared by all three models and 
is a flat disk with an inclination of 39\degree~(again in good agreement 
with the other two metrics); however, the disk P.A.~which gives the 
smallest peak residual is 126\degree.  
The residual histograms and maps for both of these models are 
shown in Fig.~\ref{figure6}. 
That the inner disk velocity structure is better fit with a shallower 
P.A.~than the outer disk, highlights the presence of a twisted warp: 
this is investigated in the subsequent section. 

\begin{table*}
\centering
\caption{Best-fit parameters for the flat and flared kinematic models. \label{table1}}
\def\arraystretch{1.5}
\begin{tabular}{lcccccccc}
\hline\hline
Model     & Metric       & Inclination & P.A.& Opening  & Pixel  & Percentage & Sum of           & Peak \\ 
          & of best-fit  &             &     & angle    & number &            & residual squares$^{1}$ & residual ($\delta v$) \\ 
\hline
Flat                & (i)   & 36\degree & 145\degree & 0\degree  & 1590 & 62.1\%  & 0.737 & 7.23 \\
                    & (ii)  & 37\degree & 142\degree & 0\degree  & 1550 & 60.5\%  & 0.665 & 6.71 \\
                    & (iii) & 39\degree & 126\degree & 0\degree  &  224 &  8.8\%  & 3.610 & 3.63 \\
Flared (lower cone) & (i)   & 38\degree & 142\degree & 13\degree & 1605 & 62.6\%  & 0.718 & 6.64 \\
Flared (upper cone) & (i)   & 36\degree & 145\degree & 9\degree  & 1665 & 65.0\%  & 0.764 & 7.28 \\
\hline
\end{tabular}
\tablefoot{$^{1}$~Scaled by the total number of unmasked pixels.}
\end{table*}

\subsection{A warped disk}
\label{warp}

The residual maps displayed in Figs.~\ref{figure5} and \ref{figure6} 
reveal two features: (i) a rotating disk within $\approx$1\farcs0 of the source 
position with an inclination angle approximately orthogonal to the line of sight, and 
(ii) a shallower position angle on small scales ($\lesssim$1\farcs0) than on larger scales.  
Both results point towards a twisted warp in the inner disk 
(see, e.g, \citealt{juhasz17} and \citealt{facchini17} and references therein).

Because the residuals are of the order of the size of the synthesised beam, 
a simple toy prescription for the warp is used.  
The inner disk is modelled as a planar disk within a fixed radius which 
possesses its own inclination and P.A., i.e., the inner disk is 
misaligned relative to the outer disk.  
This is similar to the approach used by \citet{rosenfeld14} to 
model the kinematics of HD~142527.  
Figure~\ref{figurea3} presents model first moment maps for a warped 
disk for a range of inclinations and position angles; the transition radius is fixed 
at 100~au, and the outer disk parameters are given values 
appropriate for the HD~100546 disk.

The outer disk velocity structure is fixed to that of the best-fit upper cone model. 
As mentioned in the previous section, recent VLT/SPHERE images of scattered light 
from HD~100546 suggest that the far side of the (flared) disk lies towards the 
north-east \citep{garufi16} in agreement with the moment maps in Fig.~\ref{figure3}.  
This results in three additional fitting parameters only: 
the inner disk inclination ([40\degree, 90\degree]), the inner disk P.A.~([40\degree, 100\degree]), and 
a transition radius marking the boundary between the inner 
and outer disks ([40, 120]~au).  
Note that, for simplicity, we assume that the inner disk velocity structure is 
described using the flat disk prescription (i.e., Equation~\ref{velocity}). 
A coarse grid with a resolution of 10\degree~and 10~au is first run to 
identify the parameter space containing the global best-fit, followed by 
a finer grid over this zoomed-in region (with a resolution of 2\degree~and 2~au).  

\begin{figure*}[]
\centering
\subfigure{\includegraphics[width=0.33\textwidth]{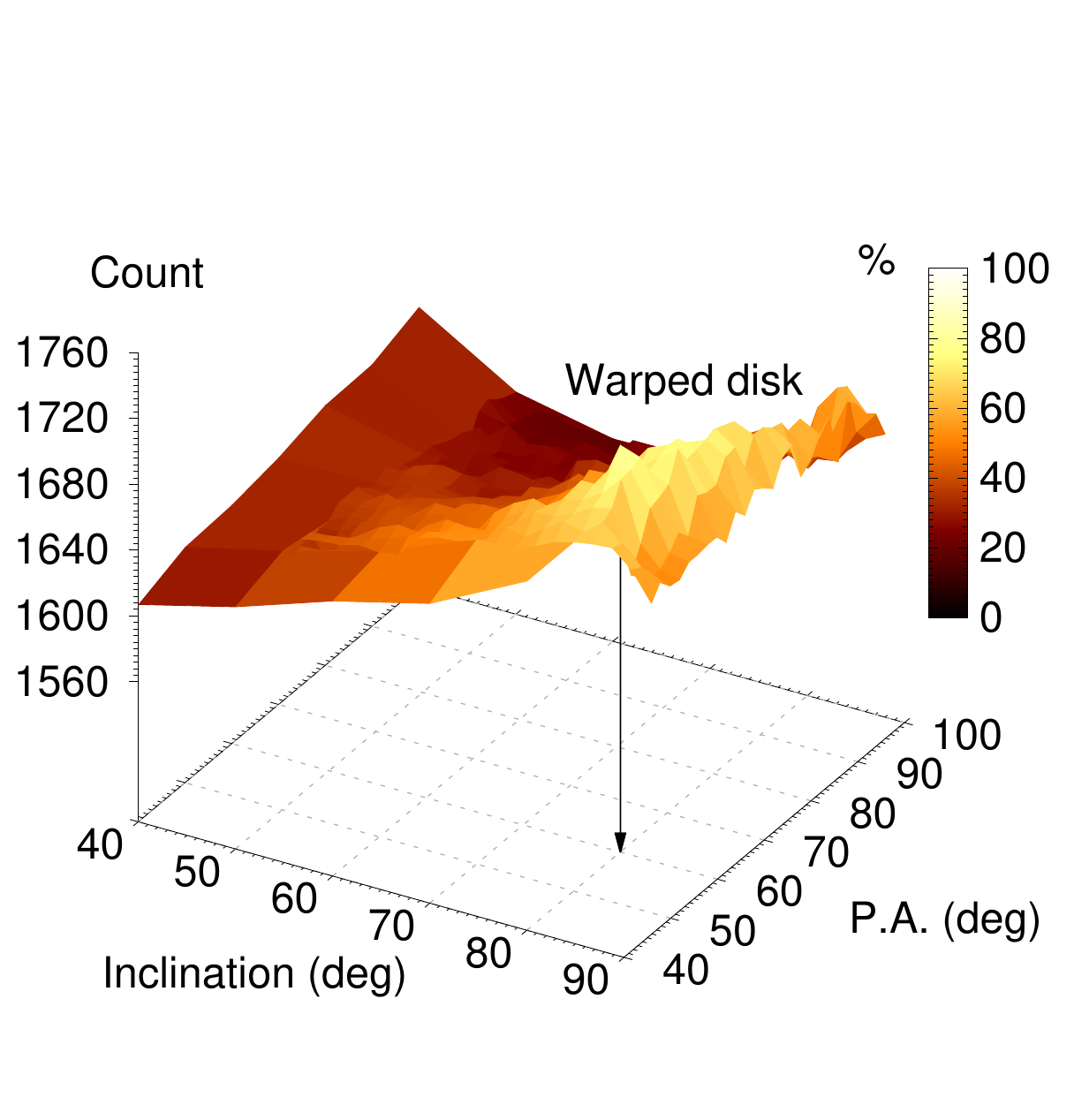}}
\hspace{2cm}
\subfigure{\includegraphics[width=0.33\textwidth]{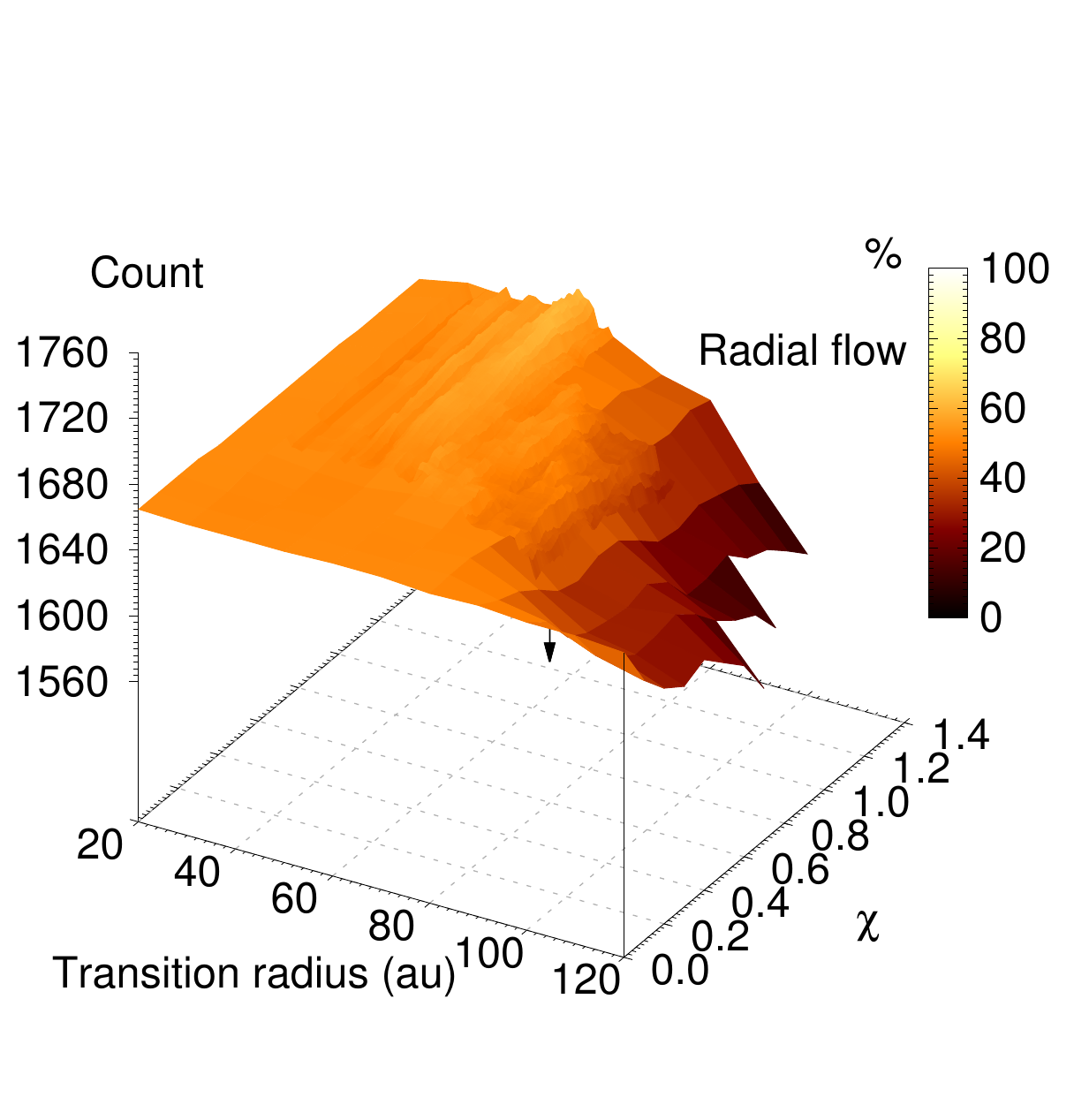}}
\caption{ {Distribution of model best-fit values using metric (i) 
for the best-fit warped disk (left) and radial flow model (right).  
The best-fit transition radius for the warped disk and using this metric is 90~au. 
The pixel count for the warped disk is given as a function of inclination and 
position angle whereas that for the radial flow model is given as a function of 
transition radius and radial velocity scaling factor, $\chi$.  
In these plots, the percentage scale corresponds to the z-axis range.}}
\label{figure7}
\end{figure*}

\begin{figure*}[]
\centering
\subfigure{\includegraphics[width=0.33\textwidth]{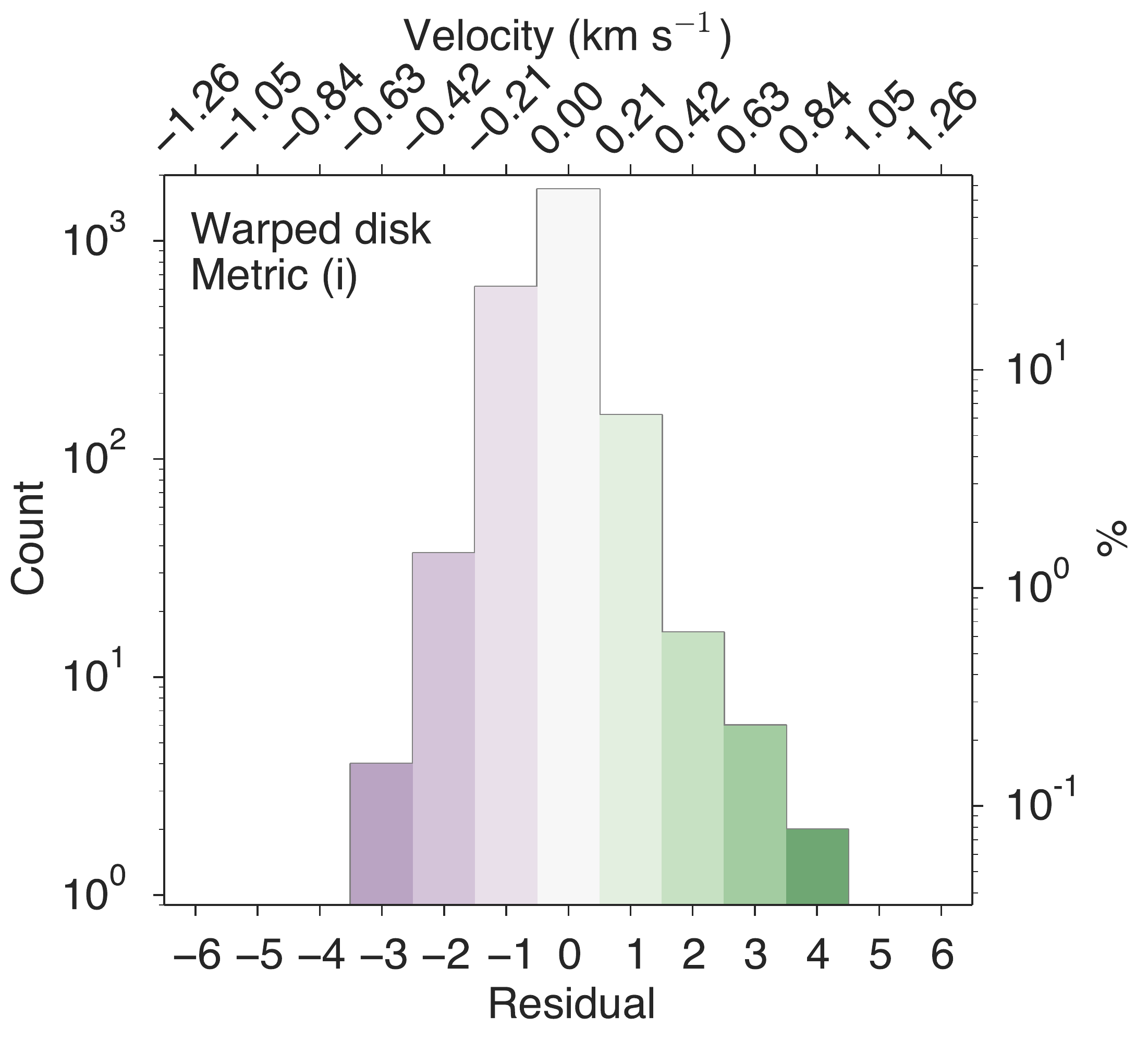}}
\subfigure{\includegraphics[width=0.33\textwidth]{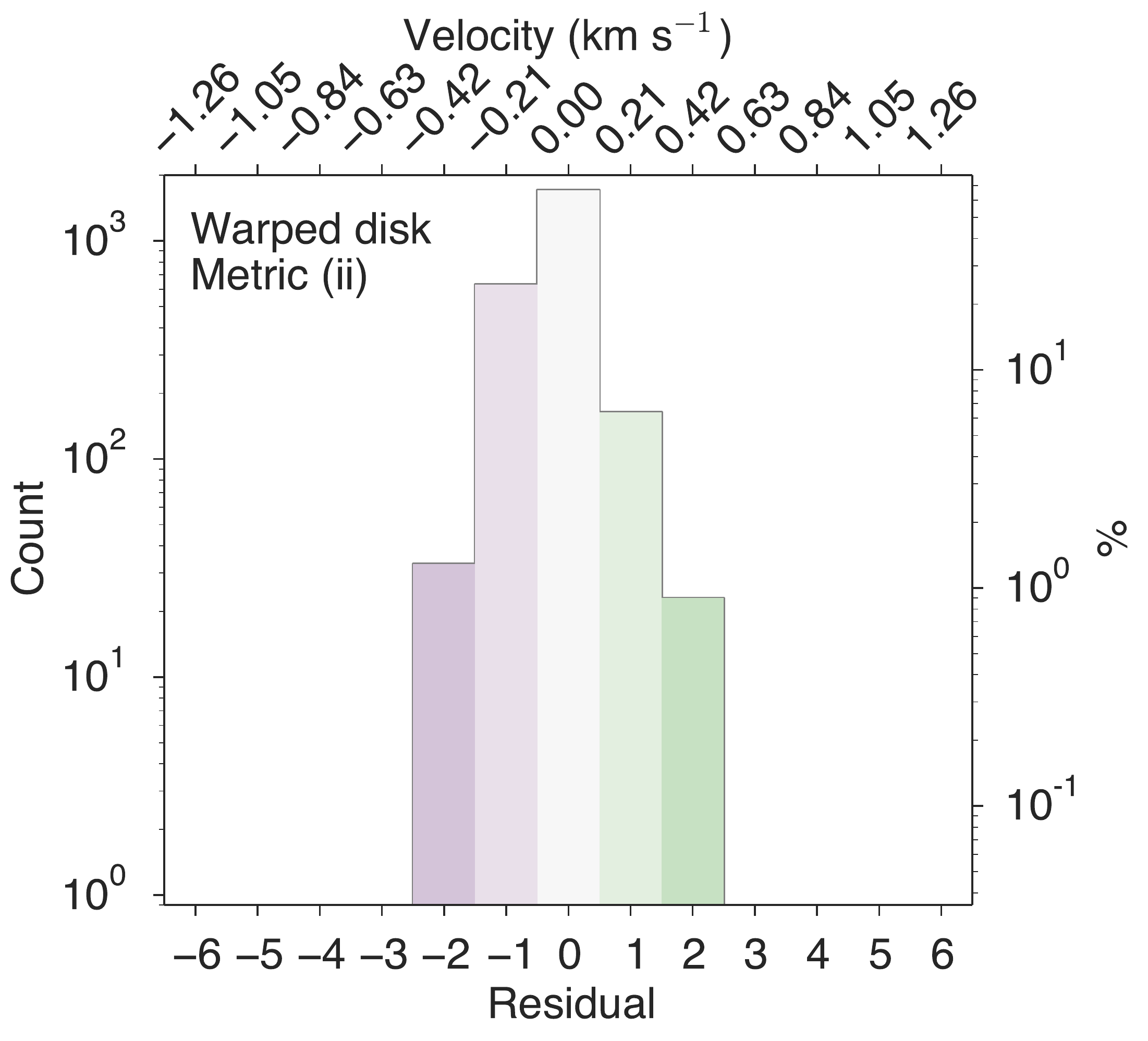}}\\
\subfigure{\includegraphics[width=0.33\textwidth]{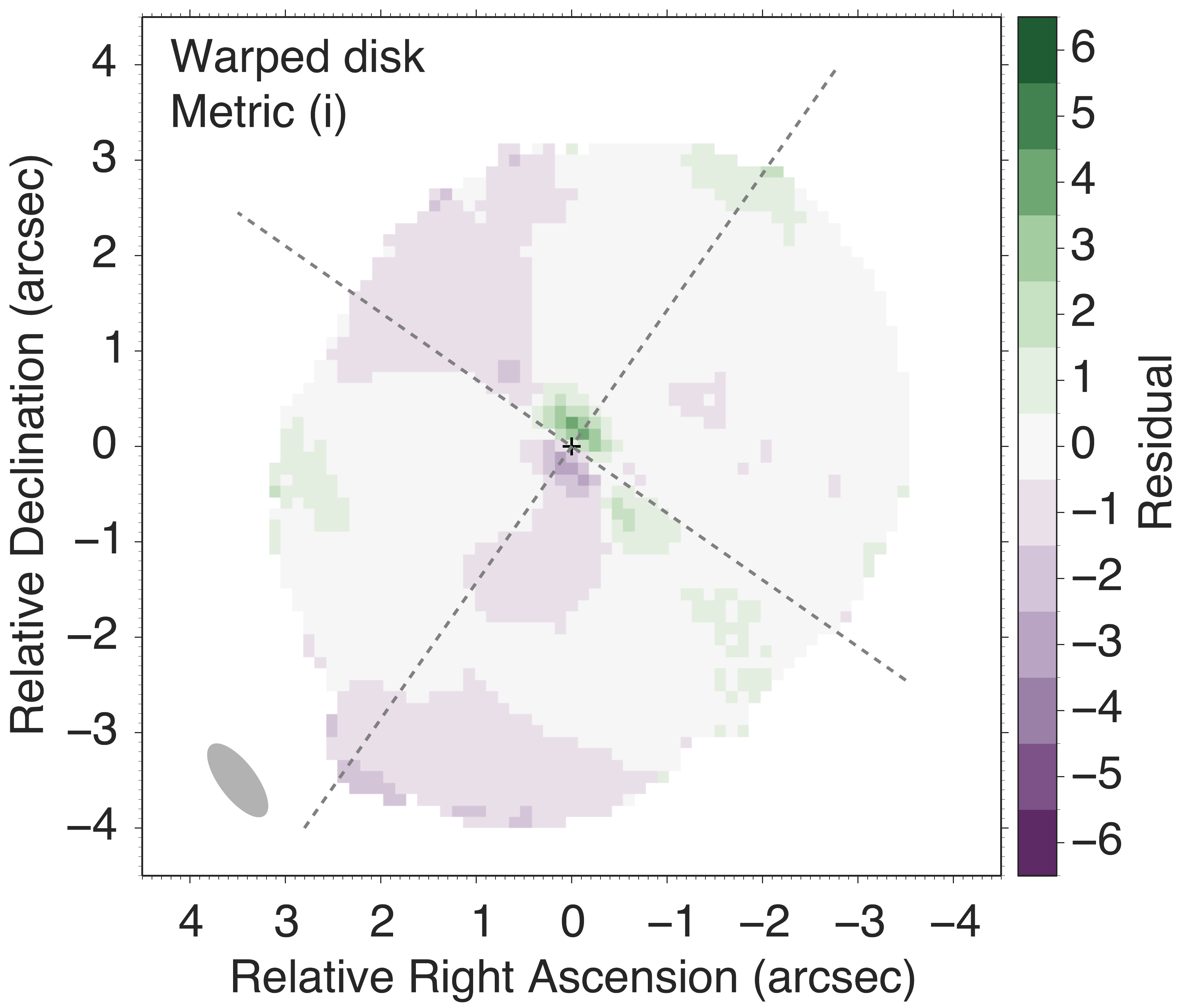}}
\subfigure{\includegraphics[width=0.33\textwidth]{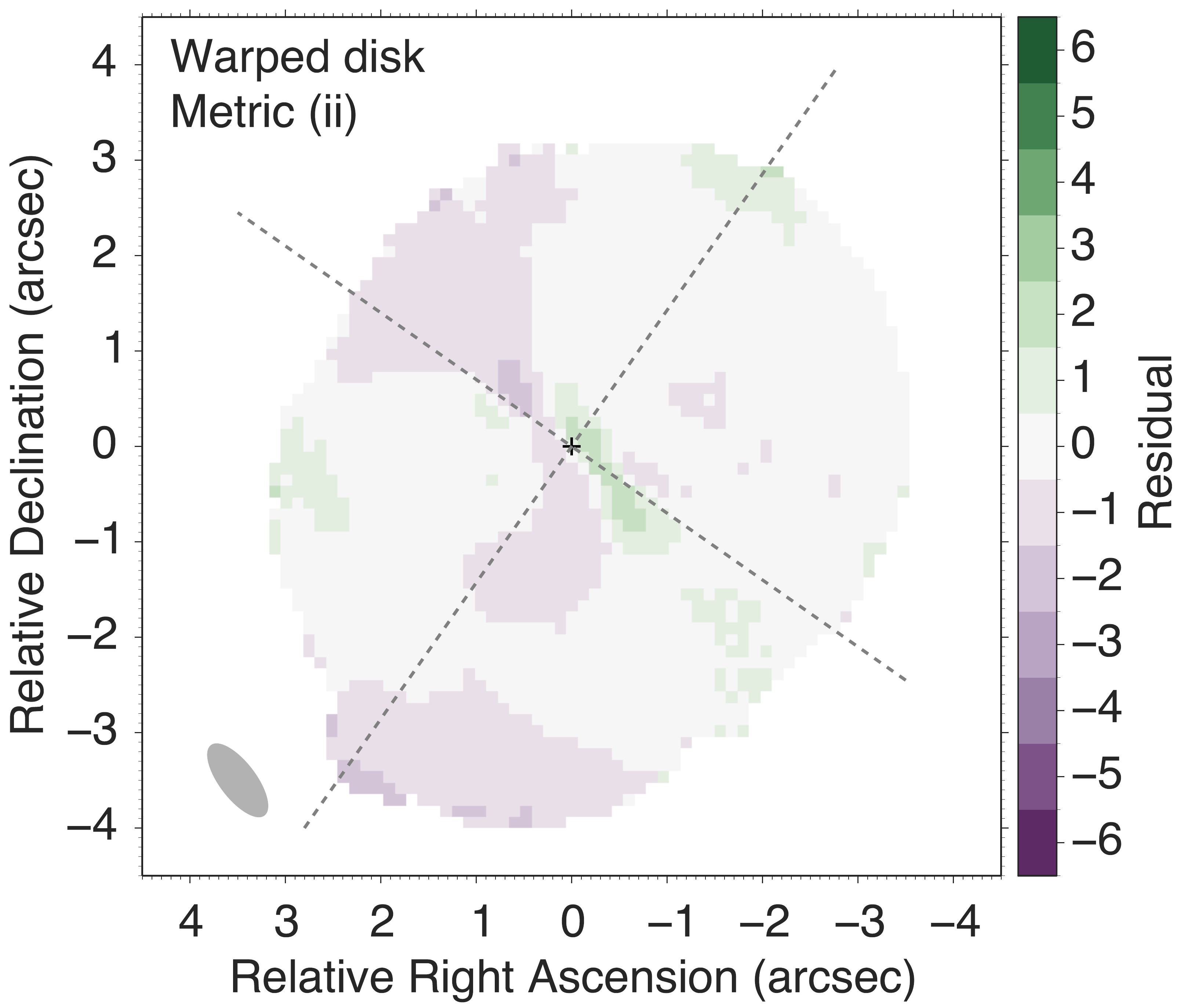}}
\caption{Residual histograms (top) and maps (bottom) for a protoplanetary disk 
with a warped inner disk using metrics (i) and (ii) as the metric of best fit.  
The histograms are displayed on a log scale to emphasise the largest residuals. 
Note that metrics (ii) and (iii) select the same warped inner disk parameters 
(see Table 2).} 
\label{figure8}
\end{figure*}

Figures~\ref{figure7} and \ref{figure8} present the statistics and residuals for
the best-fit warped disk, respectively.  
Metric (i) favours a model with an inner disk that is almost ``edge-on'' 
($i=80\degree$) to the line of sight, almost orthogonal to the outer disk 
major axis (P.A.$= 60\degree$), and with a transition radius of 90~au 
(see Fig.~\ref{figure7} and Table~\ref{table2}).  
These values are consistent with the morphology of the residuals of both 
the flat and flared models (see Figs.~\ref{figure5} and \ref{figure6}).  
The magnitude of the peak residual of this model is significantly smaller than 
the previous models selected using metric (i), $4 \delta v$ versus $7 \delta v$.  
Metrics (ii) and (iii) select the same model (see Table~\ref{table2}) 
with parameters similar to those using metric (i); an inclination of 
84\degree, a P.A. of 64\degree, and a transition radius of 100~au.  
Comparing the residual histograms and maps for these two models (shown in 
Fig.~\ref{figure8}), highlights how a small change in inclination and/or 
position angle can significantly reduce the magnitude of the peak residual.  
This latter model results in a peak residual of only $2.4 \delta v$ and has 
the smallest dispersion of residuals: 98\% of pixels match the data within 
$\pm 0.315$~\kms~and 100\% of pixels match within $\pm 0.525$ \kms. 

\begin{figure}[!h]
\centering
\includegraphics[width=0.5\textwidth]{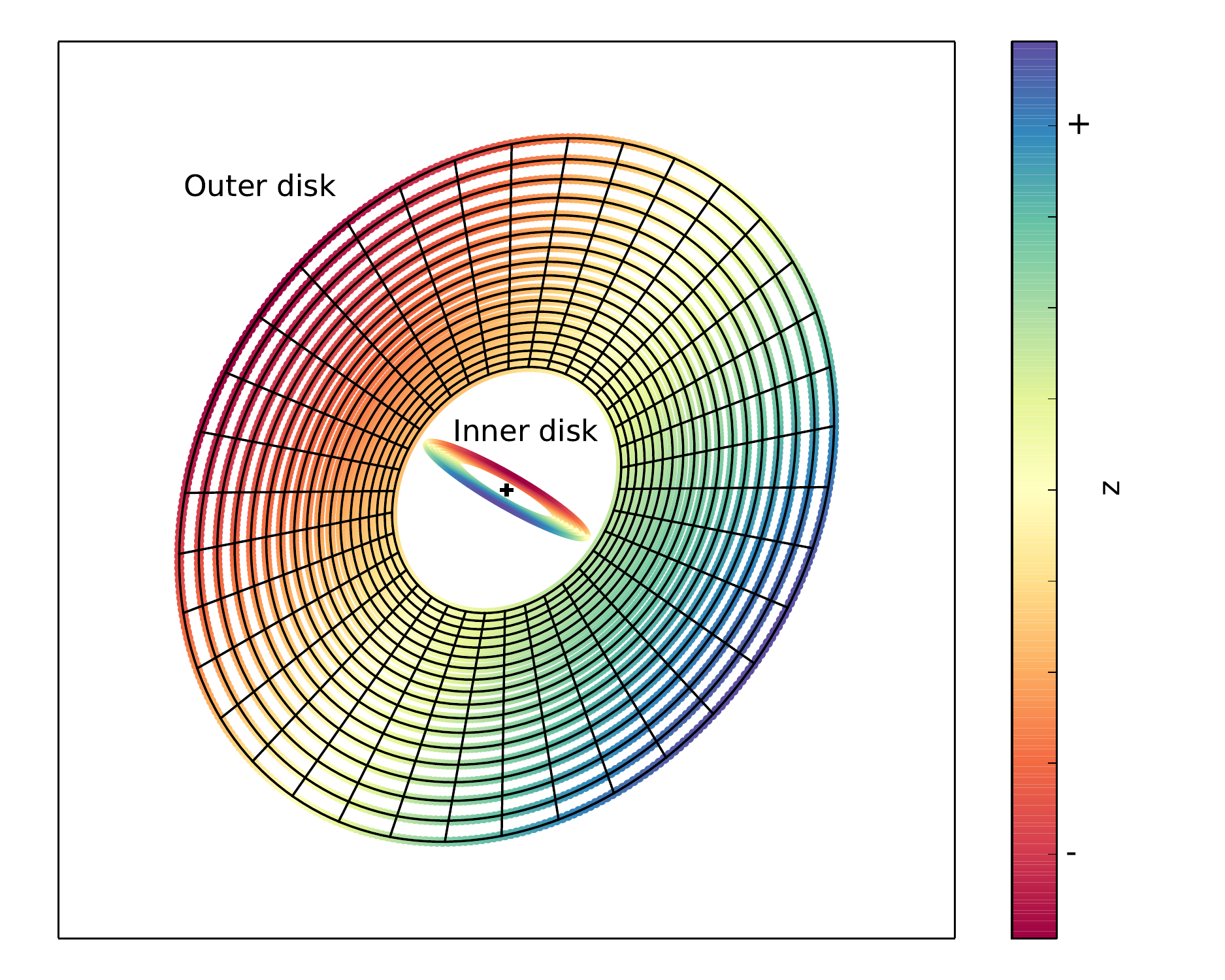}
\caption{Cartoon of the HD~100546 protoplanetary disk.  
The colour scale indicates the $z$ coordinate of the inner and outer 
disks relative to the sky plane ($z=0$).  
Note that the morphology of the intermediate region between the inner and outer 
disks is not yet known and so it is intentionally left blank.} 
\label{figure9}
\end{figure}

Figure~\ref{figure9} shows an idealised model of the HD~100546 
protoplanetary disk as proposed by the best-fit warp model.  
The morphology of the intermediate region between the inner and outer 
disks is not yet known and so is intentionally left blank in the cartoon.  
However, such ``broken disks'' are predicted by SPH models of disks around 
binary systems \citep[see, e.g.,][]{facchini17}.

\begin{table*}
\centering
\caption{Best-fit parameters for the warped kinematic models. \label{table2}}
\def\arraystretch{1.5}
\begin{tabular}{lcccccccc}
\hline\hline
Model     & Metric       & Inclination & P.A.& Transition  & Pixel  & Percentage & Sum of           & Peak \\ 
          & of best-fit  &             &     & radius (au) & number &            & residual squares$^{1}$ & residual  ($\delta v$)\\ 
\hline
Warped  & (i)            & 80\degree & 60\degree &  90 & 1722 & 67.2\% & 0.387 & 4.06 \\
        &  (ii) \& (iii) & 84\degree & 64\degree & 100 & 1710 & 66.4\% & 0.350 & 2.44 \\
\hline
\end{tabular}
\tablefoot{$^{1}$~Scaled by the total number of unmasked pixels.}
\end{table*}

\subsection{A radial flow}
\label{radialflow}

The modelling presented in the previous section demonstrated how a 
misaligned and Keplerian gas disk within 100~au can reproduce the velocity structure 
in this region; however, scattered light images taken with VLT/SPHERE and MagAO/GPI 
with a spatial resolution of 0\farcs02 and 0\farcs01, respectively, reveal no evidence 
of a severely misaligned dust disk beyond $\approx10$~au \citep{garufi16,follette17}.   
However, both datasets do suggest the presence of 
spirals in the inner disk within 50~au, and resolve the inner 
edge of the outer dust disk traced in small grains (11-15~au, depending on wavelength).  
Spiral arm features are also seen in larger-scale scattered light images 
\citep[$\gtrsim 100$~au;][]{grady01,ardila07,boccaletti13}.
Further, modelling of recent VLTI/PIONIER interferometric observations of HD~100546 
reported in the survey by \citet{lazareff17}, and which have a spatial resolution of 
order $\sim 1$~au, 
suggests that the very innermost dust disk has a similar position 
angle (152\degree) and inclination (46\degree) as the outer disk traced in 
sub-mm emission \citep[146\degree~and 44\degree, respectively;][]{walsh14}.  
Hence, if the velocity structure of the CO gas is indeed 
caused by an extremely misaligned Keplerian disk inwards of $\sim 100$~au, in light 
of these new observations it appears that the small dust grains may 
have a different morphology than the gas in the inner disk.  
However, there is no known physical mechanism that would 
decouple the small dust grains and molecular gas to such a severe extent. 
It should be noted that the disk parameters 
derived from the VLTI/PIONIER interferometric data are 
dependent upon the model used to reproduce the data and other 
morphologies may be possible.  

Despite the warp model providing a good fit to the velocity structure, we now consider 
an alternative explanation for deviations from global Keplerian motion: 
radial flows \citep{casassus13,rosenfeld14,casassus15,loomis17}.  
As already highlighted by \citet{rosenfeld14}, warps and radial flows can be difficult 
to distinguish using the first moment map alone.  
Examination of the morphology of the residuals in Figs.~\ref{figure5} and \ref{figure6} 
support a potential radial component to the gas motion.  
For an inclined and rotating disk, the line-of-sight projected velocity of 
the radial component will be zero along the major axis of the disk, and will reach 
its maximum value along the minor axis of the disk.  
Further, this radial component will be blue-shifted in velocity for material falling inwards 
from the far side of the disk (from the north east) and red-shifted in velocity for 
material falling inwards from the near side of the disk (from the south west).

We now investigate whether a radial velocity component to the gas 
can reproduce the velocity structure of the CO emission from HD~100546.  
We adopt the same approach as the toy models presented in 
\citet{rosenfeld14} and assume that the radial velocity component is a fixed fraction of 
the local Keplerian velocity, $v_r = \chi v_\mathrm{K}$ where $v_\mathrm{K} = \sqrt{(GM_\star/\rho)}$.  
The line-of-sight projected velocity is thus given by,
\begin{equation}
v(x',y') = v_\mathrm{K}\sin i \sin\theta - \chi v_\mathrm{K}\sin i \cos\theta
\end{equation}
where $\theta$ is defined as in Sect.~\ref{flat} and $\chi$ 
is a scaling factor we assume can vary between 0 (no radial 
flow) and $\sqrt{2}$ \citep[i.e., radial infall at the local free-fall velocity, see e.g.,][]{murillo13}.  
The disk velocity structure is fixed to that of the best-fit upper cone model for the outer disk, 
and we define the transition radius as the radius within which the radial 
component for the velocity field is non-zero for $\chi > 0$.  
Given the symmetry in the residuals, we assume that inwards of the 
transition radius, the disk is flat (i.e., $\alpha=0\degree$, see Sect.~\ref{flared}).
Hence, there are only two fitting parameters, the transition 
radius ([10,120]~au) and the scaling factor $\chi$ ([0:$\sqrt{2}$]).  
As before, we first run a coarse grid at a resolution of 10~au and 
0.1, followed by a finer grid with a resolution of 1~au and 0.01 over 
a zoomed in region covering the region of global best fit.  
Figure~\ref{figurea4} presents model first moment maps for a disk 
with a radial component to the gas velocity for a range 
of values of $\chi$ and a transition radius fixed at 100~au.  
The outer disk parameters 
are values appropriate for the HD 100546 protoplanetary disk.

\begin{figure*}[]
\centering
\subfigure{\includegraphics[width=0.33\textwidth]{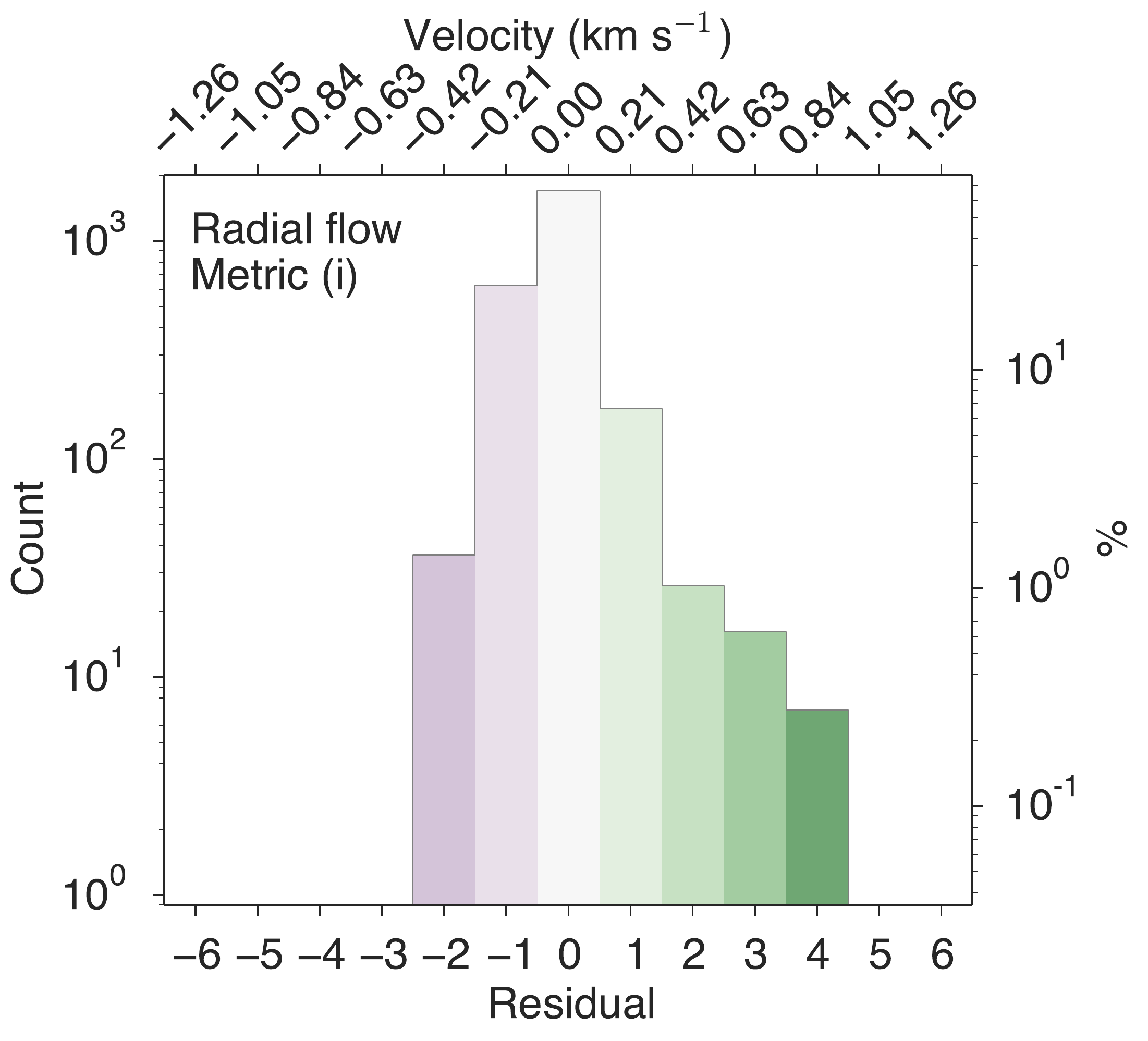}}
\subfigure{\includegraphics[width=0.33\textwidth]{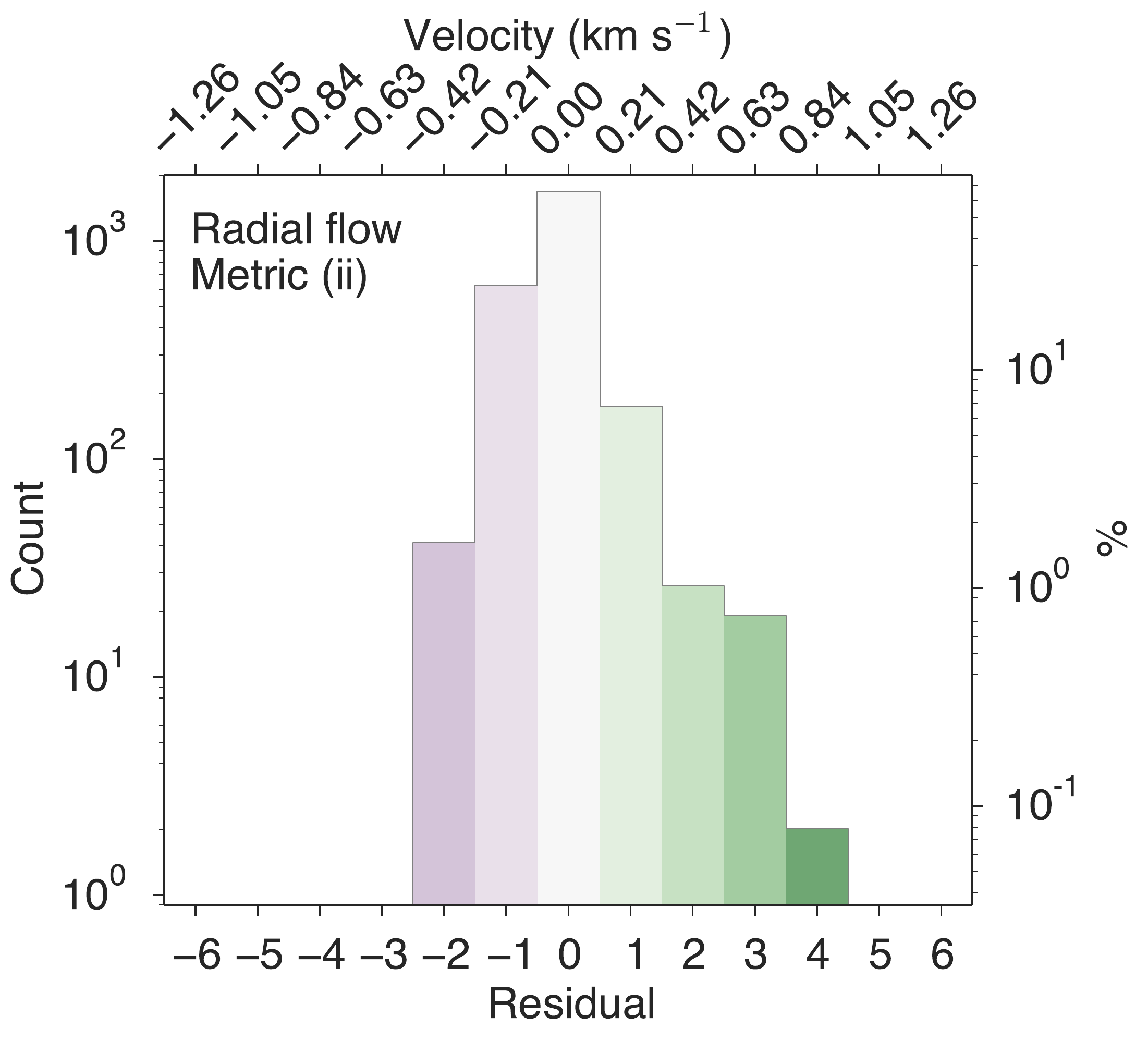}}
\subfigure{\includegraphics[width=0.33\textwidth]{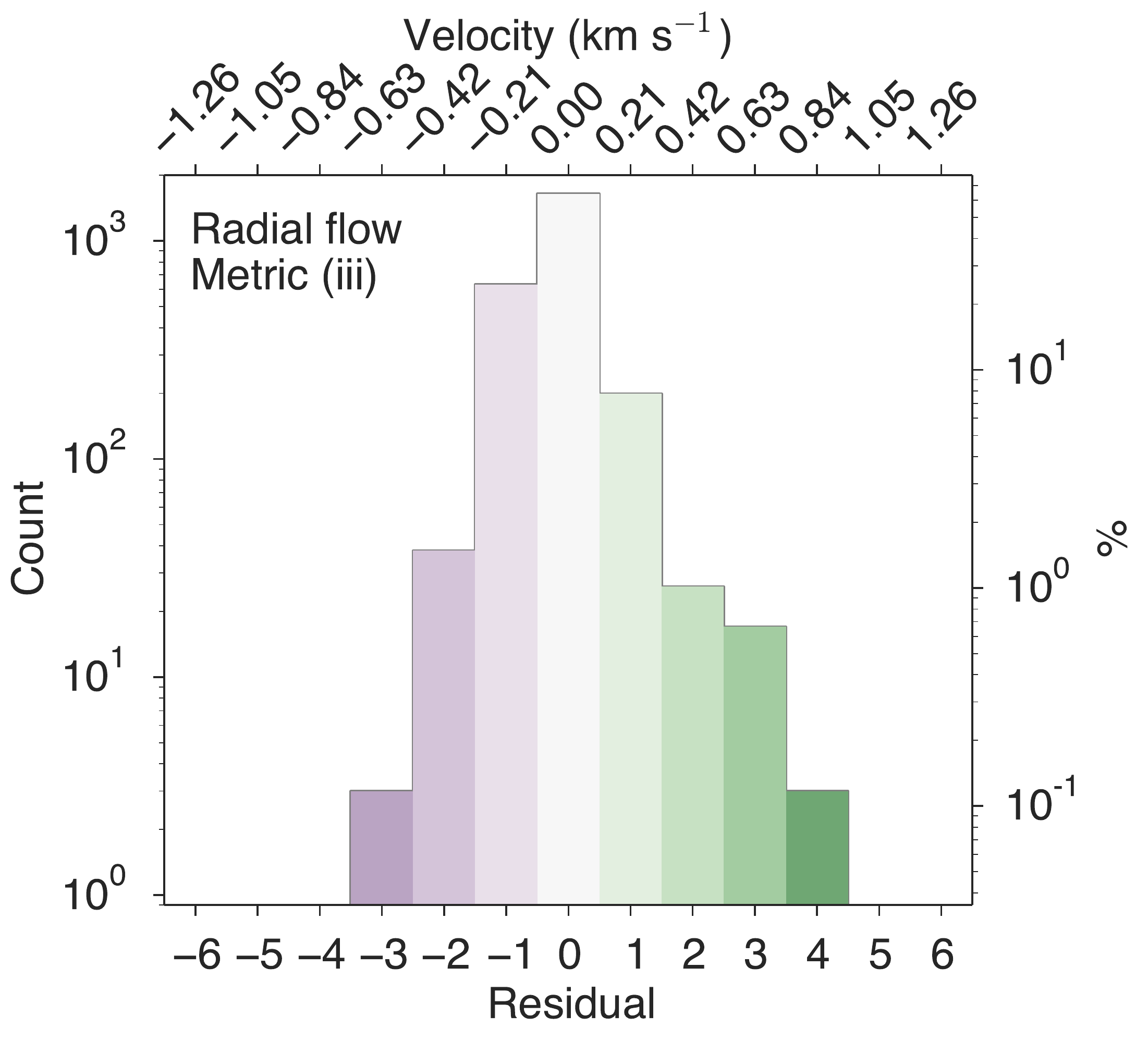}}\\
\subfigure{\includegraphics[width=0.33\textwidth]{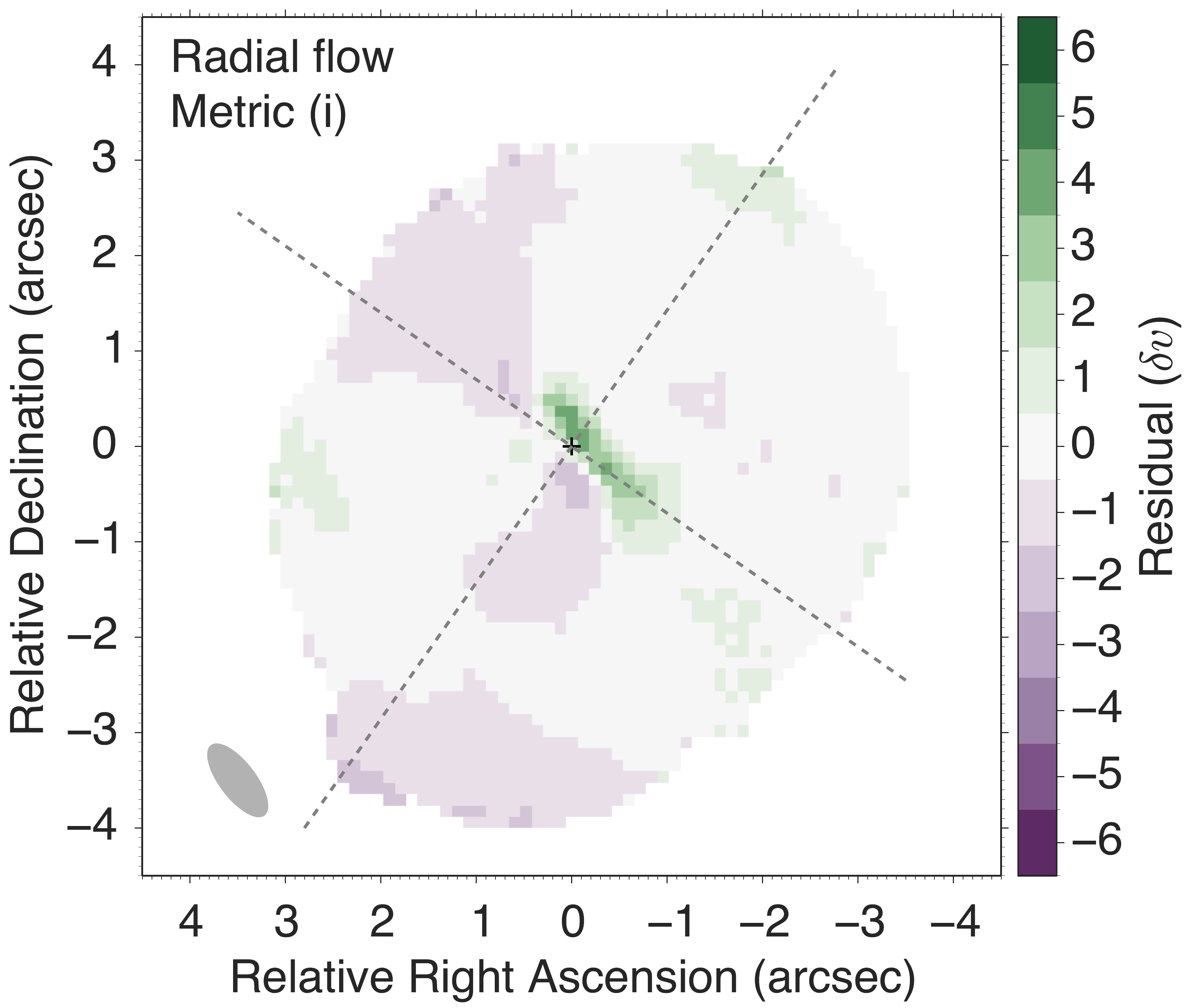}}
\subfigure{\includegraphics[width=0.33\textwidth]{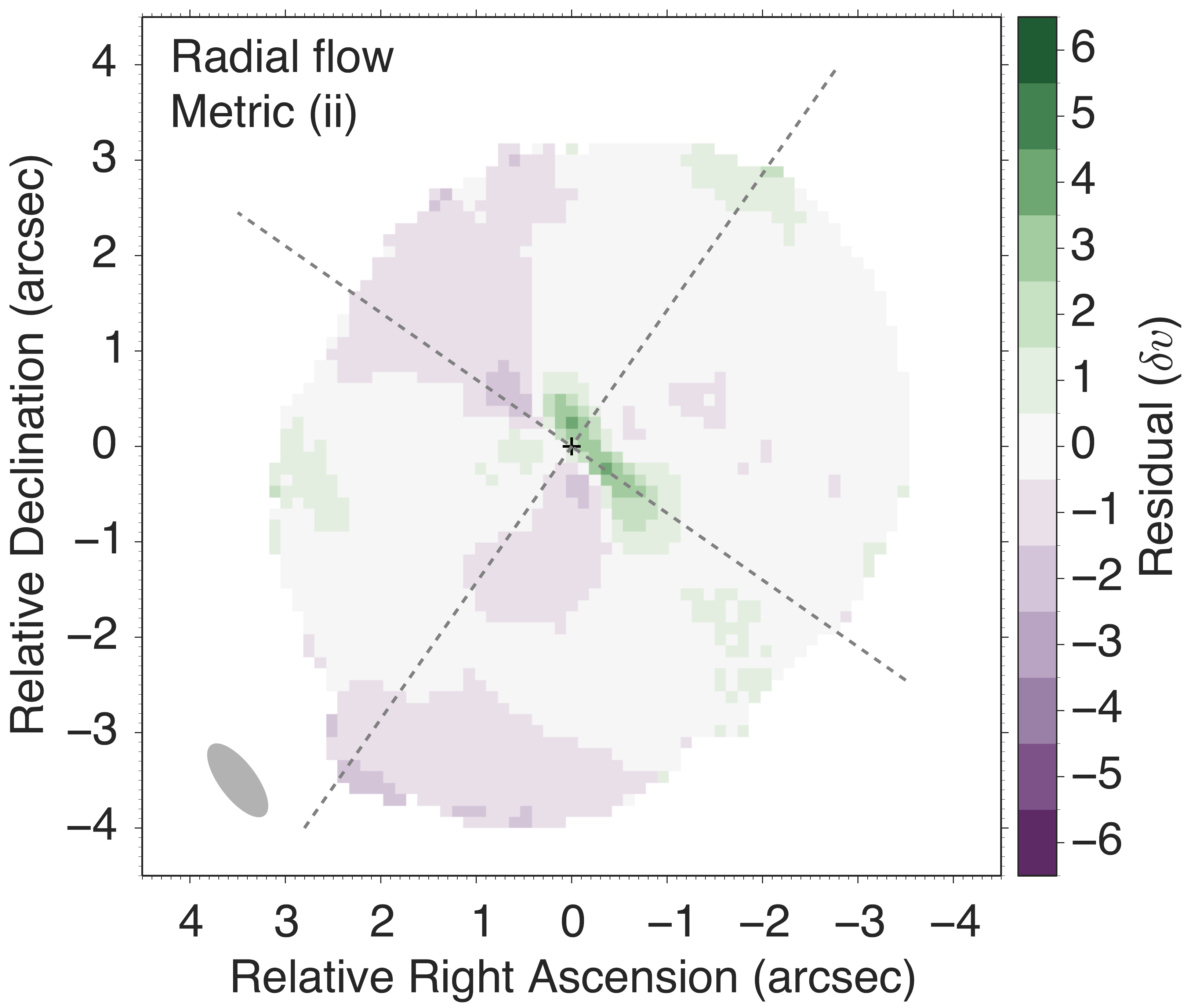}}
\subfigure{\includegraphics[width=0.33\textwidth]{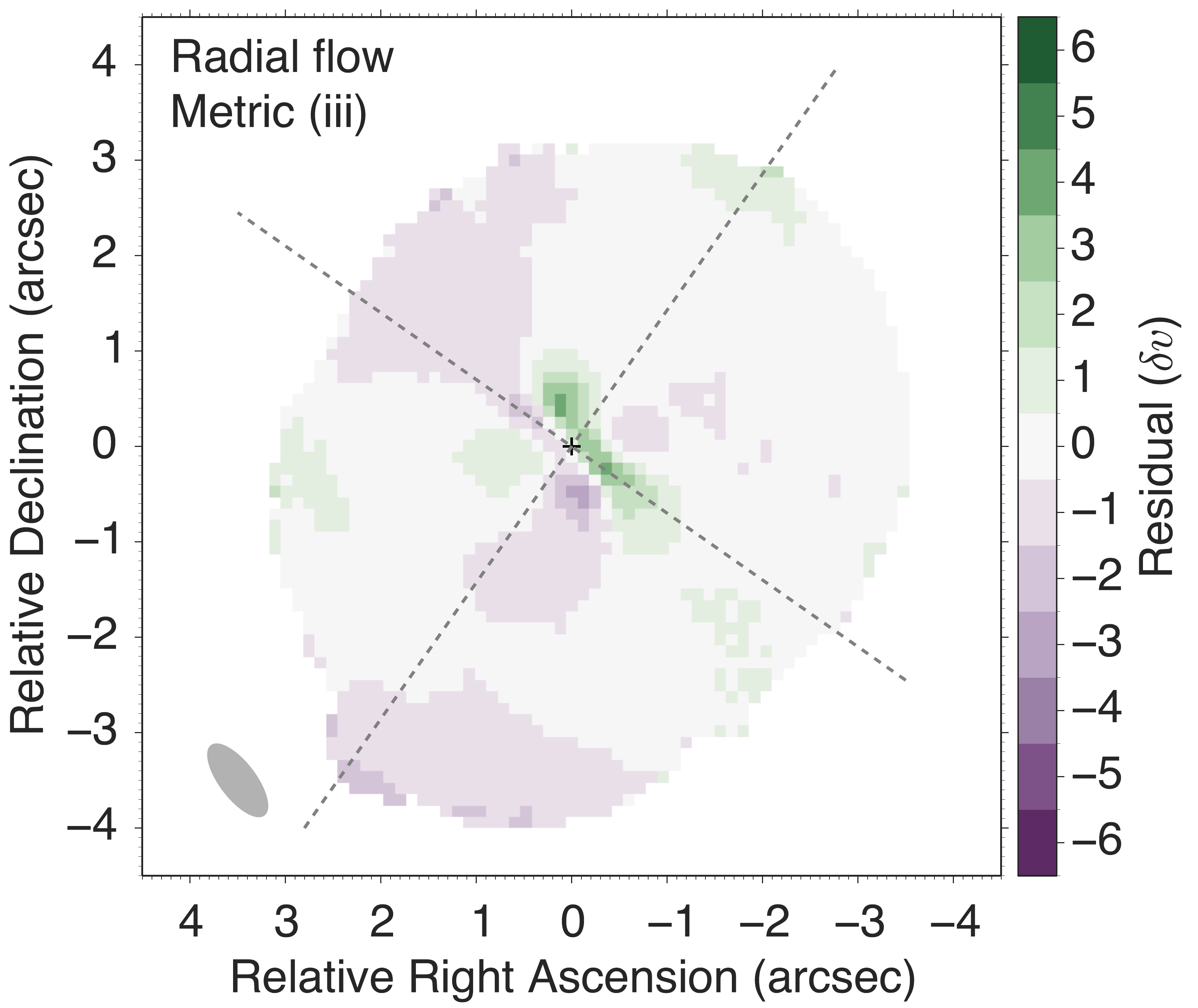}}
\caption{{Residual histograms (top) and maps (bottom) for a protoplanetary disk 
with a radial velocity component for each metric of best fit.  
The histograms are displayed on a log scale to emphasise the largest residuals.}}
\label{figure10}
\end{figure*}

Figures~\ref{figure7} and \ref{figure10} and 
Table~\ref{table3} present the statistics and residuals for the 
best-fit radial flow model.  
Metric (i) favours a radial component to the flow within a 
radius of 54~au and a velocity scaling factor of $\chi = 1.23$ which
is close to the free-fall velocity.  
Metric (ii) selects a model with very similar parameters 
(52~au and $\chi=1.40$) whereas metric 
(iii) favours a larger transition radius (84~au) and a slower 
flow ($\chi=0.63$).  
Inspection of the residuals in Fig.~\ref{figure10} shows that all three metrics 
have a very similar distribution of residuals. 
However, the radial flow model does not reproduce the velocity structure 
as well as the warped disk: using metric (iii) the peak residual for the former 
model is 3.7$\delta v$ versus 2.4$\delta v$ for the latter. 

\begin{table*}
\centering
\caption{{Best-fit parameters for the radial flow model.} \label{table3}}
\def\arraystretch{1.5}
\begin{tabular}{lccccccc}
\hline\hline
Model     & Metric       & Transition     & Scaling    & Pixel  & Percentage & Sum of           & Peak \\ 
          & of best-fit  &  radius (au)   &   factor   & number &            & residual squares$^{1}$ & residual ($\delta v$) \\ 
\hline
Radial flow        & (i)   & 54 & 1.23 & 1685 & 65.8\% & 0.461 & 4.49 \\
                   & (ii)  & 52 & 1.40 & 1678 & 65.5\% & 0.440 & 3.92 \\
                   & (iii) & 84 & 0.63 & 1644 & 64.2\% & 0.454 & 3.74 \\
\hline
\end{tabular}
\tablefoot{$^{1}$~Scaled by the total number of unmasked pixels.}
\end{table*}

\section{Discussion}
\label{discussion}

\subsection{Is the inner disk warped?}

The analysis of the kinematics of CO emission from 
the HD~100546 protoplanetary disk presented here has suggested the 
presence of a misaligned molecular gas disk within 100~au of the central star.  
The best-fit warp model suggests that the inner gas disk has a position angle that is almost 
orthogonal to that of the outer disk, 
and an inclination almost edge-on to the line-of-sight (see Fig.~\ref{figure9}).   
As discussed in Sect.~\ref{radialflow}, 
scattered light images with a spatial resolution down to $1-2$~au 
suggest no misalignment between the inner and outer {\em dust} disks 
\citep{garufi16,follette17,lazareff17}. 
Thus, if scattered light observations are to be reconciled with the predicted warp, 
the small dust grains must be severely decoupled from the gas. 
At this time, we are not aware of any mechanism that could produce 
such a phenomenon.

\begin{figure}[]
\centering
\includegraphics[width=0.5\textwidth]{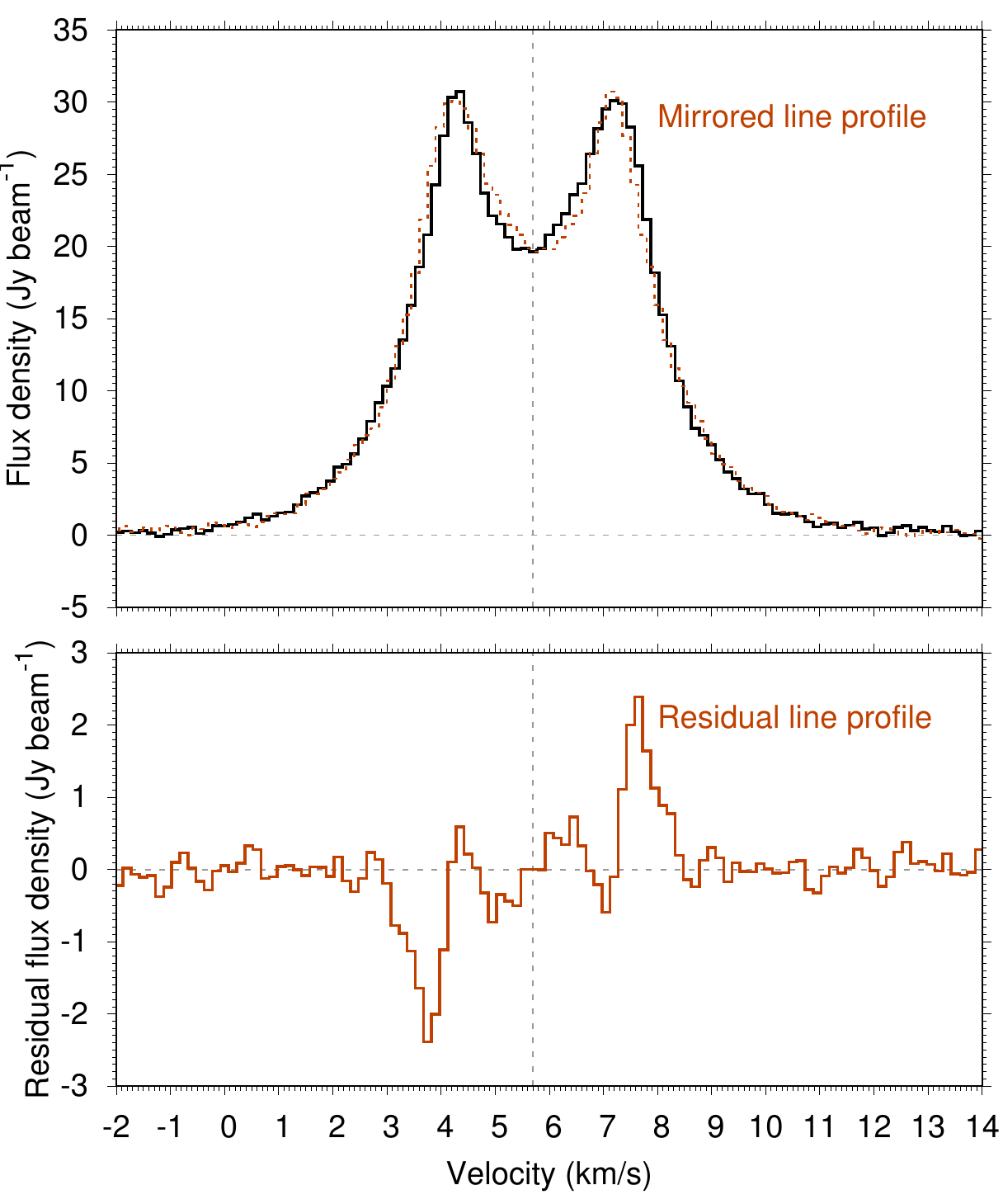}
\caption{CO $\mathrm{J}=3-2$ line profile extracted from within the 
$3\sigma$ contour of the integrated intensity (see Fig.~\ref{figure3}).  
The bottom panel shows the residual line profile highlighting the asymmetry in the 
red and blue peaks.} 
\label{figure11}
\end{figure}

However, the presence of a warp in the inner disk 
does help to explain several observations: 
asymmetries in intensity of the CO emission in both the single-dish 
data \citep{panic10} and the ALMA data presented here 
(see Figs.~\ref{figure3} and \ref{figure11}), 
and the ``dark wedge" seen in the scattered light images \citep{garufi16} is 
predicted by scattered light models including a misaligned inner disk 
\citep{marino15,facchini17}.  
This ``dark wedge'' aligns well with the ``dark lane'' in the 
CO eighth-moment map in Fig.~\ref{figure3}.

The presence of a disk warp in HD~100546 was originally 
proposed by \citet{panic10} and based on asymmetries seen 
in the red and blue peaks of single-dish spectra 
observed with APEX.  
The ALMA data presented here show no such asymmetries in the 
peaks of the line profile (see Fig.~\ref{figure11}); however, 
the mirrored line profile does highlight that the red and blue 
lobes of emission have different shapes when integrated over the disk.  
The bottom panel in Fig.~\ref{figure11} shows the residuals following 
subtraction of the average of the original and mirrored line profiles from the 
original.
This is consistent with the morphology seen in the channel maps 
and described in Sect.~\ref{almaimages}.    

A warp in the inner disk 
could shadow the outer disk and lead to an asymmetric temperature structure, 
because one side of the disk is more illuminated by the central star 
than the opposite side (see figure 3 in \citealt{panic10}).  
Because the $^{12}$CO emission is optically thick, it traces 
the gas temperature, and thus any perturbations, in the emitting layer, usually 
the disk atmosphere.  
The APEX $^{12}$CO $\mathrm{J}=3-2$ line profile from \citet{panic10} 
shows a stronger blue peak.  
The dates of the observations taken with 
APEX and ALMA are November 2008 \citep[][]{panic10} and 
November 2012 \citep[][]{walsh14}, respectively.  
There also exist multiple APEX observations of 
the $^{12}$CO $\mathrm{J}=6-5$ transition.  
The line profile presented in 
\citet{panic10} shows an even higher-contrast asymmetry in the blue and red 
peaks than in the $\mathrm{J}=3-2$ line.  
More recent observations taken 
between April and July 2014 and presented in \citet{kama16} show a similar asymmetry; however, 
the ratio between the peaks is within the flux density uncertainties 
of the observations (M.~Kama, priv.~commun.).   
Thus, it is currently difficult to draw concrete conclusions on any periodicity 
in the shapes of the line profiles.   
Conservatively assuming that the warp has rotated $\approx 90\degree$ between 
acquirement of the original APEX data (with the stronger blue peak) and the ALMA data 
(with no asymmetry in the line profile)
suggests a warp precession period of the order of $16$~yr.  

The presence of a warp indicates the presence of a perturbing companion.  
A massive ($\approx 20 M_\mathrm{J}$) planetary companion has been proposed to reside 
at $\approx 10$~au in HD~100546.
This is thought to be responsible for clearing the inner cavity in both 
the dust and gas \citep{mulders13,panic14,walsh14,pinilla15,wright15,garufi16},  
as well as inducing dynamical perturbations in [OI], OH, and CO line emission 
from the inner region \citep{acke06,vanderplas09,brittain14}.   
In addition, a second companion, which may still be in the act of formation, 
has been proposed to reside at $\approx50$~au with a mass estimate of $\approx 15 M_\mathrm{J}$ 
\citep{quanz13,walsh14,quanz15,currie15,pinilla15}.  
However, as mentioned in the introduction, the identification 
of this point source as a (proto)planet has been disputed in a recent 
analysis of MagAO/GPI observations showing close-to-zero proper motion of 
the point source on a 4.6~yr timescale at the $2\sigma$ level \citep{rameau17}. 
Given that the presence of the outer ring of (sub)mm dust emission is 
difficult to explain in the absence of an outer companion \citep{walsh14,pinilla15}, 
we proceed for the remainder of the discussion with the assumption that this disk 
``feature'' indeed arises due to a (proto)planet.  
{It should be noted that the planet masses derived 
from previous analysese have all assumed that the planet orbital path and disk 
are co-planar.}

The precession period of a warp induced by a perturbing companion can be estimated 
using equation (2) in \citet{debes17}, used for the case of TW~Hya, 
and which was originally derived by \citet{lai14},  
\begin{equation}
P \propto (\mu_c \cos i_c)^{-1}(M_\ast)^{-1/2}(r_\mathrm{disk})^{-3/2}(a_c)^{3}
\end{equation}
where $\mu_c$ is the mass ratio of the companion to the central star, 
$i_c$ is the angle between the star-planet orbital plane 
and disk plane (here 80\degree), 
$r_\mathrm{disk}$ is the radius of the outer edge of the inner disk, and 
$a_c$ is the orbital radius of the planet.  
Assuming that the 
protoplanet at 50~au is the perturbing body and setting 
$\mu_c = 0.006$, $r_\mathrm{disk} = $40~au, and 
$a_c=50$~au results in a precession period of 1940~yr.  
Hence, it is unlikely that the outer companion has 
perturbed the inner disk creating a warp that explains 
the change in shape in line profile seen in the single-dish data and 
these ALMA data on a 4-year baseline.  
Assuming instead that the inner companion is triggering the 
warp, and that the inner disk cavity is not completely devoid of gas 
($\mu_c = 0.008$, $r_\mathrm{disk} = $10~au, $i_c = 80$\degree, and 
$a_c=10$~au), leads to a precession period of 111~yr. 

Keeping all other parameters constant, a companion mass of 
$\approx 140 M_\mathrm{J}$ ($\approx 0.1 M_\odot$) is required for 
a warp precession period of $\approx 16$~yr.  
There is no evidence that HD~100546 is in 
a binary system with an M-dwarf star.
Alternatively, a closer-in companion of only 
$\approx 4.5 M_\mathrm{J}$ at 1~au and assuming an inner disk radius of 1~au, 
leads to a similar precession period.
However, these ALMA data suggest that the non-Keplerian gas motion lies significantly beyond 
1~au since the highest velocity gas detected only probes down to a radius of 8-16~au depending 
upon the assumed disk inclination.  
Longer baseline (in time) observations are required to determine if any 
clear periodicity in the line profile shape exists, and to then 
relate this more concretely to the precession period of an inner disk warp.  

Similar analyses have revealed warps in disks around other stars.  
\citet{rosenfeld12} showed that the kinematic structure of TW Hya 
could be reproduced using a parametric model of a disk warp 
with a moderate inclination ($\approx~8\degree$) at a radius of 5~au.  
\citet{facchini14} demonstrated the observed warp amplitude  
could be induced by a misaligned close-in companion as massive as 
$\approx 14 M_\mathrm{J}$ orbiting within 4~au of the central star.  
New scattered light data from HST/STIS, coupled with 
archival data over a 17-year baseline, reveal an orbiting 
azimuthal brightness asymmetry, with a 
period of $\approx 16$~yr \citep{debes17}.  
The authors argue that this is consistent with partial 
shadowing of the outer disk by a misaligned inner 
disk interior to 1~au and that is precessing due to 
the presence of a roughly Jupiter-mass planetary companion.  

\subsection{Is a radial flow contributing to the CO kinematics?}

The presence of a 
severely misaligned gas disk 
in the inner $\sim 100$~au of HD~100546 is contradicted 
by multiple scattered light observations at a resolution of $~\sim 1-2$~au.  
On the other hand, a radial flow does not require any misalignment between the 
gas and dust disks.

Recent observations of Br$\gamma$ emission 
with VLTI/AMBER have revealed that the accretion rate of HD~100546 
is of order $10^{-7}$~M$_\odot$~yr$^{-1}$ \citep{mendigutia15} which suggests that 
inner disk material ($\lesssim 1$~au) is replenished from the outer disk despite the presence of 
a CO gas cavity within 11~au \citep{vanderplas09}.    
Further, very recent observations of HD~100546 with SPHERE/ZIMPOL have revealed 
an intriguing bar-like structure within the 10~au dust cavity and at 
a similar position angle as our residuals \citep{mendigutia17}.  
The authors have speculated that this feature arises due to small dust 
grains being dragged along by a channelled gas flow through the cavity.  
However, the ALMA data presented here are not sensitive to these spatial scales.  
Modelling of the ALMA data including a radial velocity component 
suggests that if such a component is present, it arises from as far 
out as $\approx 50$~au.
Taken together, the ALMA data and SPHERE/ZIMPOL data suggest an 
intriguing connection between the gas kinematics on larger ($\gg 10$~au) 
scales and the dust and gas morphology within the cavity.

Despite this intriguing connection, 
the model including only a radial component does not fit the data as well 
as the warp model.  
One reason for this is that the radial flow model enforces that the 
maximum line-of-sight velocity due to the flow aligns with the disk minor axis 
as defined by the position angle of the outer disk.  
On the other hand, the warp model suggests that the maximum line-of-sight velocity 
lies at a position angle of 64\degree: this is 9\degree~rotated from the disk minor axis.
It remains possible that the inner $\lesssim 100$~au 
of the HD~100546 protoplanetary disk is host to a both a warp and a radial flow.  
In that case, the warp parameters (i.e., the inclination and twist angles) 
may not be as extreme as suggested here.  
However, fitting a more complex and multiple-component velocity structure 
requires guidance from higher spatial resolution data.

Thanks to ALMA, evidence is mounting that radial flows may be a common 
feature of transition disks.
HD~142527 is another protoplanetary disk in which  
gas motion traced by \ce{HCO+} 
($\mathrm{J}=4-3$) and $^{12}$CO ($\mathrm{J}=6-5$) 
emission has been observed to deviate from purely Keplerian rotation 
\citep{casassus13,rosenfeld14,casassus15}.  
HD~142527 is a transition disk which possesses the largest known
dust cavity in sub(mm) emission 
\citep[$\approx 140$~au;][]{casassus13,fukagawa13}. 
The observed gas motions were postulated to arise from fast 
(near free-fall) radial flows across the cavity and 
accretion onto the central star via a severely misaligned inner disk 
\citep{casassus13,casassus15,marino15}.  
\citet{marino15} propose that the inner and outer disk are misaligned 
by 70\degree.  
Also, ``hot-off-the-press'' ALMA observations of 
\ce{HCO+} ($\mathrm{J}=3-2$) line emission from AA~Tau at an angular 
resolution of 0\farcs2 ($\approx 30$~au), also show 
deviations from global Keplerian rotation due 
to either a warp or fast radial inflow in the inner regions \citep{loomis17}.  
Similar to here, this was revealed by a clear twist in the first 
moment map of the molecular line emission.  
 
\section{Conclusion}
\label{conclusion}

The data and analyses shown here demonstrate that 
the kinematic structure of the gas disk around HD~100546 cannot be 
described by a purely Keplerian velocity profile with a universal inclination 
and position angle.  
Given the current evidence for the presence of (at least) 
one massive planetary companion orbiting within 10~au of the central 
star, the presence of a disk warp, albeit with extreme parameters, 
has been our favoured explanation.  
However, recent near-IR interferometric observations 
suggest that the very inner dust disk is not misaligned 
relative to the outer dust disk traced in (sub)mm emission 
and scattered light images.
This suggests that the gas and dust disks may have different morphologies 
on radial scales of $\sim 1$ to 10's of au{; however, there currently 
exists no known physical mechanism that would lead to such extreme decoupling 
between the gas and small dust grains.}  
To determine whether the alternative explanation of fast radial flows could be 
responsible for the observed kinematic structure of HD~100546, higher spatial 
{\em and} spectral resolution data are required to resolve gas emission across 
the inner dust ($\lesssim 25$~au) and gas ($\lesssim 10$~au) cavities 
and explore any spatial association with the proposed planetary candidate(s). 

\begin{acknowledgements}
We thank the referee, Dr Ruobing Dong, whose comments helped improve the 
discussion in the paper.  
We also thank Dr Ignacio Mendigut\'{i}a and Professor Rene Oudmaijer 
for sharing the results of their SPHERE/ZIMPOL observations of HD~100546 
in advance of publication.  
This paper makes use of the following ALMA data: 
ADS/JAO.ALMA\#2011.0.00863.S. ALMA is a partnership of ESO 
(representing its member states), NSF (USA) and NINS (Japan), together with NRC (Canada), 
NSC and ASIAA (Taiwan), and KASI (Republic of Korea), in 
cooperation with the Republic of Chile. 
The Joint ALMA Observatory is operated by ESO, AUI/NRAO and NAOJ.  
This work was supported by the Netherlands Organisation for Scientific Research 
(NWO: program number 639.041.335).  
C.W.~also thanks the University of Leeds for financial support.  
S.F.~is supported by the CHEMPLAN project, grant agreement 291141 
funded by the European Research Council under ERC-2011-ADG.
A.J.~is supported by the DISCSIM project, grant agreement 341137 
funded by the European Research Council under ERC-2013-ADG.

\end{acknowledgements}

\begin{appendix}

\section{Model first moment maps}

\begin{figure*}[]
\subfigure{\includegraphics[width=0.33\textwidth]{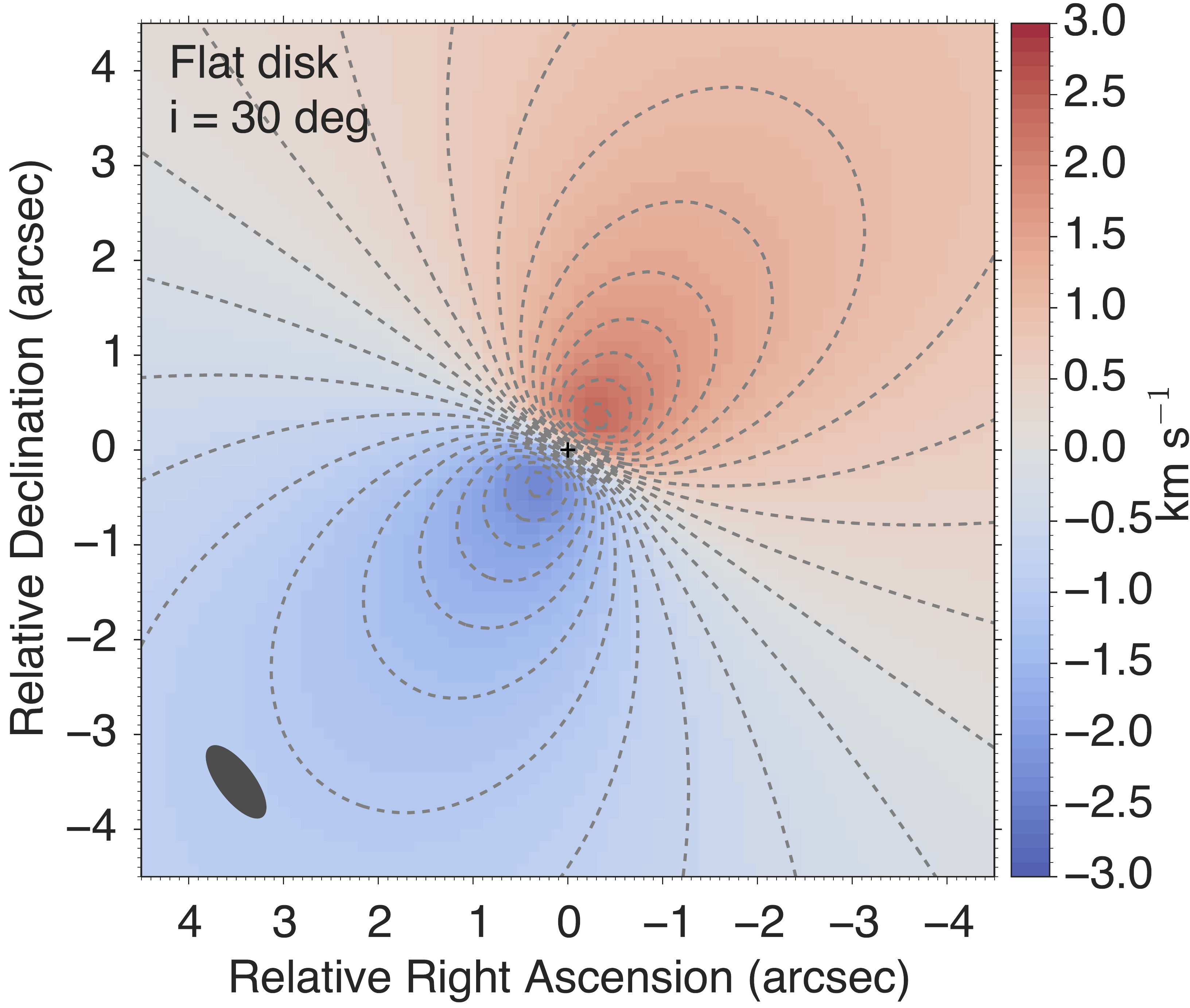}}
\subfigure{\includegraphics[width=0.33\textwidth]{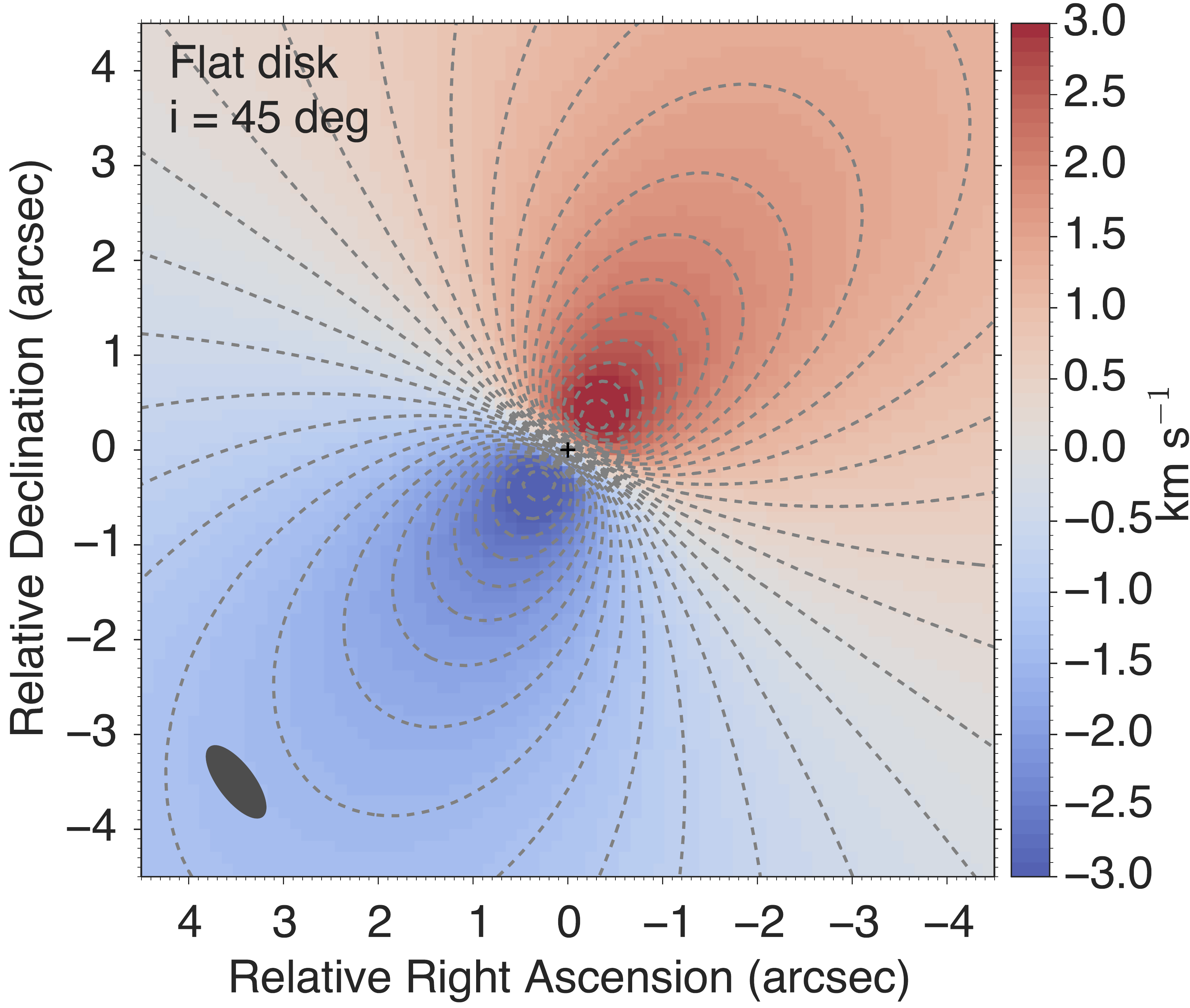}}
\subfigure{\includegraphics[width=0.33\textwidth]{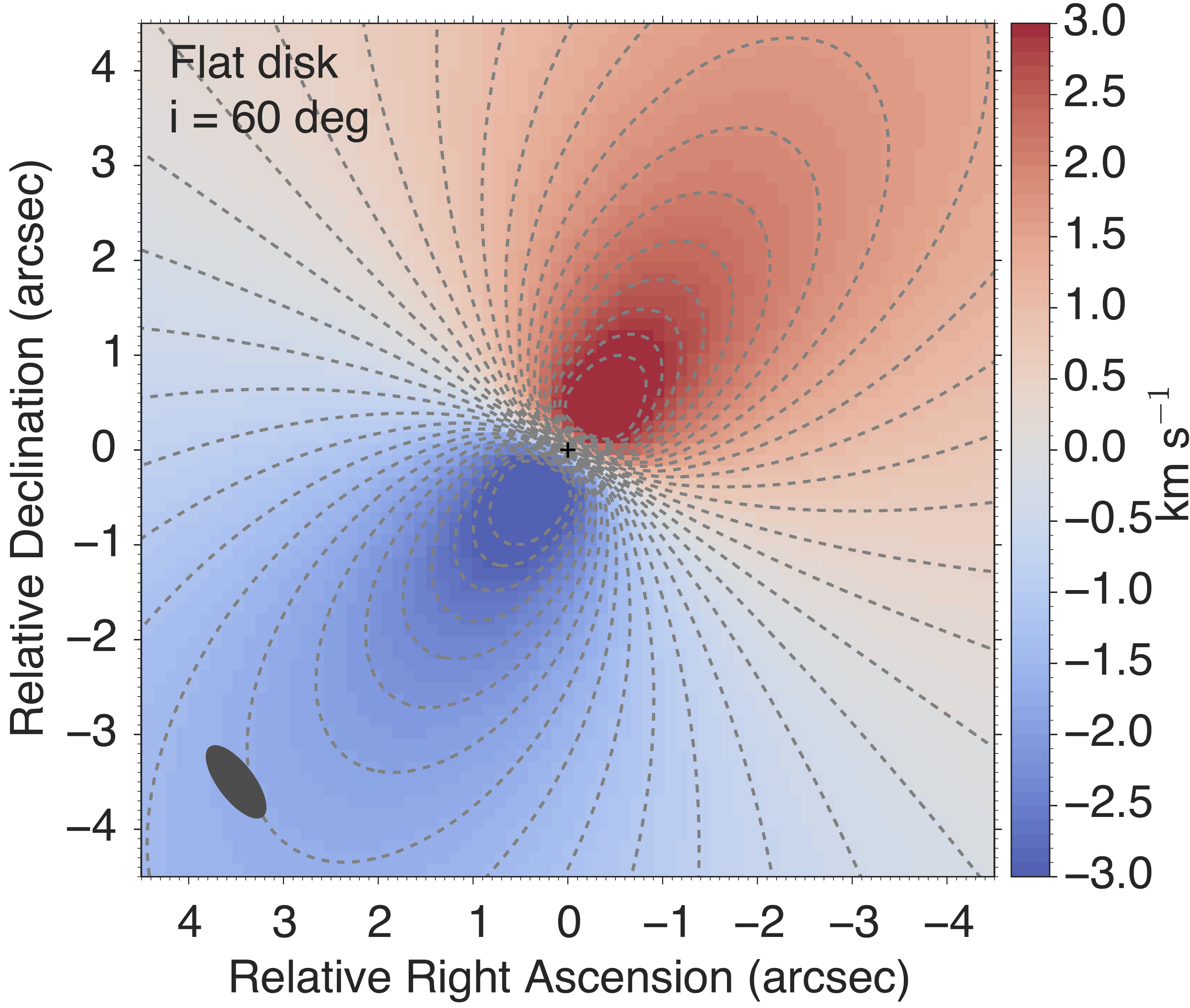}}
\caption{Model first moment maps for a geometrically flat disk with a P.A.~of 
145\degree~(representative of the HD~100546 disk) at an inclination of 30\degree~(left), 
45\degree~(middle), and 60\degree~(right).  
The contours are in units of a single spectral resolution element (0.21 \kms).} 
\label{figurea1}
\end{figure*}

\begin{figure*}[]
\subfigure{\includegraphics[width=0.33\textwidth]{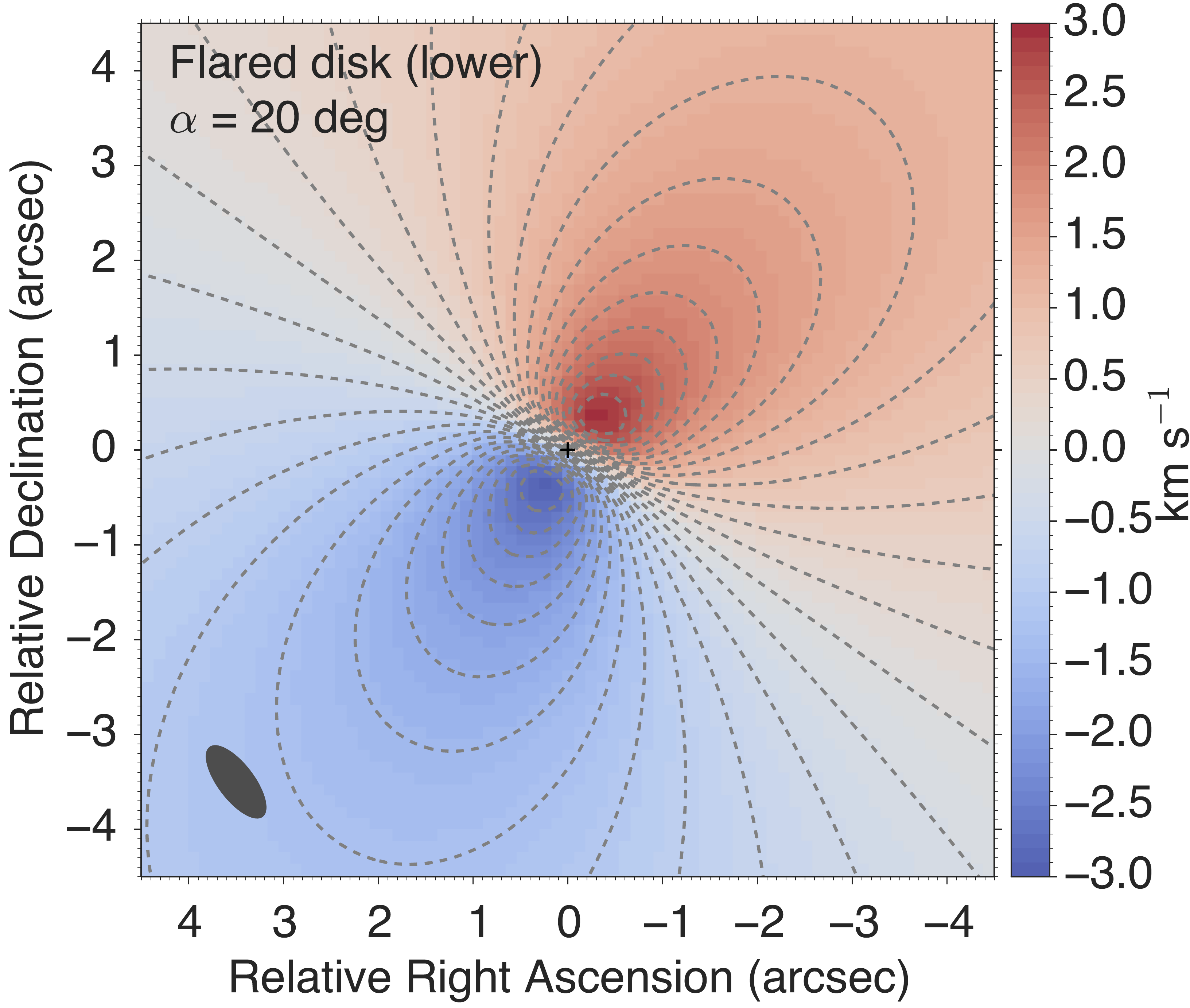}}
\subfigure{\includegraphics[width=0.33\textwidth]{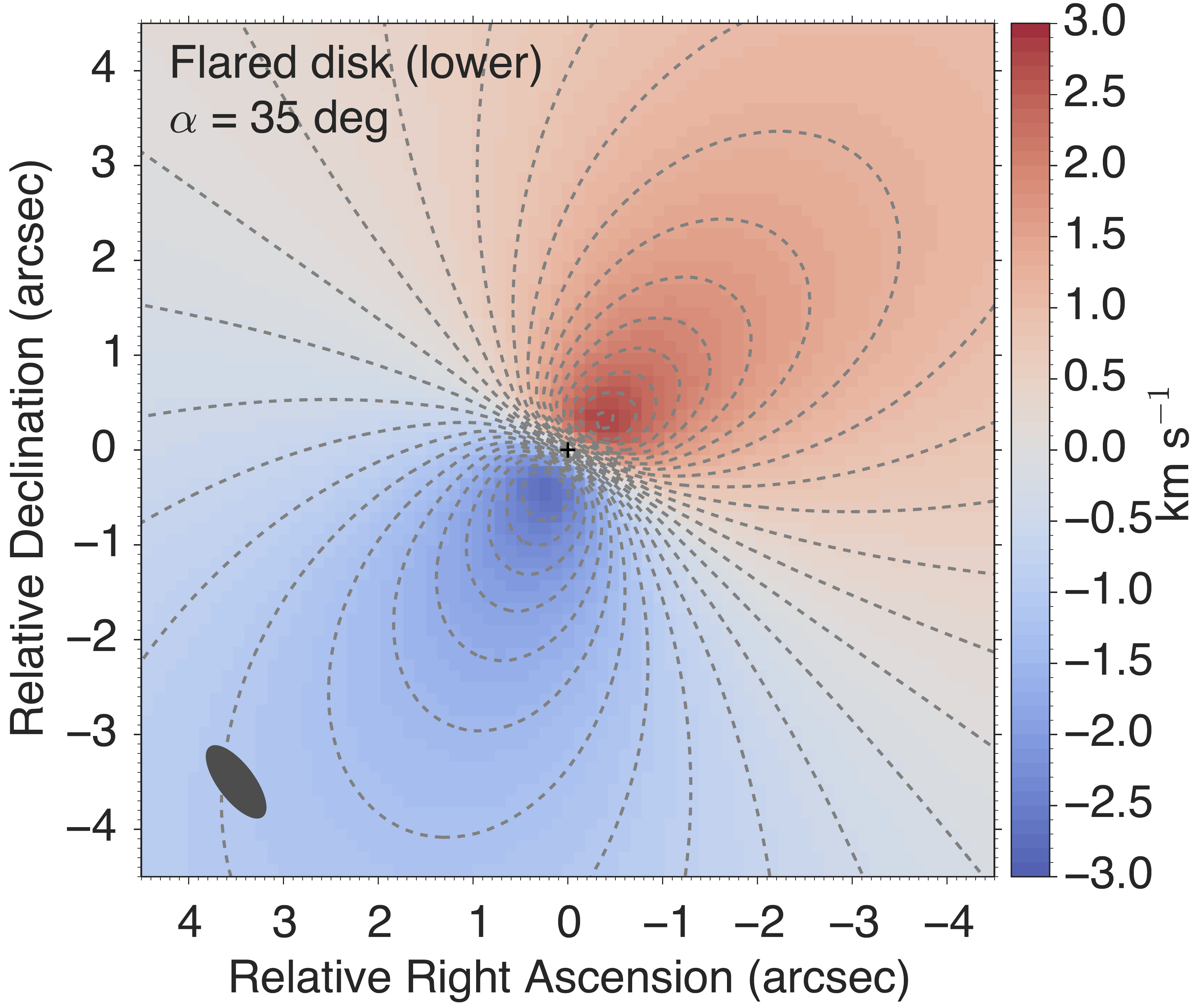}}
\subfigure{\includegraphics[width=0.33\textwidth]{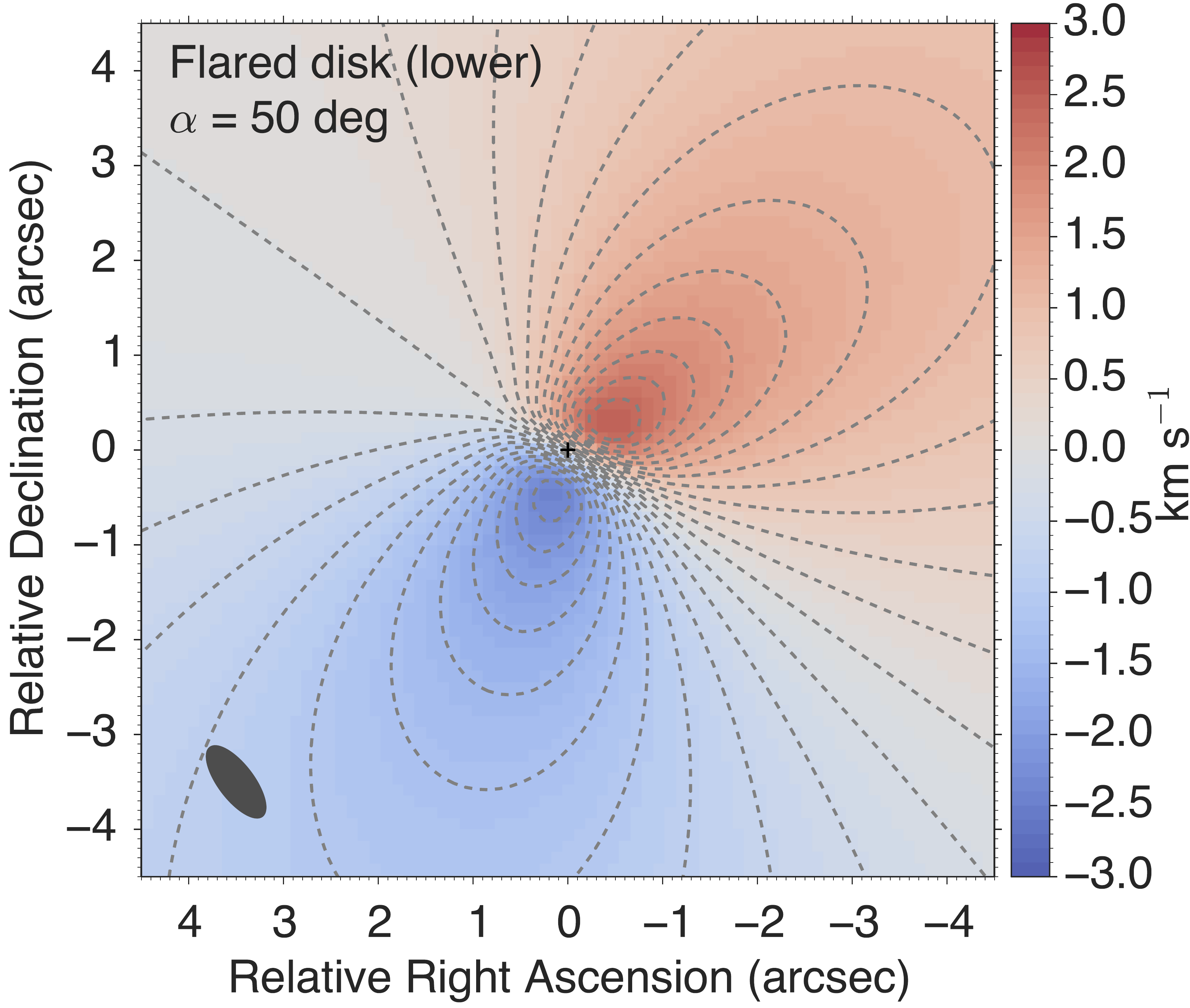}}
\subfigure{\includegraphics[width=0.33\textwidth]{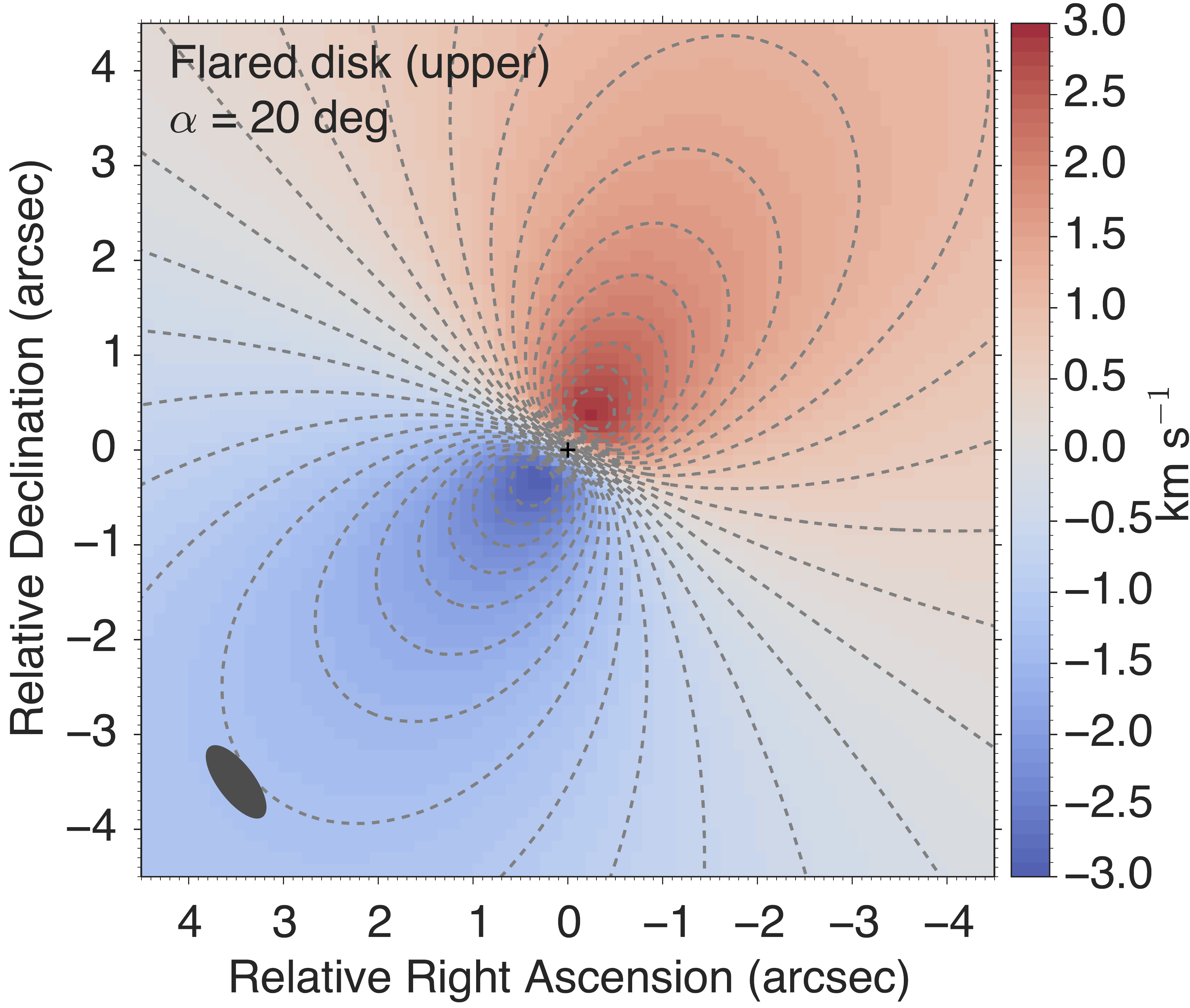}}
\subfigure{\includegraphics[width=0.33\textwidth]{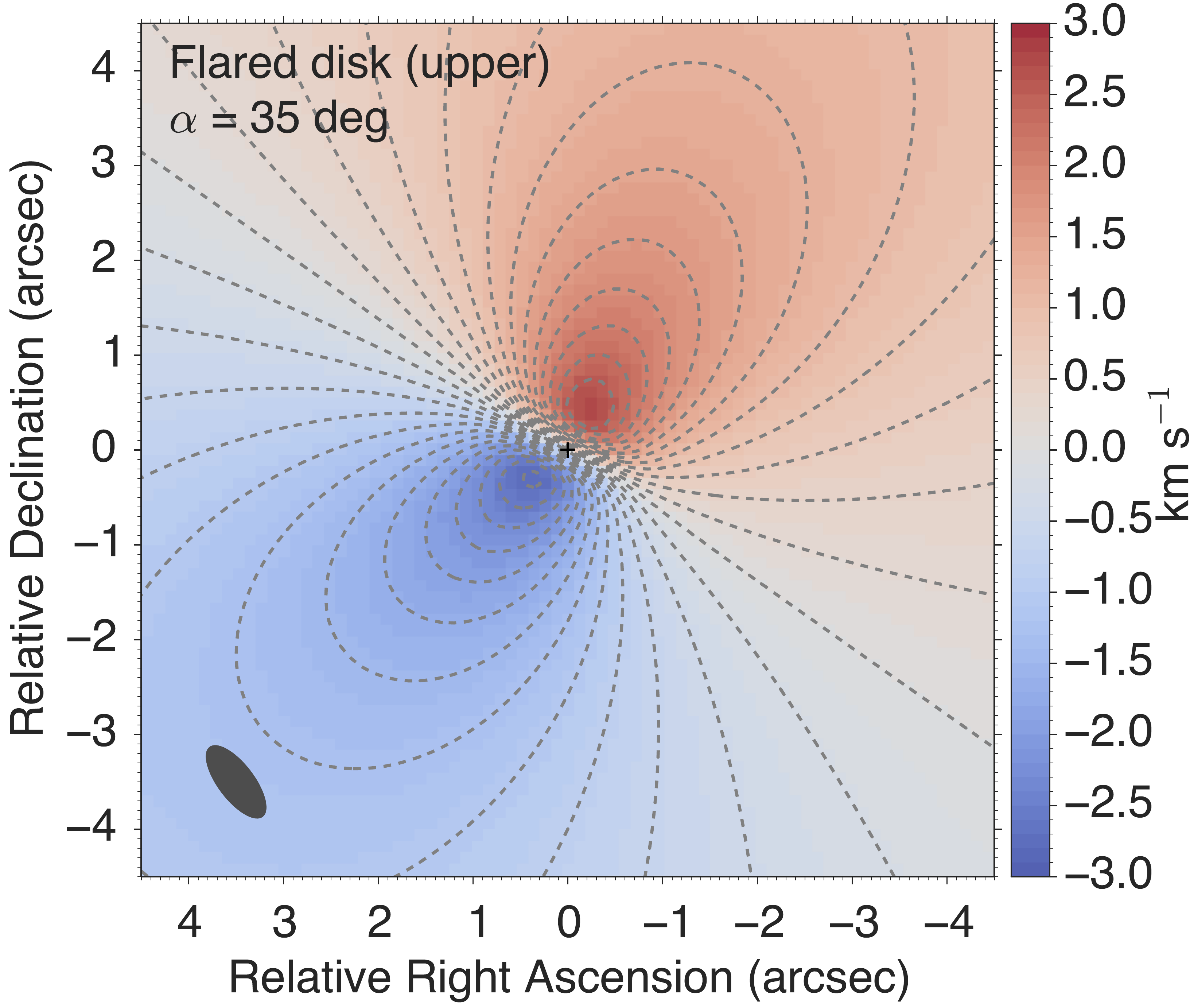}}
\subfigure{\includegraphics[width=0.33\textwidth]{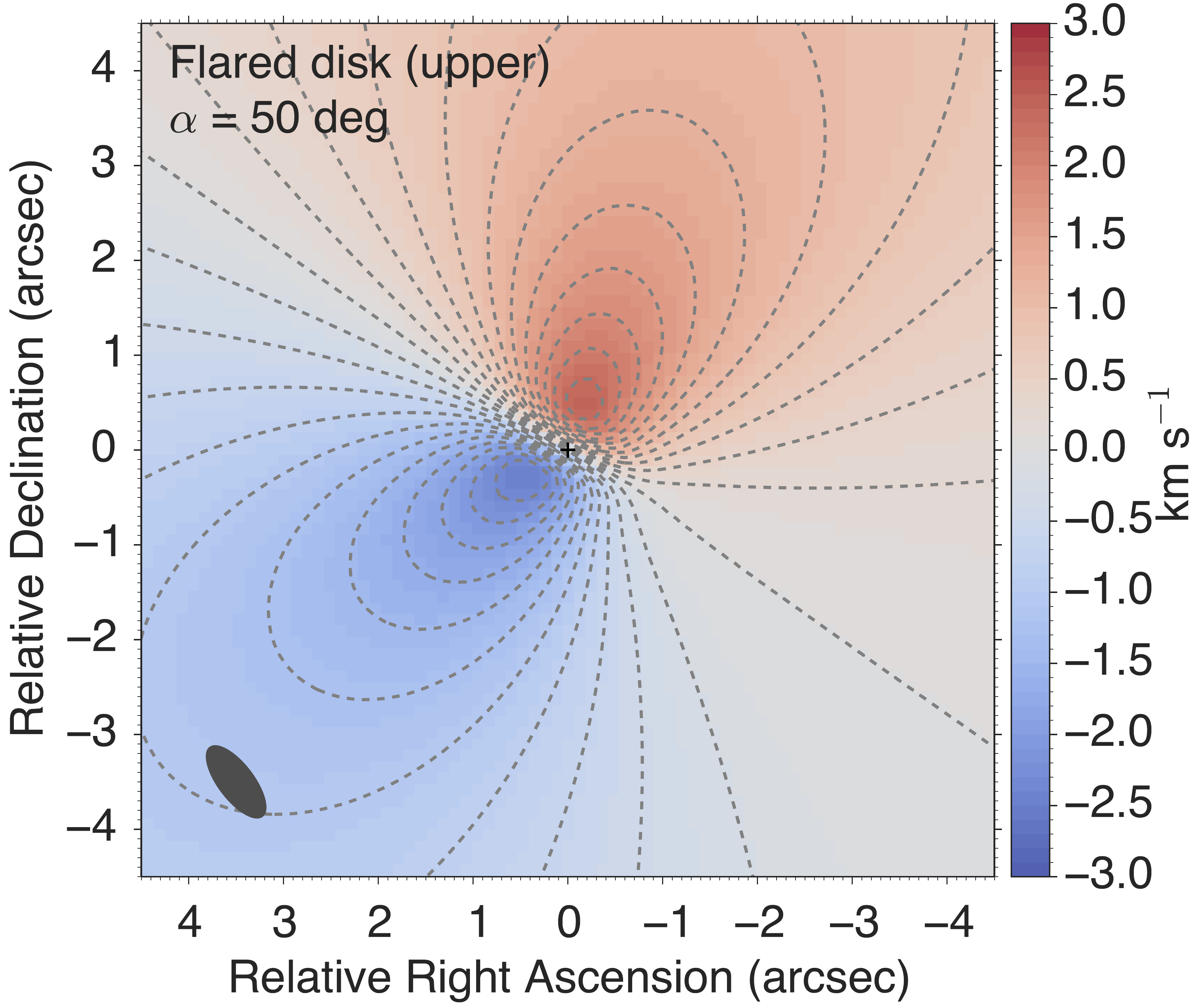}}
\caption{Model first moment maps for a flared disk (lower and upper cones) with a P.A.~of 
145\degree~and an inclination of 40\degree~(representative of the HD~100546 disk), 
with an opening angle of 20\degree~(left), 35\degree~(middle), and 50\degree~(right).  
The contours are in units of a single spectral resolution element (0.21 \kms).} 
\label{figurea2}
\end{figure*}

\begin{figure*}[]
\subfigure{\includegraphics[width=0.33\textwidth]{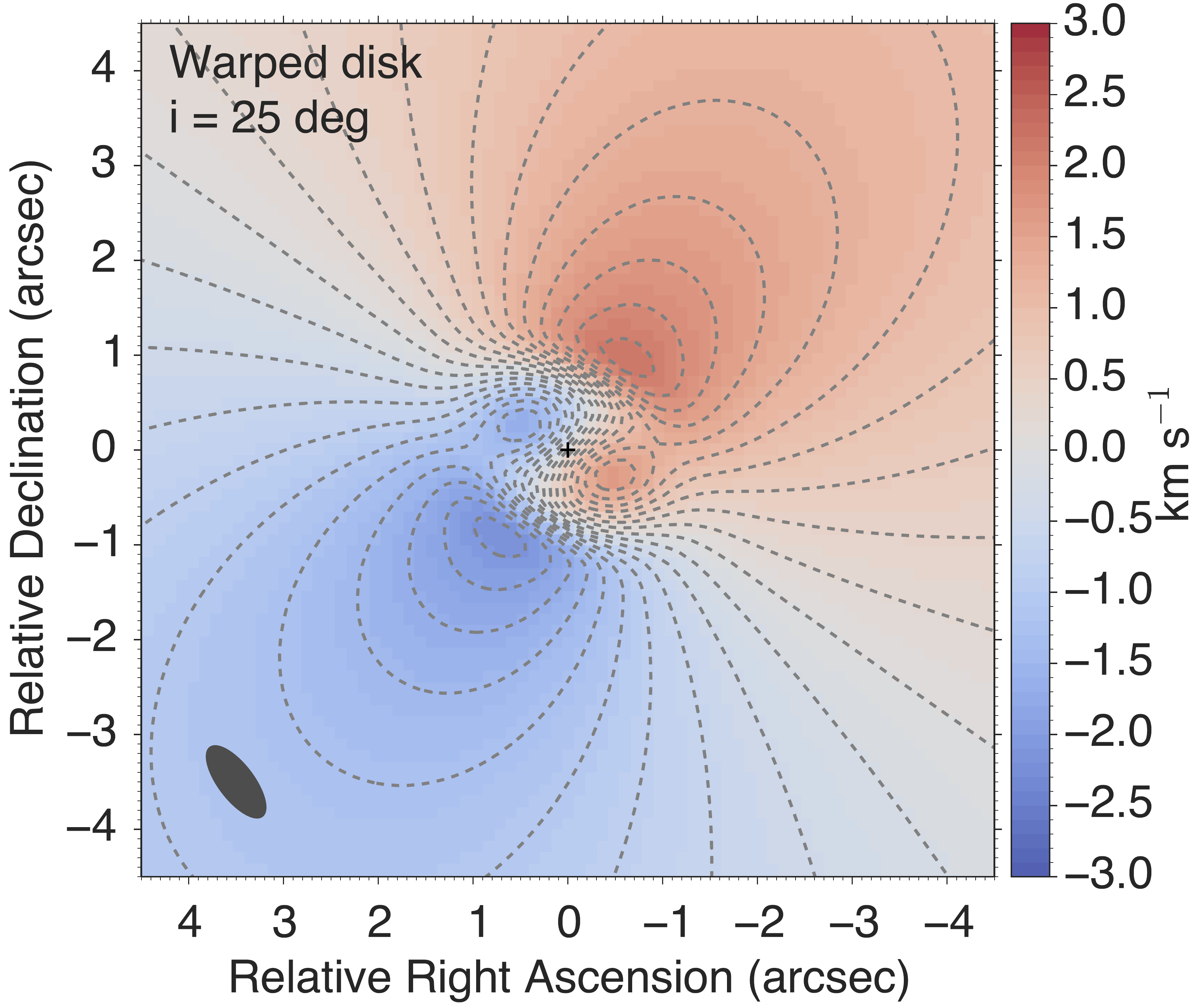}}
\subfigure{\includegraphics[width=0.33\textwidth]{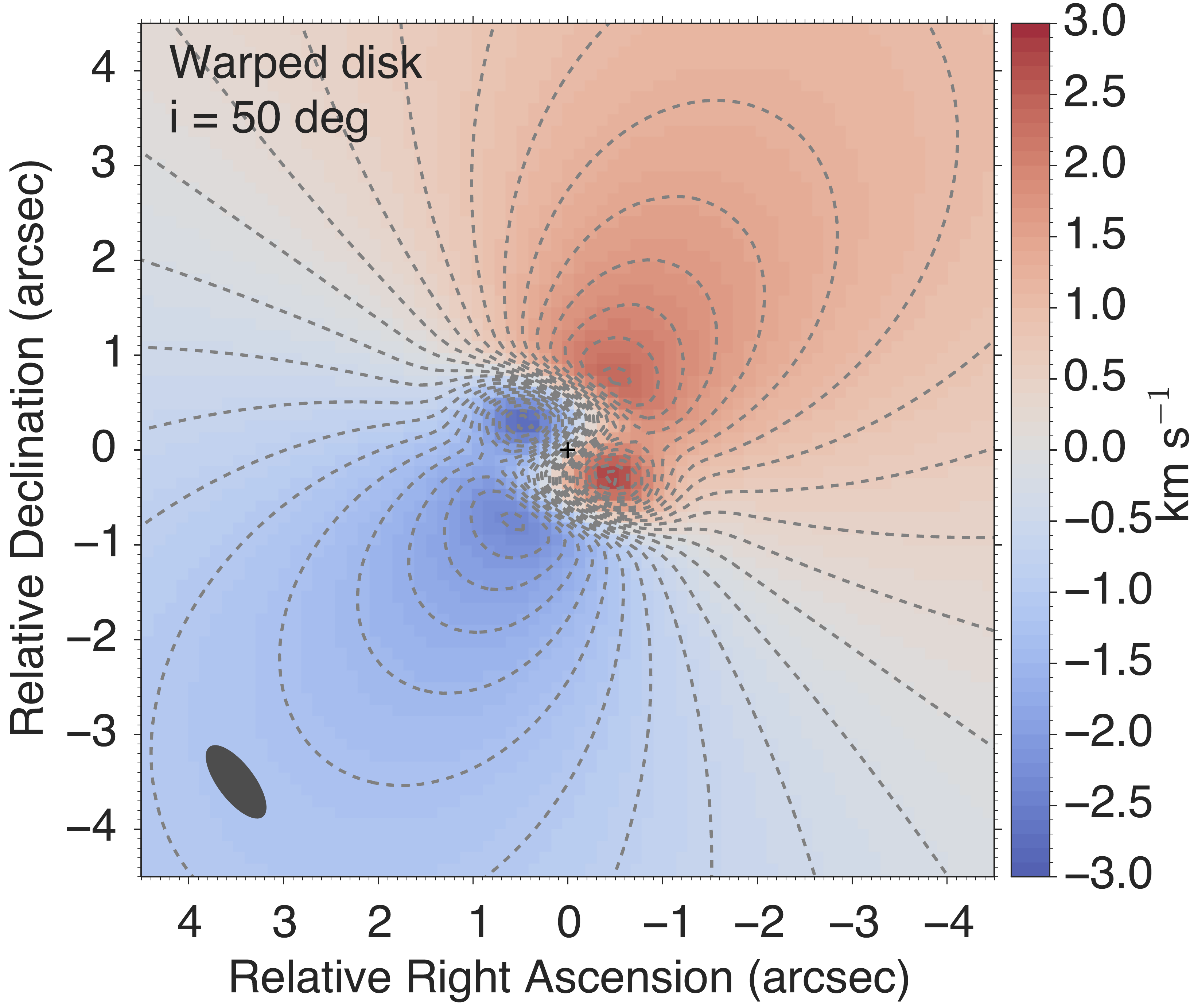}}
\subfigure{\includegraphics[width=0.33\textwidth]{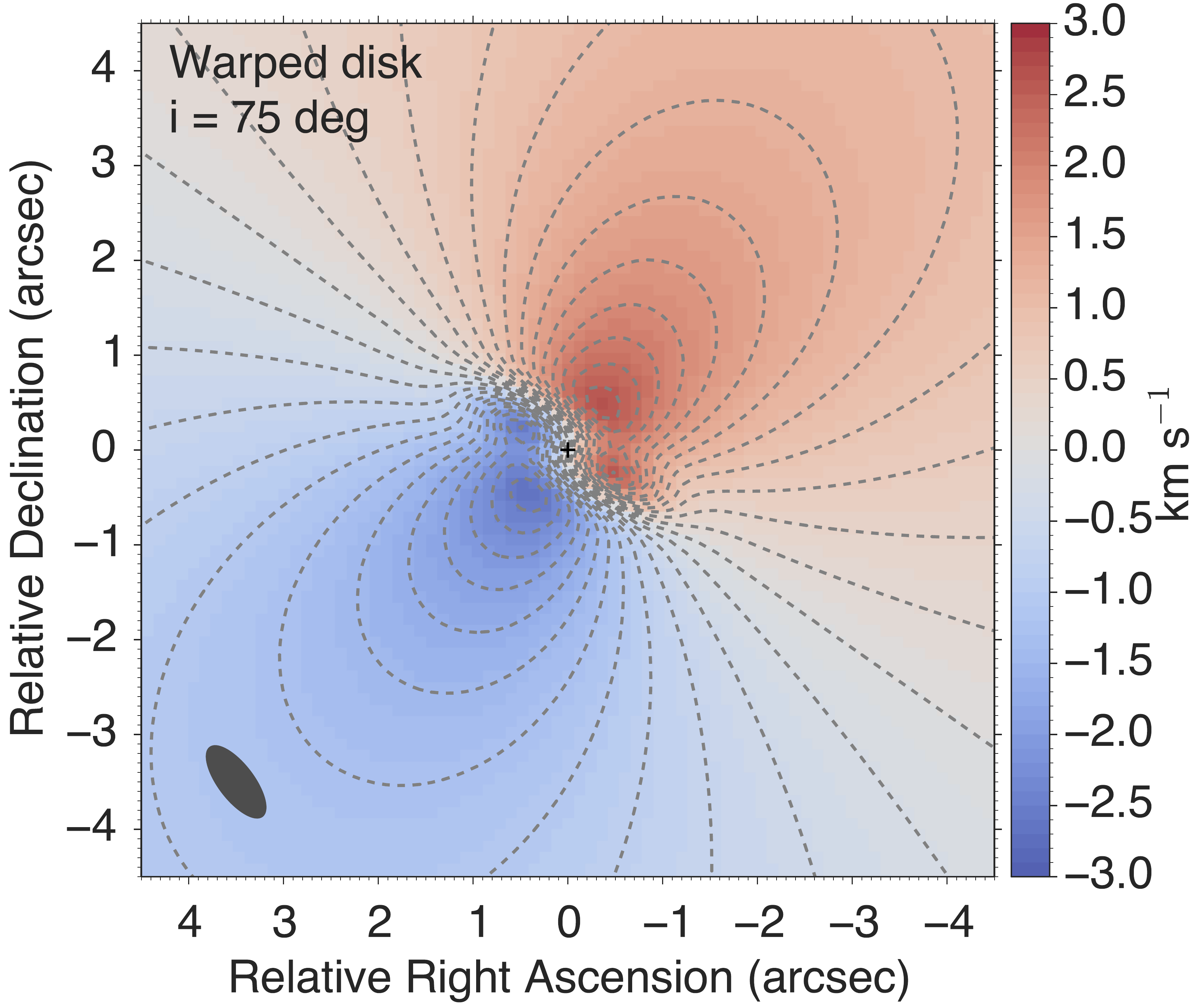}}
\subfigure{\includegraphics[width=0.33\textwidth]{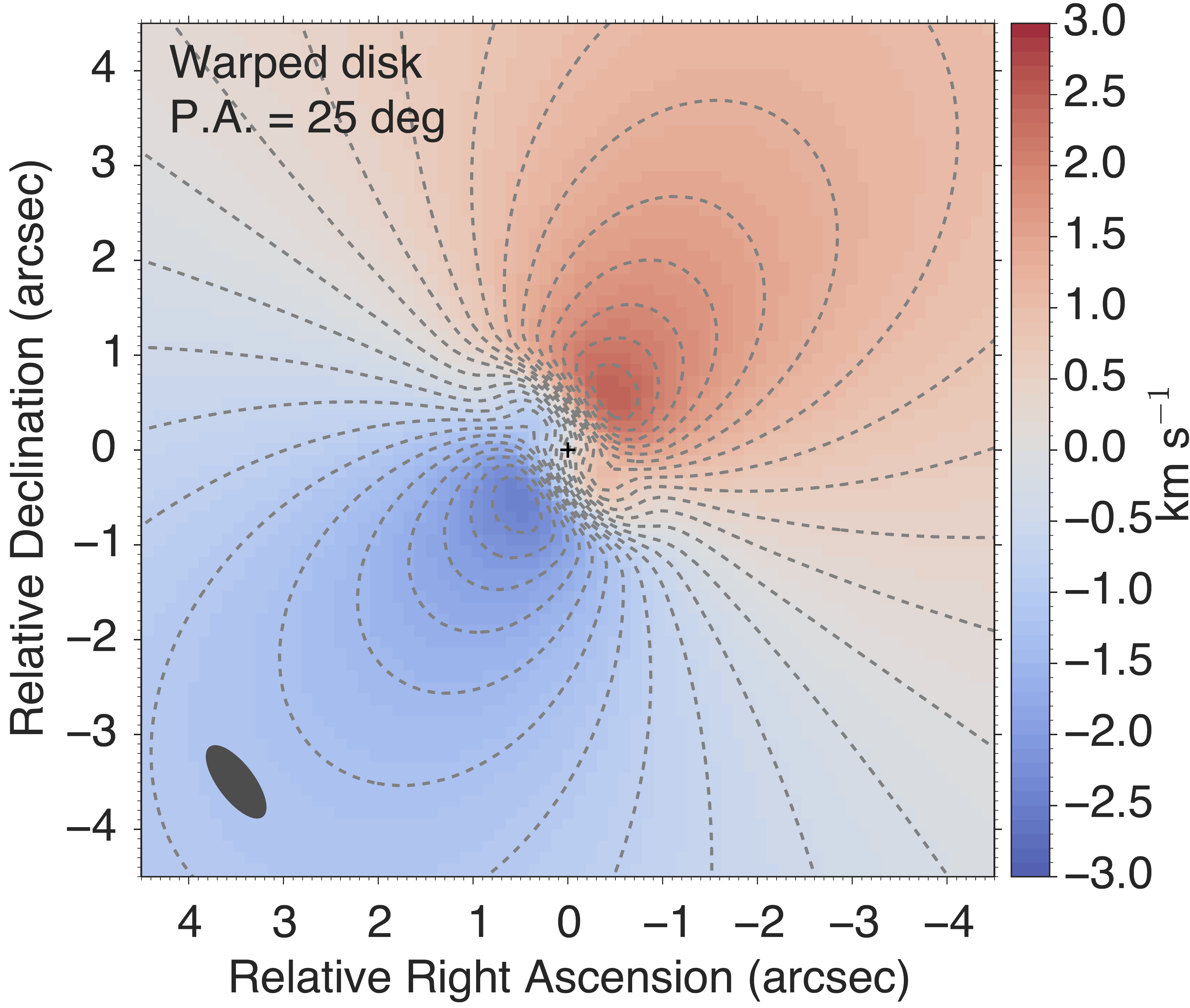}}
\subfigure{\includegraphics[width=0.33\textwidth]{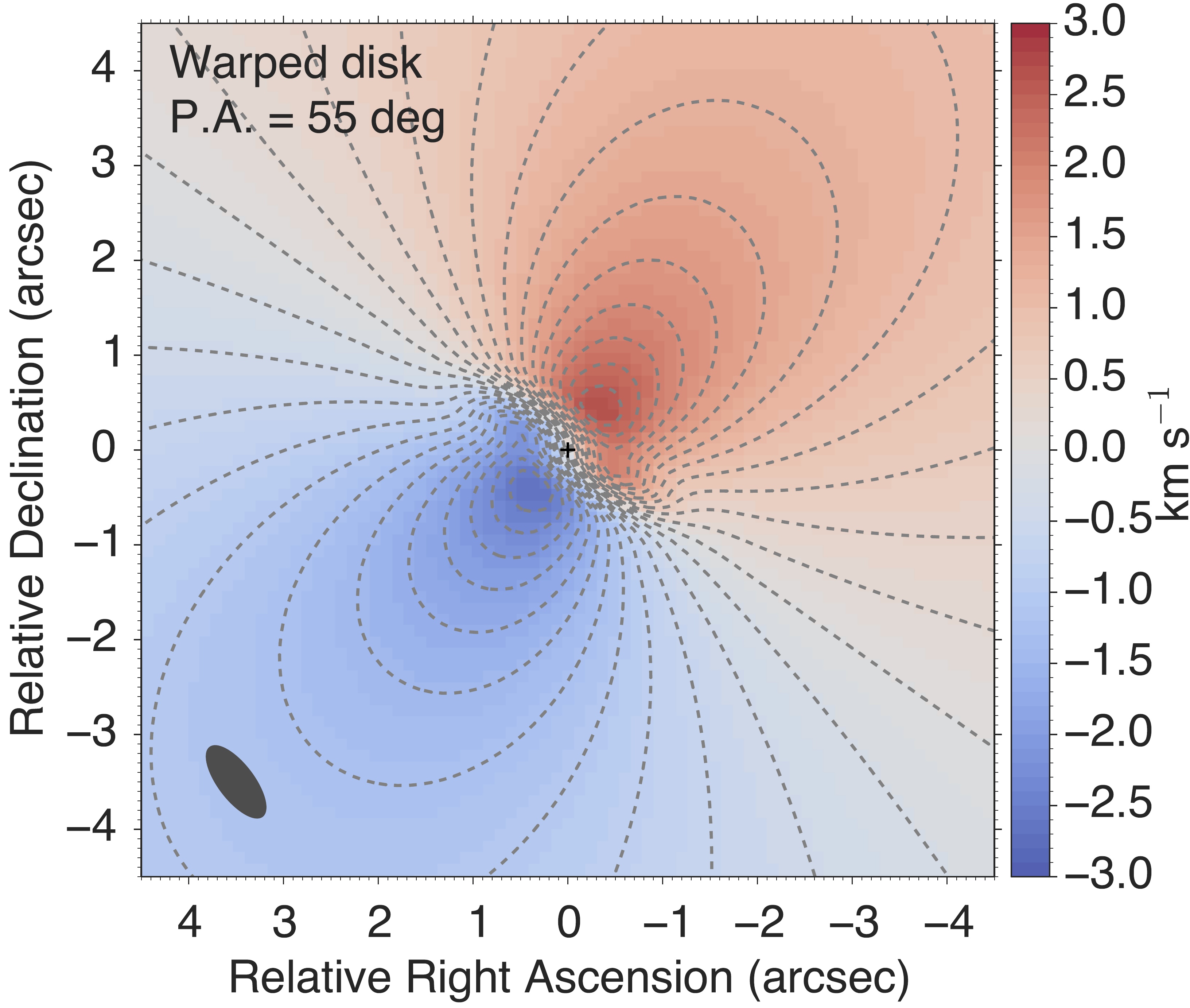}}
\subfigure{\includegraphics[width=0.33\textwidth]{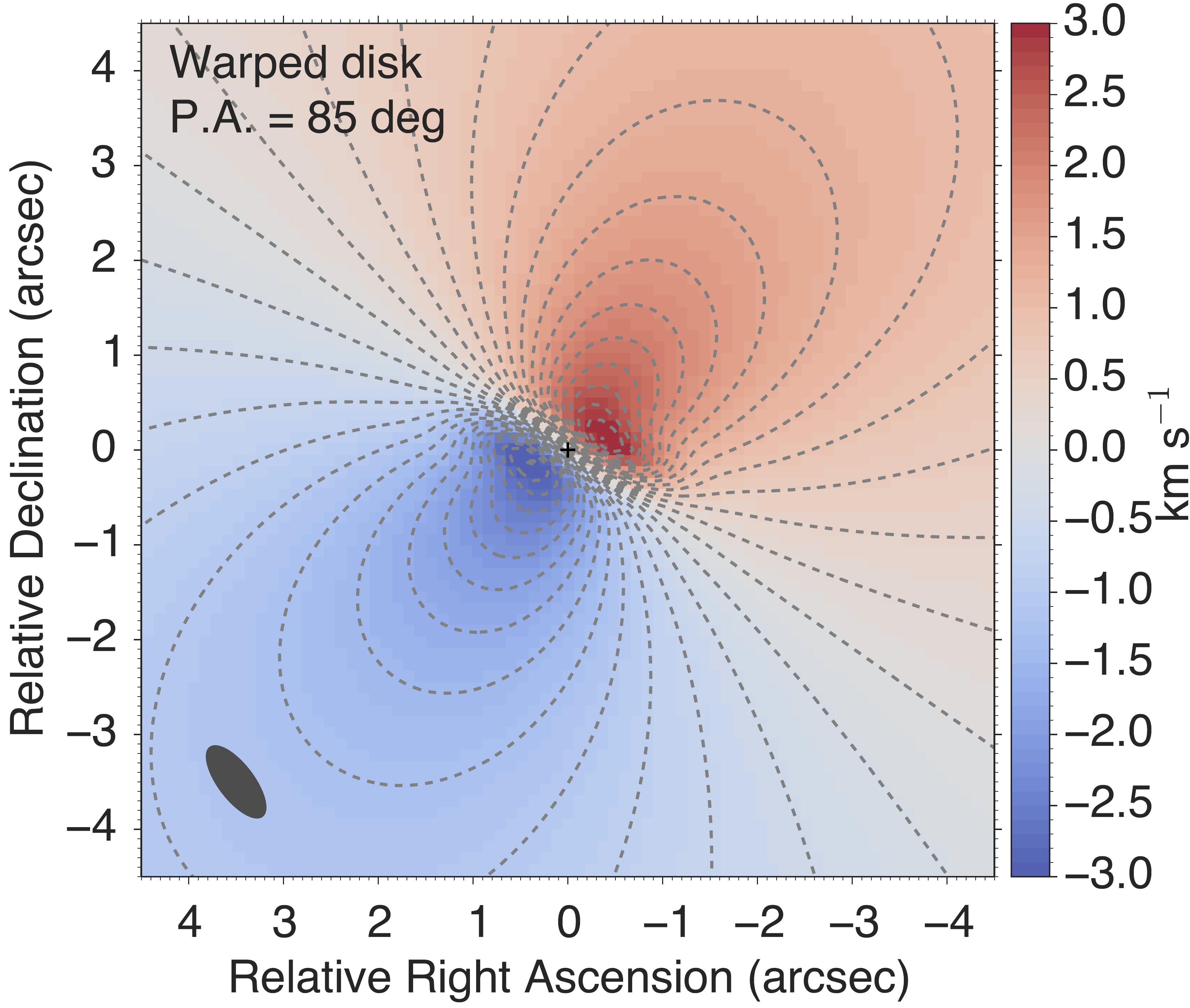}}
\caption{Model first moment maps for a warped disk with an outer disk P.A.~of 
145\degree, an inclination of 36\degree, and an opening angle of 9\degree.  
The top row of models have an inner warp with a fixed P.A.~of 55\degree~and 
and inclination of 25\degree~(left), 50\degree~(middle), and 75\degree.  
The bottom row of models have an inner warp with a fixed inclination of 80\degree~  
(representative of that in the HD~100546 disk) and a position angle of 25\degree~
(left), 55\degree~(middle), and 85\degree~(right).   
The transition radius in all models is 100~au. 
The contours are in units of a single spectral resolution element (0.21 \kms).} 
\label{figurea3}
\end{figure*}

\begin{figure*}[]
\subfigure{\includegraphics[width=0.33\textwidth]{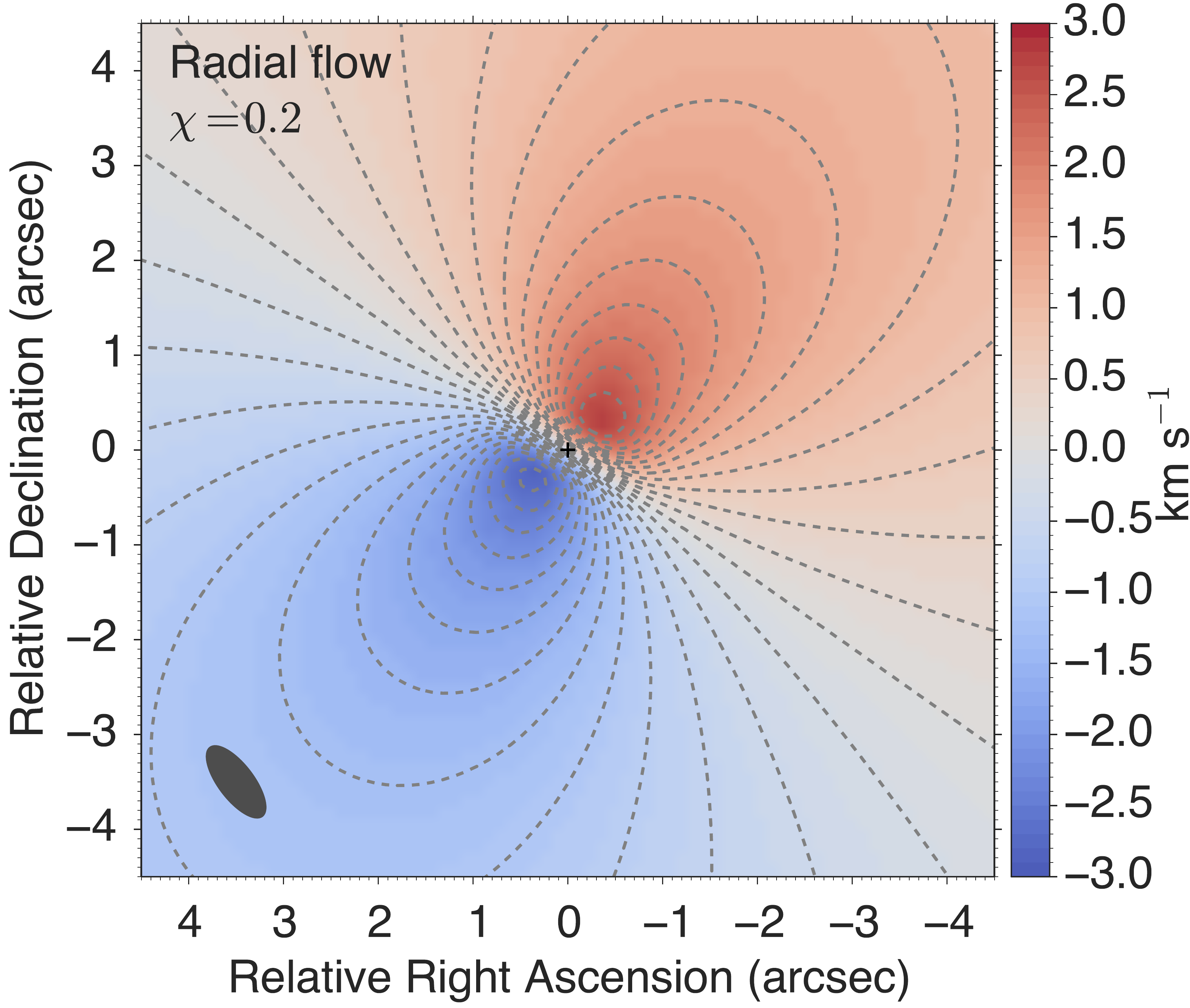}}
\subfigure{\includegraphics[width=0.33\textwidth]{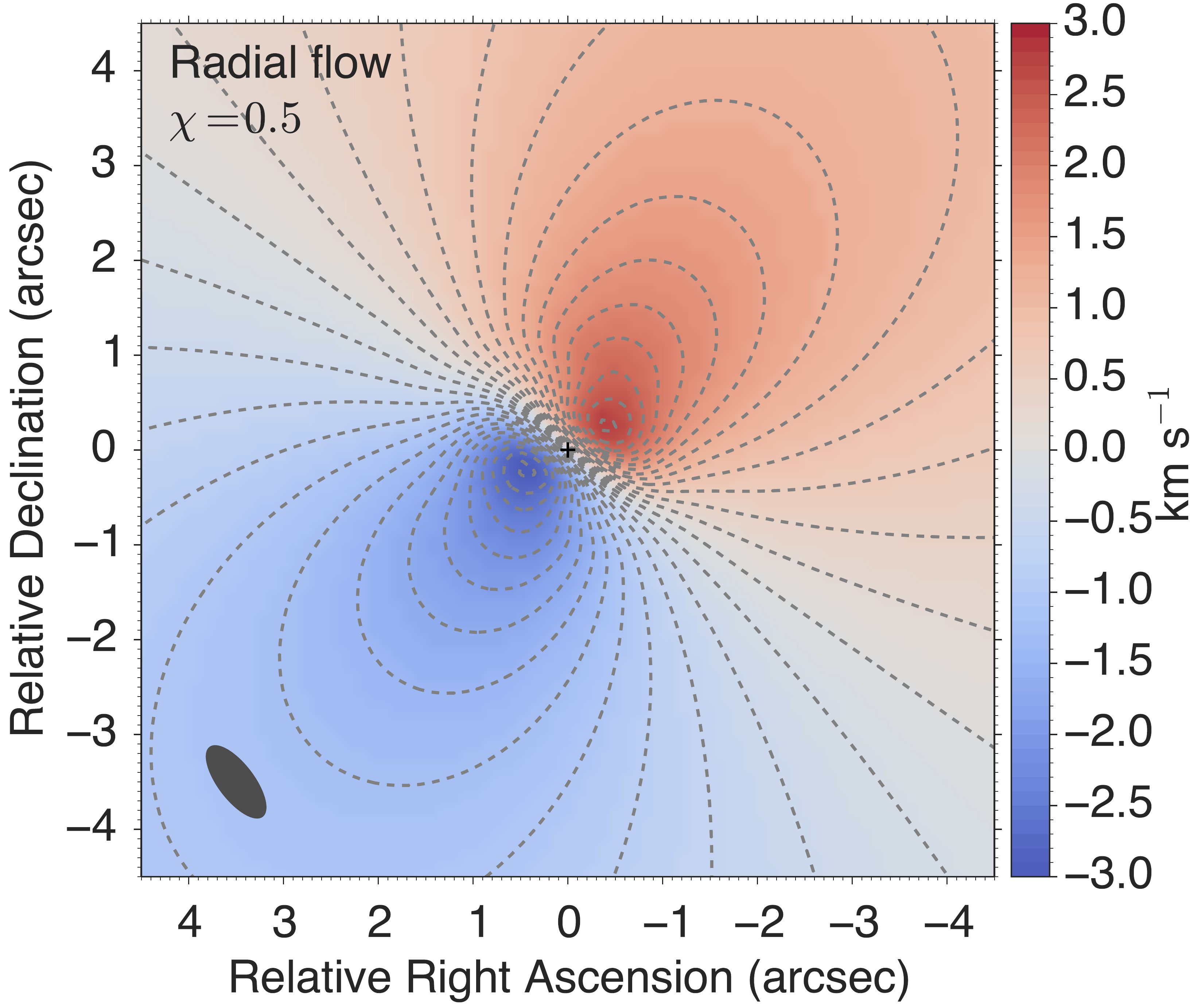}}
\subfigure{\includegraphics[width=0.33\textwidth]{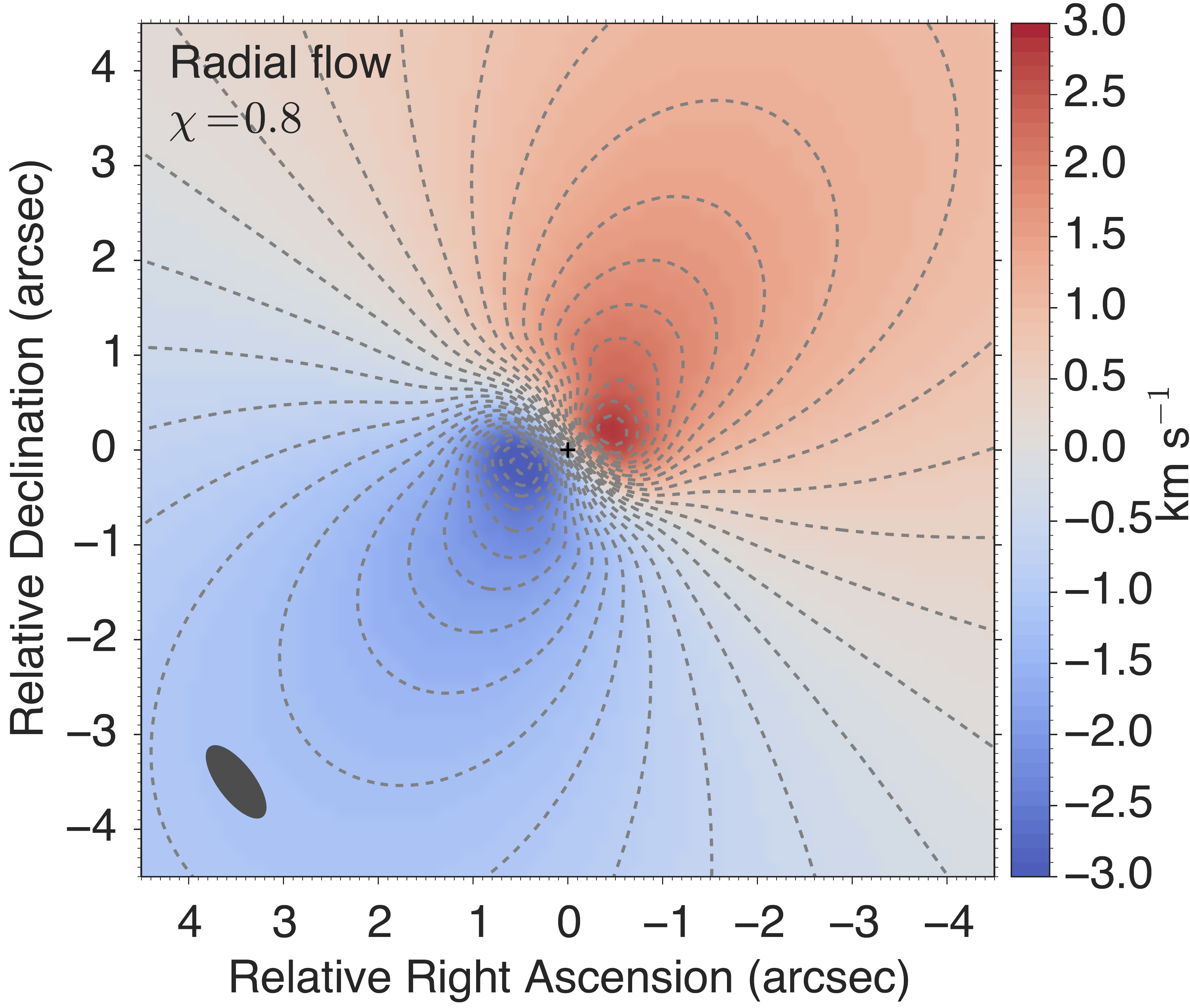}}
\caption{Model first moment maps for a disk with an outer disk P.A.~of 
145\degree, an inclination of 36\degree, and an opening angle of 9\degree~and which has a
radial flow component to the gas velocity 
with a scaling factor for the radial flow $\chi$, of 0.2 (left), 0.5 (middle), and 
0.8 (right).  
The transition radius in all models is 100~au. 
The contours are in units of a single spectral resolution element (0.21 \kms).} 
\label{figurea4}
\end{figure*}

Figure~\ref{figurea1} shows the first moment maps for a geometrically 
flat protoplanetary disk for a range of disk inclinations.  
Figure~\ref{figurea2} shows model first moment maps for a flared 
disk with a fixed inclination for a range opening angles, $\alpha$, of the emitting 
surface. 
Figure~\ref{figurea3} shows model first moment maps for a warped inner disk 
with a fixed transition radius of 100~au, and a range 
of warp inclinations and position angles.  
The outer disk parameters are those from the best-fit upper cone model 
(i.e., an inclination of 36\degree, a P.A.~of 145\degree, and an opening angle 
of 9\degree).
Figure \ref{figurea4} shows model first moment maps for 
a disk with a radial flow component to the velocity within a transition 
radius of 100~au, and for a range of velocity scaling factors, $\chi$ 
($v_r = \chi v_\mathrm{K}$).  
The outer disk parameters are fixed to those of the best-fit upper cone model.

\end{appendix}


\begin{thebibliography}{}
\bibitem[Acke \& van den Ancker(2006)]{acke06} Acke, B. \& van den Ancker, M.~E.~2006, \aap, 449, 267
\bibitem[Andrews(2015)]{andrews15} Andrews, S.~M. 2015, \pasp, 127, 961
\bibitem[Ardila et al.(2007)]{ardila07} Ardila, D.~R., Golimowski, D.~A., Krist, J.~E., et al.~2007, \apj, 665, 512
\bibitem[Boccaletti et al.(2013)]{boccaletti13} Boccaletti, A., Pantin, E., Lagrange, A.-M., et al.~2013, \aap, 560, A20
\bibitem[Booth et al.(2017)]{booth17} Booth, A., Walsh, C., Kama, M., et al.~2017, \aap, submitted
\bibitem[Brittain et al.(2009)]{brittain09} Brittain, S.~D., Najita, J.~R., \& Carr, J.~S. 2009, \apj, 702, 85
\bibitem[Brittain et al.(2014)]{brittain14} Brittain, S.~D., Carr, J.~S., Najita, J.~R., Quanz, S.~P., \& Meyer, M.~R. \apj, 791, 136
\bibitem[Bruderer et al.(2012)]{bruderer12} Bruderer, S., van~Dishoeck, E.~F., Doty, S.~D., \& Herczeg, G.~J. 2012, \aap, 541, 91
\bibitem[Bruderer(2013)]{bruderer13} Bruderer, S. 2013, \aap, 559, A46
\bibitem[Casassus et al.(2013)]{casassus13} Casassus, S., van der Plas, G., P\'{e}rez, S.~M., et al.~2013, Nature, 493, 191
\bibitem[Casassus et al.(2015)]{casassus15} Casassus, S., Marino, S., P\'{e}rez, S., et al.~2015, \apj, 811, 92
\bibitem[Christiaens et al.(2014)]{christiaens14} Christiaens, V., Casassus, S., Perez, S., van der Plas, G., \& M\'{e}nard, F. 2014, \apjl, 785,L12
\bibitem[Currie et al.(2015)]{currie15} Currie, T., Cloutier, R., Brittain, S., et al.~2015, \apjl, 814, L27
\bibitem[Debes et al.(2017)]{debes17} Debes, J.~H., Poteet, C.~A., Jang-Condell, H., et al.~2017, \apj, 835, 205
\bibitem[de Gregorio Monsalvo et al.(2013)]{degregorio13} de Gregorio Monsalvo, I., M\'{e}nard, F., Dent, W., et al.~2013, \aap, 557, 133
\bibitem[Dutrey et al.(2014)]{dutrey14a} Dutrey, A., Semenov, D., Chapillon, E., et al.~2014a, in Protostars and Planets VI, ed. H. Beuther et al.~(Tucson, AZ: Univ. Arizona Press), 317
\bibitem[Dutrey et al.(2014)]{dutrey14b} Dutrey, A., di~Folco, E., Guilloteau, S., et al.~2014b, Nature, 514, 600
\bibitem[Eistrup et al.(2016)]{eistrup16} Eistrup, C., Walsh, C., \& van Dishoeck, E.~F. 2016, \aap, 595, A83
\bibitem[Espaillat et al.(2014)]{espaillat14} Espaillat, C., Muzerolle, J., Najita, J., et al.~2014, in Protostars and Planets VI, ed. H. Beuther et al.~(Tucson, AZ: Univ. Arizona Press), 497
\bibitem[Facchini et al.(2014)]{facchini14} Facchini, S., Ricci, L., \& Lodato, G. 2014, \mnras, 442, 3700
\bibitem[Facchini et al.(2017)]{facchini17} Facchini, S., Juh\'{a}sz, A., \& Lodato, G. 2017, \mnras, in press (arXiv:1709.08369)
\bibitem[Follette et al.(2017)]{follette17} Follette, K.~B., Rameau, J., Dong, R., et al.~2017, \aj, 153, 264
\bibitem[Fukagawa et al.(2013)]{fukagawa13} Fukagawa, M., Tsukagoshi, T., Momose, M., et al.~2013, \pasj, 65, L14
\bibitem[Furuya \& Aikawa(2014)]{furuya14} Furuya, K. \& Aikawa, Y. 2014, \apj, 790, 97
\bibitem[Gaia Collaboration(2016a)]{gaia16a} Gaia Collaboration (Prusti, T., et al.)~2016a, \aap, 595, A1
\bibitem[Gaia Collaboration(2016b)]{gaia16b} Gaia Collaboration (Brown, A.~G.~A., et al.)~2016b, \aap, 595, A2
\bibitem[Garufi et al.(2016)]{garufi16} Garufi, A., Quanz, S.~P., Schmid, H.~M., et al. 2016, \aap, 588, A8 
\bibitem[Grady et al.(2001)]{grady01} Grady, C.~A., Polomski, E.~F., Henning, Th., et al.~2001, \aj, 122, 3396  
\bibitem[Grady et al.(2015)]{grady15} Grady, C., Fukagawa, M., Maruta, Y., et al.~2015, \apss, 355, 253
\bibitem[Helling et al.(2014)]{helling14} Helling, C., Woitke, P., Kamp, I., et al.~2014, Life, 4, 142
\bibitem[Juh\'{a}sz \& Facchini(2017)]{juhasz17} Juh\'{a}sz, A. \& Facchini, S. 2017, \mnras, 466, 4053
\bibitem[Kama et al.(2016)]{kama16} Kama, M., Bruderer, S., van~Dishoeck, E.~F., et al.2016, \aap, 592, A83
\bibitem[Lai(2014)]{lai14} Lai, D. 2014, \mnras, 440, 3532
\bibitem[Lazareff et al.(2017)]{lazareff17} Lazareff, B., Berger, J.-P., Kluska, J., et al. 2017, \aap, 599, A85
\bibitem[Loomis et al.(2017)]{loomis17} Loomis, R.~A., \"{O}berg, K.~I., Andrews, S.~M., \& MacGregor, M.~A. 2017, \apj, 840, 23
\bibitem[Marino et al.(2015)]{marino15} Marino, S., P\'{e}rez, S., \& Casassus, S. 2015, \apjl,  798, L44 
\bibitem[Meeus et al.(2001)]{meeus01} Meeus, G., Waters, L.~B.~F.~M., Bouwman, J., et al.~2001, \aap, 365, 476 
\bibitem[Mendigut\'{i}a et al.(2015)]{mendigutia15} Mendigut\'{i}a, I., de Wit, W.~J., Oudmaijer, R., et al.~2015, \mnras, 453, 2126
\bibitem[Mendigut\'{i}a et al.(2017)]{mendigutia17} Mendigut\'{i}a, I., Oudmaijer, R., Garufi, A., et al.~2017, \aap, in press
\bibitem[Miotello et al.(2014)]{miotello14} Miotello, A., Bruderer, S., \& van~Dishoeck, E.~F.~2014, \aap, 572, A96
\bibitem[Miotello et al.(2016)]{miotello16} Miotello, A., van~Dishoeck, E.~F., Kama, M., \& Bruderer, S.~2016, \aap, 594, A85 
\bibitem[Mulders et al.(2013)]{mulders13} Mulders, G.~D., Paardekooper, S.-J., Pani\'{c}, O., et al. 2013, 557, 68
\bibitem[Murillo et al.(2013)]{murillo13} Murillo, N.~M., Lai, S.-P., Bruderer, S., Harsono, D., \& van~Dishoeck, E. F.~2013, \aap, 560, A103
\bibitem[Nomura et al.(2016)]{nomura16} Nomura, H., Tsukagoshi, T., Kawabe, R., et al.~2016, \apjl, 819, L7
\bibitem[Pani\'{c} et al.(2010)]{panic10} Pani\'{c}, O., van~Dishoeck, E.~F., Hogerheijde, M.~R., et al.~2010, \aap, 519, A110
\bibitem[Pani\'{c} et al.(2014)]{panic14} Pani\'{c}, O., Ratzka, Th., Mulders, G.~D., et al. 2014, 562, A101
\bibitem[Perez et al.(2015)]{perez15} Perez, S., Dunhill, A., Casassus, S., et al.~2015, \apj, 811, L5
\bibitem[Pineda et al.(2014)]{pineda14} Pineda, J.~E., Quanz, S.~P., Meru, F., et al.~2014, \apjl, 788, L34  
\bibitem[Pinilla et al.(2015)]{pinilla15} Pinilla, P., Birnstiel, T., \& Walsh, C.~2015, \aap, 580, A105
\bibitem[van der Plas et al.(2009)]{vanderplas09} van~der~Plas, G., van de Ancker, M. E., Acke, B., et al. 2009, \aap, 500, 1137 
\bibitem[van der Plas et al.(2017)]{vanderplas17}  van~der~Plas, G., Wright, C.~M., M\'{e}nard, F., et al.~2017, \aap, 597, A32
\bibitem[Quanz et al.(2013)]{quanz13} Quanz, S.~P., Amara, A., Meyer, M.~R., Kenworthy, M.~A., Kasper, M., \& Girard, J.~A.~2013, \apjl, 766, L1
\bibitem[Quanz et al.(2015)]{quanz15} Quanz, S.~P., Amara, A., Meyer, M.~R., et al.~2015, \apj, 807, 64   
\bibitem[Rameau et al.(2017)]{rameau17} Rameau, J., Follette, K.~B., Pueyo, L., et al.~2017, \aj, 153, 244
\bibitem[Reboussin et al.(2015)]{reboussin15} Reboussin, L., Wakelam, V., Guilloteau, S., Hersant, F., \& Dutrey, A.~2015, \aap, 579, A82
\bibitem[Rosenfeld et al.(2012)]{rosenfeld12} Rosenfeld, K.~A., Qi, C., Andrews, S.~M., et al.~2012, \apj, 757, 129
\bibitem[Rosenfeld et al.(2013)]{rosenfeld13} Rosenfeld, K.~A., Andrews, S.~M., Hughes, A.~M., Wilner, D.~J., \& Qi, C.~2012, \apj, 774, 16
\bibitem[Rosenfeld et al.(2014)]{rosenfeld14} Rosenfeld, K.~A., Chiang, E., \& Andrews, S.~M. 2014, \apj, 782, 62
\bibitem[Schwarz et al.(2016)]{schwarz16} Schwarz, K.~R., Bergin, E.~A., Cleeves, L.~I., et al.~\apj, 823, 91
\bibitem[Semenov et al.(2008)]{semenov08} Semenov, D., Pavlyuchenkov, Ya., Henning, Th., Wolf, S., \& Launhardt, R. 2008, \apjl, 673, L195  
\bibitem[Sicilia-Aguilar et al.(2016)]{sicilia-aguilar16} Sicilia-Aguilar, A., Banzatti, A., Carmona, A., et al.~2016, \pasa, 33, 59
\bibitem[Simon et al.(2000)]{simon00} Simon, M., Dutrey, A., \& Guilloteau, S. 2000, \apj, 545, 1034
\bibitem[Tang et al.(2012)]{tang12} Tang, Y.-W., Guilloteau, S., Pi\'{e}tu, V., et al. 2012, \aap, 547, A84
\bibitem[van den Ancker et al.(1998)]{vandenancker98} van~den~Ancker, M.~E., de~Winter, D., Tijn A Djie, H.~R.~E., \aap, 330, 145
\bibitem[Visser et al.(2009)]{visser09} Visser, R., van~Dishoeck, E.~F., \& Black, J.~H. 2009, \aap, 503, 323
\bibitem[Williams \& Cieza(2011)]{williams11} Williams, J. P. \& Cieza, L.~A. 2011, \araa, 49, 67
\bibitem[Walsh et al.(2014)]{walsh14} Walsh, C., Juh\'{a}sz, A., Pinilla, P., et al.~2014, \apjl, 791, L6
\bibitem[Walsh et al.(2015)]{walsh15} Walsh, C., Nomura, H., \& van~Dishoeck, E.~F. 2015, \aap, 582, A88
\bibitem[Walsh et al.(2016)]{walsh16} Walsh. C., Juh\'{a}sz, A., Meeus, G., et al.~2016, \apj, 831, 200
\bibitem[Wright et a.(2015)]{wright15} Wright, C.~M., Maddison, S.~T., Wilner, D.~J., et al. 2015, \mnras, 453, 414
\bibitem[Yu et al.(2016)]{yu16} Yu, M., Willacy, K., Dodson-Robinson, S.~E., Turner, N.~J., \& Evans II, N.~J. 2016, \apj, 822, 53  
\bibitem[Yu et al.(2017)]{yu17} Yu, M., Evans II, N.~J., Dodson-Robinson, S.~E., Willacy, K., \& Turner, N.~J. 2017, \apj, 841, 39
\bibitem[Zhang et al.(2017)]{zhang17} Zhang, K., Bergin, E.~A., Blake, G.~A., Cleeves, L.~I., \& Schwarz, K.~R.~2017, Nature Astronomy, 1, 130
\end{thebibliography}
\end{document}